\documentclass[10pt,letterpaper]{article}
\usepackage[top=0.85in,left=1.16in,right=1.16in, bottom= 0.85in, footskip=0.4in]{geometry}

\usepackage{newfloat}
\DeclareFloatingEnvironment[name={Supplementary Figure},fileext=lsf,listname={List of Supplementary Figures}]{suppfigure}
\usepackage{amsmath,amssymb}
\usepackage{csquotes} % provides the \enquote command

\usepackage{changepage}

\usepackage[utf8]{inputenc}

\usepackage{textcomp,marvosym}

\usepackage{nameref,hyperref}

\usepackage[right]{lineno}

\usepackage{microtype}
\DisableLigatures[f]{encoding = *, family = * }

\usepackage[table]{xcolor}
\usepackage{float}

\usepackage{lineno}
\usepackage[aboveskip=1pt,labelfont=bf,labelsep=period,singlelinecheck=off]{caption}
\usepackage[labelfont={bf,large}, labelformat = simple]{subcaption} % applies options to all figure, tables etc.
\usepackage[
style=plos2015,
sortcites,
sorting=none,
url=false,
isbn=false,
doi=true,
eprint=false,
date=year,
citestyle=numeric-comp,
bibstyle=numeric-comp,
  giveninits   = true,
  hyperref     = true,
]{biblatex}
\bibliography{main}
\AtEveryBibitem{
 \clearlist{language}
 \clearlist{editor}
 \clearfield{note}
}

\makeatletter
\renewcommand{\@biblabel}[1]{\quad#1.}
\makeatother

\renewbibmacro{in:}{%
  \ifentrytype{article}{}{\printtext{\bibstring{in}\intitlepunct}}}

\newcommand{\beginsupplement}{%
 \setcounter{table}{0}
 \renewcommand{\thetable}{S\arabic{table}}%
 \setcounter{figure}{0}
 \renewcommand{\thefigure}{S\arabic{figure}}%
}

\usepackage{graphicx}
\usepackage{authblk}
\usepackage{pdfpages} % include sm cover page
\makeatletter
\AtBeginDocument{\let\LS@rot\@undefined}
\makeatother
\usepackage{multicol}

\usepackage{calc}

\usepackage{graphbox} % allows to add keys to \includegraphics
\usepackage{xcolor}
\usepackage{graphicx}
\usepackage{tcolorbox}
\tcbuselibrary{skins, theorems}
\usepackage{nameref}            % only needed if you use \nameref

\definecolor{boxTitle}{HTML}{fff79a}
\definecolor{boxBackground}{HTML}{fffce0}
\definecolor{boxFrame}{HTML}{f1e2b8}

\tcbset{my box/.style={
    enhanced, fonttitle=\bfseries,
    colback=boxBackground, colframe=boxFrame,
    coltitle=black, colbacktitle=boxTitle,
    attach boxed title to top left={xshift=0.3cm,
                                    yshift*=-\tcboxedtitleheight/2},
    boxed title style={
    }
  }
}

\newtcolorbox[auto counter, number within=section]{infobox}[2][]{%
  my box, title={Infobox~\thetcbcounter: #2}, #1}

\title{Signatures of hierarchical temporal processing\\ in the mouse visual system} %\\
\author[1,*]{Lucas Rudelt}
\author[1]{Daniel Gonz\'alez Marx}
\author[1]{F.~Paul Spitzner}
\author[2]{Benjamin Cramer} % benjamins old mail probably not working anymore: benjamincramer@web.de
\author[1]{Johannes Zierenberg}
\author[1,3,4,*]{Viola Priesemann}
\affil[1]{\small Max Planck Institute for Dynamics and Self-Organization, Göttingen, Germany}
\affil[2]{Kirchhoff-Institute for Physics, Heidelberg University, Germany}
\affil[3]{Institute for the Dynamics of Complex Systems, University of Göttingen, Germany}
\affil[4]{Bernstein Center for Computational Neuroscience (BCCN), Göttingen, Germany}
\affil[*]{\small Corresponding authors: lucas.rudelt@ds.mpg.de, viola.priesemann@ds.mpg.de}

\date{}

\begin{document}

\maketitle
\renewcommand{\thefootnote}{\arabic{footnote}}
% ABSTRACT
% word limit for abstract in Science Advances is 150 words
\begin{abstract}
\noindent
A core challenge for the brain is to process information across various timescales.
This could be achieved by a hierarchical organization of temporal processing through intrinsic mechanisms (e.g., recurrent coupling or adaptation), but recent evidence from spike recordings of the rodent visual system seems to conflict with this hypothesis.
Here, we used an optimized information-theoretic and classical autocorrelation analysis to show that information- and intrinsic timescales of spiking activity increase along the anatomical hierarchy of the mouse visual system, while information-theoretic predictability decreases.
Moreover, the timescale hierarchy was invariant to the stimulus condition, whereas the decrease in predictability was strongest under natural movie stimulation. We could reproduce this effect in a basic recurrent network model with correlated sensory input.
Our findings suggest that the rodent visual system indeed employs intrinsic mechanisms to achieve longer integration for higher cortical areas, while simultaneously reducing predictability for an efficient neural code. \\

% \vspace{.5cm}
\end{abstract}

% \textbf{Keywords}\\
% hierarchy of timescales,
% intrinsic timescale,
% autocorrelation,
% information theory,
% mutual information,
% predictability,
% redundancy reduction,
% efficient coding,
% mouse visual cortex,

% ethics statement, copied from siegle
% Mice were maintained in the Allen Institute for Brain Science animal facility and used in accordance with protocols approved by the Allen Institute's Institutional Animal Care and Use Committee.

% code repo
% https://github.com/Priesemann-Group/mouse_visual_timescales

% data repo
% https://gin.g-node.org/pspitzner/mouse_visual_timescales

% \section*{Teaser}
% Information-theoretic and autocorrelation analyses suggest a hierarchy of integration and adaptation in mouse visual cortex.

% \newpage

\section*{Introduction}
The brain has the ability to seamlessly process and integrate information on vastly different timescales.
In primates, past work suggested that this may be supported by two essential features of neocortex:
the highly recurrent architecture of cortical networks~\cite{chaudhuri_2015}, and an organization of different cortical areas into a temporal processing hierarchy~\cite{kiebel2008hierarchy, hasson2015hierarchical,murray_2014}.
It was found that early sensory areas specialize on fast processing of sensory inputs~\cite{buracas1998efficient, uchida2003speed, yang2008millisecond}, whereas higher (transmodal) areas perform temporal processing with long timescales --- combining new information with past information that is maintained over extended periods~\cite{hasson_2008, honey2012slow}.

This hierarchy is reflected by an increase in the \emph{intrinsic timescale} of neural activity, as measured by the decay rate of autocorrelation~\cite{murray_2014, gao2020neuronal, raut2020hierarchical, spitmaan2020multiple}.
In addition, intrinsic timescales were found to be indicative of the specialization for behaviorally relevant computations~\cite{cavanagh_2018, wasmuht_2018, wilting_2018, manea2023neural}.
Finally, there exist gradients in intra-areal properties across the anatomical cortical hierarchy, all of which point to a stronger recurrent coupling for areas specialized on long timescales~\cite{huntenburg2018large, wang_2020}.
In particular, higher cortical areas have an increased dendritic spine density for pyramidal neurons~\cite{elston2007specialization, chaudhuri_2015}, overall excitation-inhibition ratio~\cite{wang2020macroscopic}, the expression of related receptor genes~\cite{gao2020neuronal, burt2018hierarchy}, gray matter myelination~\cite{gao2020neuronal, glasser2011mapping}, and the strength of functional connectivity~\cite{raut2020hierarchical, wasmuht_2018, hart2020recurrent, safavi2018nonmonotonic}.
From modelling studies, a stronger recurrence is known to enable stronger and longer-lasting activity fluctuations~\cite{wilting_2018, wilting_2019, zeraati2023intrinsic, wilting_2019a}, consistent with the increase of timescales for higher areas.
Overall, this led to the understanding that in primates, temporal processing is organized hierarchically~\cite{hasson2015hierarchical}, and that specializations along that hierarchy are likely governed by differences in recurrent coupling~\cite{mauk2004neural, wilting_2019a}.
Yet, it is still open how a temporal hierarchy shaped by recurrence fulfills requirements of neural coding and information processing, and whether it manifests as a general organization principle in mammals.

Here, we investigate mouse cortex, because it was found to differ from primate cortical organization:
First, there seems to be no clear \emph{global} processing hierarchy from sensory to transmodal areas~\cite{huntenburg2021gradients}, as more of mouse cortex is used for sensory processing of specific modalities, instead of integrating information across modalities, or across longer timescales for cognitive processing.
Second, although there exists evidence for gradients in interneuron numbers and intra-cortical connectivity from sensory to transmodal areas~\cite{fulcher2019multimodal}, the degree of interareal variation of microstructural properties in mice~\cite{ballesteros2006density, benavides2006dendritic} is far less pronounced than in the highly differentiated primate cortex~\cite{elston2014pyramidal, gilman2017area, hsu2017comparative, luebke2017pyramidal}.

Yet, within \emph{specific} sensory pathways (e.g., visual), mice exhibit hierarchical feedforward–feedback projection-patterns~\cite{harris2019hierarchical}, which are paralleled by differences in the recruitment of inhibitory and excitatory neurons~\cite{d2016recruitment}, and a functional hierarchy that follows the anatomical hierarchy~\cite{siegle_2021}.
In addition, the rat analog of the ventral stream shows a hierarchy of temporal scales, with higher areas encoding visual information more persistently~\cite{piasini2021temporal}.
Finally, population codes vary between association and sensory cortices~\cite{runyan2017distinct}, and impairments to cortical frontal (transmodal) areas have a greater impact on evidence accumulation over long timescales than posterior (sensory) areas, which also exhibit shorter activity timescales during evidence accumulation~\cite{pinto2022multiple}.
This raises the question whether a temporal hierarchy shaped by recurrence also characterizes mice, or whether the strong focus on sensory processing requires a different, coding-optimized organization altogether.

To address this question, it is important to note that a coding perspective entails a similar trade-off as the temporal-processing perspective (long integration vs.~fast relay):
In order to increase the signal-to-noise ratio, \emph{robust coding} requires integrating information over time~\cite{atick_1992}, whereas, for low noise, \emph{efficient coding} of sensory information requires temporal decorrelation to reduce redundancies~\cite{barlow_2012, rieke1999spikes, pozzorini_2013}.
This trade-off can be characterized using the \emph{predictability} $R$, which quantifies the proportion of information in current neural spiking that can be predicted from the recent past~\cite{rudelt_2021}.
This predictable information reflects temporal redundancy, and facilitates, for instance, active information storage (maintaining past input to combine it with present input~\cite{lizier_2013, wibral_2014,wibral_2015}) and associative learning~\cite{barlow_2001}.
In addition, the closely related \emph{information timescale} $\tau_R$ gives the timescale over which past information has to be integrated for prediction. 
Together, predictability and information timescales provide a broad view into the neural code, quantifying both, the amount and the timescale of redundancy in neural spiking.

Here, we will use the information timescale, as well as the intrinsic timescale, to probe for hierarchical temporal processing in mouse visual cortex.
Moreover, we will test whether higher cortical areas show an increase in predictability, in line with robust coding, or a decrease in predictability, in line with efficient coding. 
Finally, we will compare results between spontaneous activity and natural stimuli, which is important to distinguish between stimulus-induced and intrinsically generated timescales and predictability.
Together, these results will clarify whether mouse visual cortex shows signatures of hierarchical temporal processing, and whether these are stimulus-induced, indicating a stronger role of feedforward processing, or rather intrinsically generated, indicating a stronger role of recurrent processing. 

% To test these hypotheses, we estimate predictability, as well as intrinsic and information timescales in highly parallel spike recordings of mouse visual cortex~\cite{siegle_2021}. In particular, we compare timescales and predictability between areas along the cortical hierarchy, and between stimulation with a natural movie and spontaneous activity, as well as visual selectivity measures.
% Finally, to see whether differences between areas could be explained by an increase in inter-areal recurrent coupling, or rather differences in sensory inputs to these areas, we compute timescales and predictability for a recurrent network model with temporally correlated external inputs.
% With this we aim to understand whether hierarchical temporal processing through variations in inter-areal recurrent coupling is a more general principle, which even extends to specific sensory processing streams in mouse visual cortex.

\section*{Results}
\begin{figure*}[t!]
    % Plot this with an adaptation of the original plot
    % One could add a plot here for the definition of tau_R and how T_0 works
    \centerline{\includegraphics[]{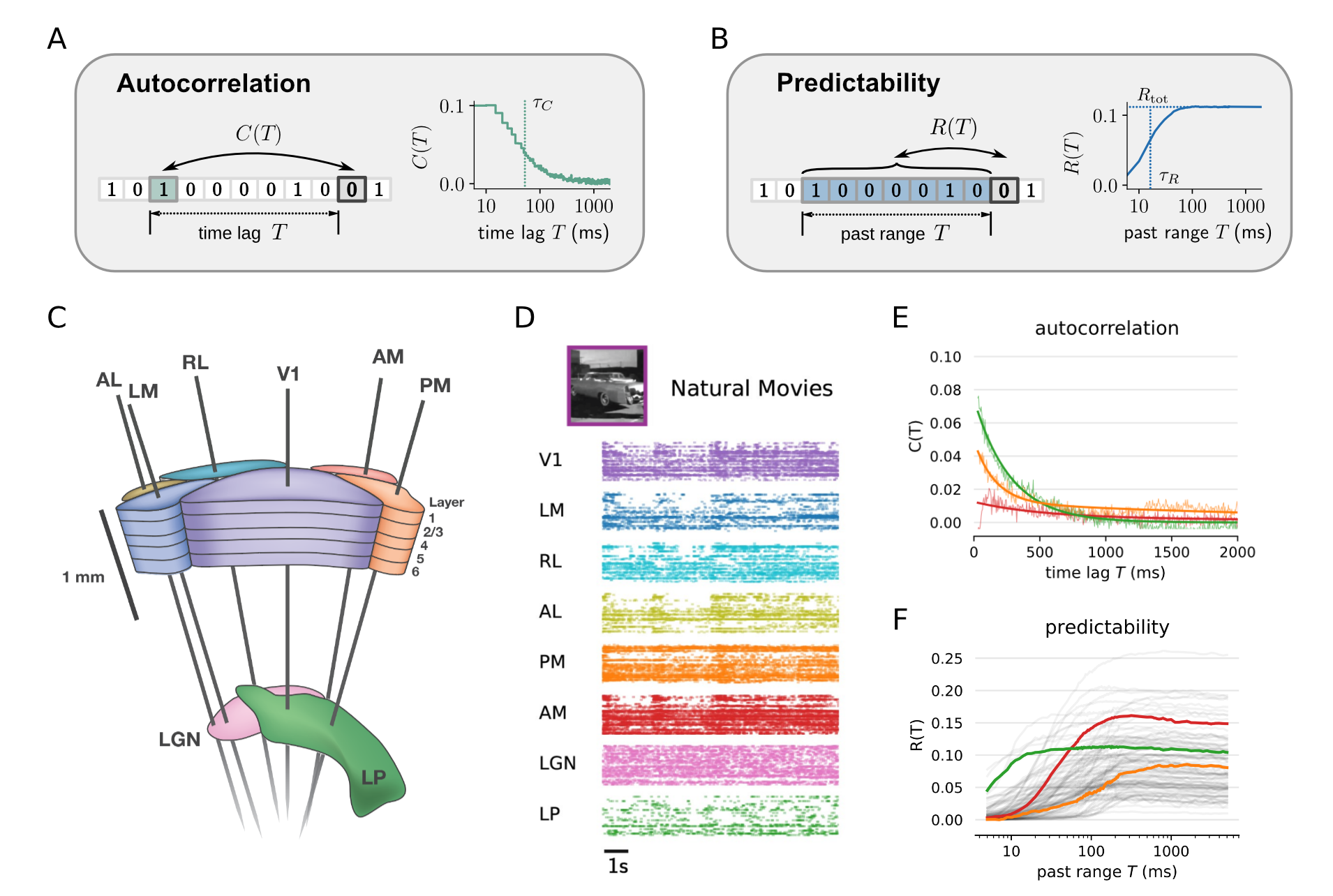}}
    \caption{\textbf{Timescales and predictability of spiking activity in the mouse visual system.} 
    %\textbf{Signatures of temporal processing in the mouse visual system.} 
    \textbf{(A)} Autocorrelation $C(T)$ quantifies the correlation between a neuron's spiking activity in two time bins with time lag $T$. 
    Typically, the measured autocorrelation (solid green line) is decaying exponentially with lag $T$ on a characteristic timescale $\tau_C$ (dashed line), termed intrinsic timescale. 
    %This timescale can be estimated from sufficiently long recordings by fitting an exponentially decaying function (solid line, dashed line indicates estimated $\tau_C$).
    \textbf{(B)} Predictability $R(T)$ gives the relative amount of spiking information in a time bin that can be predicted from spikes in all time bins in a past range $T$.
    $R(T)$ increases monotonously with $T$ until it saturates when reaching the neuron's total predictability $R_{\text{tot}}$ (horizontal dashed line), i.e., when adding past bins does not yield additional information.
    The information timescale $\tau_R$ is defined as the characteristic rise time until $R_{\text{tot}}$ is reached (vertical dashed line, c.f.~main text).
    % Again, $\tau_R$ was computed using only past ranges $T$ larger $T_{\mathrm{min}} = 30\,\mathrm{ms}$ (grey dashed line) to exclude short effects like refractoriness. 
    \textbf{(C)}~Neuropixel-electrodes simultaneously record from up to six visual cortical areas (V1, LM, RL, AL, PM, AM) and two thalamic areas (LGN and LP) (image credit: Allen Institute~\cite{allen_github_url}). 
    \textbf{(D)}~Example of spiking activity for a random subset of units from different brain areas during stimulation with a natural movie. 
    \textbf{(E)}~Examples of $C(T)$ for LP, PM and AM (green, orange, red, respectively). $\tau_C$ is estimated by fitting $C(T)$ (thin lines) with an exponential decay (thick lines).
    %The intrinsic timescale $\tau_C$ is estimated by fitting the recorded autocorrelation $C(T)$ (thin lines) with an exponentially decaying function (thick lines), here shown for three example units from LP (green), PM (orange) and AM (red). 
    \textbf{(F)} Examples of $R(T)$, same units as in E.}
    % Since $R(T)$ only increases for non-redundant past information, the $\tau_R$ is almost an order of magnitude smaller than $\tau_C$ (main text and \cite{rudelt_2021}).
    % \caption{\textbf{Single neuron autocorrelation and predictability yield complementary signatures of temporal processing.} \textbf{(A)}~ \textbf{(B)}~
    \label{fig:measures}
\end{figure*}

% General approach and data set
To identify systematic differences in temporal processing between sensory processing stages, we analyzed a dataset from simultaneous Neuropixels recordings of the mouse visual system in vivo~\cite{deVries2020large, siegle_2021} (Fig.~\ref{fig:measures}C).
This dataset contains spike trains collected from $n=57$ experimental sessions in adult mice under different stimulus conditions, and for thousands of neurons from six cortical areas [primary visual cortex (V1), lateromedial area (LM), anterolateral area (AL), rostrolateral area (RL), anteromedial area (AM) and posteromedial area (PM)] and two thalamic areas [lateral geniculate nucleus (LGN) and lateral posterior nucleus (LP)].
We focused on a stimulus condition from the \textit{Functional Connectivity} experimental sessions, where a (repeated) natural movie was used as stimulus (Fig.~\ref{fig:measures}D; see~\nameref{sec:methods} for details on the recording and data selection).

% \subsection*{Signatures of temporal processing}
\subsection*{Timescales and predictability of neural spiking activity}
% \footnotetext{\url{https://github.com/AllenInstitute/neuropixels_platform_paper}}
To analyze signatures of temporal processing in the recorded spike trains, we quantify single neuron autocorrelation and predictability (\nameref{sec:methods}). 
% These measures capture how long and how strongly spiking at one point in time depends on spikes that have been emitted in the past. 
% Here, autocorrelation $C(T)$ considers only the linear dependence to a single point in time with time lag $T$ (Fig.~\ref{fig:measures}A), whereas predictability $R(T)$ is based on the mutual information between current spiking and the entire past spiking in a past range $T$, and thus captures all linear and non-nonlinear dependencies in the entire past range $T$~\cite{rudelt_2021} (Fig.~\ref{fig:measures}B). 
Autocorrelation $C(T)$ considers only the linear dependence to a single point in time with time lag $T$ (Fig.~\ref{fig:measures}A). In contrast, predictability $R(T)$ gives the proportion of spiking information $R(T)$ that can be predicted from past spiking in an entire past range $T$ (Fig.~\ref{fig:measures}B), and thus captures all linear and non-nonlinear dependencies in that range $T$~\cite{rudelt_2021}. 
Using the autocorrelation, we estimate (i) the intrinsic timescale of spiking activity, which is computed as the decay time $\tau_C$ of an exponentially decaying autocorrelation~\cite{murray_2014, wilting_2018a} (Fig.~\ref{fig:measures}A), and serves as a proxy for how long information is stored in neural activity~\cite{wilting_2018}. 
In contrast, the predictability $R(T)$ increases monotonously with $T$, because more past information can only increase predictability. 
From this, we estimate (ii) the total predictability $R_\text{tot}$ of a unit as the value where $R(T)$ saturates for large $T$, as well as (iii) the information timescale $\tau_R$, which can be interpreted as a rise time of the predictability, and indicates a typical timescale on which past activity is informative, i.e. adds to the predictability of current spiking (Fig.~\ref{fig:measures}B). 

In the data, we found that single unit autocorrelation is generally well approximated by an exponentially decaying function (Fig.~\ref{fig:measures}E, Supplementary Fig.~\ref{fig:autocorrelation_functions_data}), except for very short time lags, which we accounted for in our fitting procedure (\nameref{sec:methods}).
Moreover, units with higher intrinsic timescale typically also have higher information timescale (Supplementary Fig.~\ref{fig:measures_vs_measures}), indicating that the different measures of timescale capture a similar trend in the data.
Consistent with previous findings~\cite{rudelt_2021}, $\tau_R$ is smaller than $\tau_C$, because $R(T)$ only increases for non-redundant past information, i.e.~information that could not be read out from a smaller past range $T$.
In contrast, the autocorrelation $C(T)$ only considers time-lagged bins, and hence also incorporates redundant contributions~\cite{rudelt_2021}.  

% correlation with other metrics 
% To gain a better intuition of the measures, we also compared them to simpler statistics of single neuron spiking. All measures are weakly correlated with the mean firing rate $\nu$ (Supplementary Fig.~\ref{fig:measures_vs_firing_statistics}): the intrinsic timescale $\tau_C$ tends to be larger for units with higher rate (Pearson correlation $r_P=0.26$), whereas $\tau_R$ and $R_{\mathrm{tot}}$ tends to be smaller ($r_P= - 0.17$ and $r_P=- 0.30$, respectively).
% Apart from that, the information timescale $\tau_R$ is weakly correlated with the median inter-spike-interval (ISI) ($r_P=  0.34$), since a longer ISI requires larger timescales to predict neural spiking. 
% Moreover, the predictability $R_{\mathrm{tot}}$ is strongly correlated with the coefficient of variation (CV) ($r_P= 0.72$), because higher predictability also means a deviation from uncorrelated Poisson spike trains with CV$=1$ towards spike trains with higher temporal variability and CV$>1$. 
% In contrast, the intrinsic timescale is hardly correlated with median ISI or CV. 
To provide a better intuition of the above measures, we can relate them to simpler statistics of single neuron spiking (Supplementary Fig.~\ref{fig:measures_vs_firing_statistics}). For instance,
$\tau_R$ is correlated with the inter-spike-interval (ISI), since a longer ISI implies larger timescales to predict neural spiking. 
As another example, predictability is correlated with the coefficient of variation (CV), because higher predictability also means a deviation from uncorrelated Poisson spike trains (CV$=1$) towards spike trains with higher temporal variability (CV$>1$).
In contrast, all measures are only weakly correlated with the mean firing rate. 
% so what:
Thus, although some relations exist, our chosen measures go beyond simple firing statistics and quantify complementary aspects of the temporal statistics of neural spiking:
% In summary, the dynamic signatures each quantify complementary aspects of the temporal statistics of neural spiking: 
the established $\tau_C$ covers linear dependencies, the more recent $\tau_R$ captures non-redundant, also non-linear dependencies, and $R_{\mathrm{tot}}$ describes the maximum information that could be predicted over the full past range. 
% TODO: Here one could mention stationarity of estimates, e.g. comparison of estimates for the two blocks of natural movies -> SI FIG 

\subsection*{Timescales and predictability differ between thalamic and cortical visual areas}

\begin{figure*}[t!]
    \centerline{\includegraphics[]{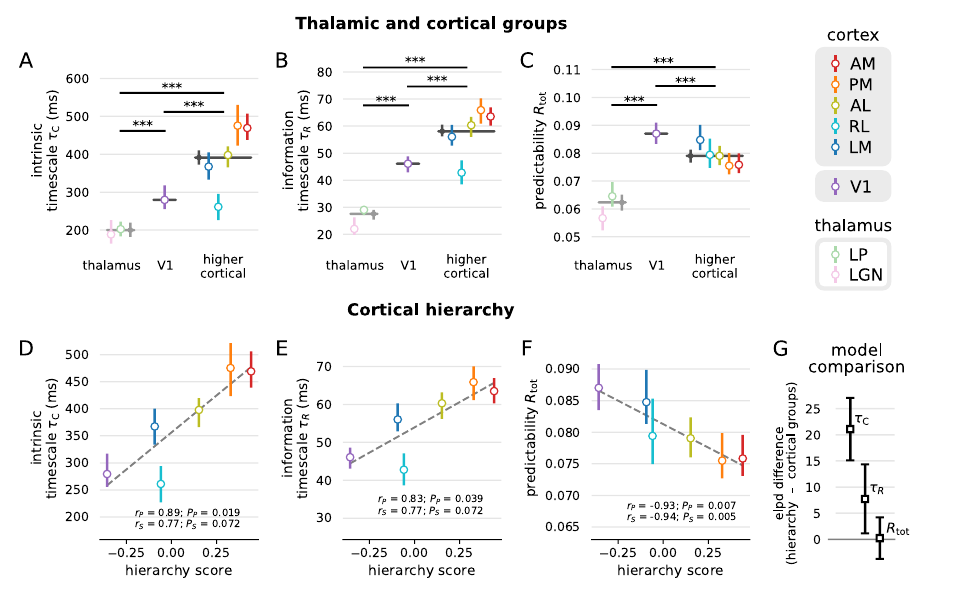}}
     \caption{\textbf{Timescales and predictability indicate a gradual hierarchy of temporal processing in mouse visual cortex.}
    \textbf{(A--C)} Hierarchy according to groups of visual areas. Gray markers indicate the median over all sorted units from different groups (thalamus, V1, higher cortical areas), colored dots indicate the median over units for individual areas. 
    Bars indicate 95 \% confidence intervals on the median obtained by bootstrapping, and p-values for comparisons of different groups were obtained by Mann-Whitney-U tests.
    \textbf{(A)} Intrinsic timescales $\tau_C$ are lowest for thalamus (areas LGN and LP) compared to primary visual cortex (V1) and higher cortical areas (LM, RL, AL, PM, AM).
    \textbf{(B)} The same trend holds for $\tau_R$, which increases from thalamic areas to V1 to higher cortical areas. 
    \textbf{(C)} $R_{\text{tot}}$ increases from thalamus to V1, but it is again smaller for higher cortical areas. 
    \textbf{(D--F)} Timescales and predictability as a function of anatomical hierarchy score~\cite{siegle_2021, harris2019hierarchical}.
    \textbf{(D)}~$\tau_C$ increases with hierarchy score of the respective cortical areas, which is well approximated by linear regression (dashed line, Pearson and Spearman correlation coefficients and p-values shown at the bottom). 
    \textbf{(E)}~The same relation holds for $\tau_R$.
    \textbf{(F)}~$R_{\text{tot}}$ in contrast decreases with hierarchy score. 
    \textbf{(G)}~Model comparison between the model based on cortical groups (V1, higher cortical in A--C), and the model based on linear relationship with hierarchy score (D--F). 
    Shown is the difference in expected log pointwise predictive density (ELPD) from leave-one-out (LOO) cross-validation~\cite{vehtari2017practical}. Boxes indicate the mean and bars the standard deviation over LOO samples. 
    The cortical hierarchy model has higher predictive power for $\tau_C$ and $\tau_R$, but models perform similarly for $R_{\mathrm{tot}}$.}
    \label{fig:main_analysis}
\end{figure*}

When comparing the estimated timescales and predictability between different brain areas, we found a significant difference between thalamic and cortical areas (Fig.~\ref{fig:main_analysis}A--C): Units from LGN and LP had much lower timescales $\tau_C %\approx 200\,\text{ms}
$ and $\tau_R$
% \approx 28\,\text{ms}$ (median over sorted units), 
compared to V1% with $\tau_C \approx 280\,\text{ms}$, $\tau_R\approx 46\,\text{ms}$
, or higher visual cortical areas. % with $\tau_C \approx 391\,\text{ms}$, $\tau_R\approx 58\,\text{ms}$. 
Moreover, predictability $R_{\text{tot}}$ is significantly smaller in thalamus %($R_{\text{tot}} \approx 6.2\,\%$) 
compared to V1 %($R_{\text{tot}} \approx 8.7\,\%$) 
or higher cortical areas. %($R_{\text{tot}} \approx 7.9\,\%$). 
These trends were confirmed with a second, independent dataset (\emph{Brain Observatory 1.1}~\cite{siegle_2021}, c.f.~Supplementary Fig.~\ref{fig:analysis_brain_observatory}A--C). 
In sum, timescales and predictability in thalamus were smaller than in cortex, which likely reflects the role of thalamus as a relay of information with fast processing, fast forgetting, and temporal decorrelation~\cite{dong_1995,cramer_2020}. 
In contrast, the higher timescales in cortical areas suggests an enhanced integration of temporal information, which might be supported by the extensive recurrent connectivity in cortex, and longer reverberations of activity in these areas~\cite{wilting_2018a}.
To better understand how this temporal processing is organized, we next focused on cortical areas.

\subsection*{Timescales and predictability indicate an organization of temporal processing along the anatomical cortical hierarchy}

% Evidence from macaques and humans suggests that temporal processing in cortex is organized hierarchically, where higher cortical areas are specialized to integrate information on longer timescales, and early areas are specialized for fast processing of sensory information~\cite{murray_2014, hasson_2008}.
% However, it is not clear whether this hierarchy is gradual for each co,rtical area, or whether there is a
% Now we disentangle two competeing hyptotheses for cortex
% we note that predicictability decresaes within cortex, while remaining consistenly higher than thalamus.
In mouse visual cortex, the intrinsic timescale as well as the information timescale and predictability are correlated with the anatomical hierarchy of cortical areas (Fig.~\ref{fig:main_analysis}E--F). 
Here, the cortical hierarchy is characterized by the anatomical hierarchy score based on inter-areal feedforward and feedback connectivity from the Allen Mouse Brain Connectivity Atlas~\cite{harris_hierarchical_2019}.
% according to the allen website this is the set of references for the brain atlast.
% https://alleninstitute.org/citation-policy/
% a significant difference in timescales and predictability between the groups of primary visual cortex (V1) and higher cortical areas (Fig.~\ref{fig:main_analysis}E--F), where timescales are higher for units from higher cortical areas (Fig.~\ref{fig:main_analysis}A,B), and the predictability is lower (Fig.~\ref{fig:main_analysis}C).
% In particular, median timescales increased with hierarchy score, whereas median predictability decreased (Pearson correlation coefficient $r_P = 0.89$, $0.83$ and $- 0.93$ for $\tau_C$, $\tau_R$ and $R_\text{tot}$, respectively).
In particular, median timescales increased with hierarchy score (Pearson correlation coefficient $r_P = 0.89$ and $0.83$ for $\tau_C$ and $\tau_R$), indicating longer integration times for higher areas. In contrast, median predictability decreased ($r_P = - 0.93$), which might reflect an increase of adaptation that attenuates responses to predictable stimuli for an efficient code.
Moreover, we carefully assessed whether the hierarchy in timescales and predictability indeed follows the anatomical hierarchy, or, alternatively, is better described by a grouping into primary, and higher, extrastriate cortical areas (V1 vs.~higher cortical areas, Fig.~\ref{fig:main_analysis}A--C).
This is important, since cortex in rodents is thought to have fewer hierarchical stages, where visual information from V1 is provided more readily (and in parallel) to multimodal interactions or motor outputs~\cite{marshel_2011}, which is supported by direct V1 input to essentially all extrastriate (higher cortical) visual areas~\cite{coogan1993hierarchical, olavarria1989organization, wang2007area}. 
% Despite the overall strong correlation between signatures and hierarchy score, it is not trivial whether 
% primary visual cortex (V1) should follow the linear trend of higher cortical areas (Fig.~\ref{fig:main_analysis}D--E), as it sits much earlier in processing pathway.
% Thus, we next investigated whether the
% temporal hierarchy (including V1) linearly follows the anatomical hierarchy score,
% or, alternatively, the hierarchy is discontinuous between structure groups (V1 vs.~higher cortical areas, Fig.~\ref{fig:main_analysis}A--C).
To this end, we used hierarchical Bayesian modelling (\nameref{sec:methods}), which allowed us to compare the group and hierarchy hypotheses. Moreover, this allowed us to disentangle differences in temporal processing from trivial differences in average firing rate and visual responsiveness (whether or not units have of a clear receptive field).
% To test whether the data rather imply a difference between structure groups (primary visual cortex and higher cortical areas, Fig.~\ref{fig:main_analysis}A--C), or a gradual increase of timescales along a cortical hierarchy (Fig.~\ref{fig:main_analysis}D--F), we used model comparison in a Bayesian regression (\nameref{sec:methods}).

% We compared two hypothesis, a grouping of areas, or a continuous, hierarchical organization. 
For the group hypothesis, we modelled timescales and predictability with a different log mean for each group (V1, higher cortical). For the cortical hierarchy hypothesis we modelled a linear relation between an area's hierarchy score and the mean timescales or predictability.
In addition, each model included average firing rate and responsiveness of each unit as predictors.%, to partial out the effect of firing rates.
% Moreover, to disentangle area differences in temporal processing from trivial differences in mean firing rate and visual responsiveness (whether or not units have of a receptive field on screen), we included predictors based on the the mean firing rate and responsiveness of each unit. 

First, the Bayesian regression confirmed our earlier results: Posteriors over log means of the groups indicated a credible increase of timescales and decrease of predictability for higher cortical areas when compared to V1 (Supplementary Fig.~\ref{fig:bayes_structure_groups_model_offset_posterior}).
Moreover, posteriors over the mean slope indicated a credible positive slope for both timescales, and a negative slope for the predictability (Supplementary Fig.~\ref{fig:bayes_hierarchy_score_model_slope_posterior}).

Second, the Bayesian regression suggested a hierarchical organization over a grouping. When comparing the two models by leave-one-out cross-validation, we found that the hierarchy-score model has more predictive power than the group model for the intrinsic and information timescale, and both models perform equally well for predictability (Fig.~\ref{fig:main_analysis}G).
Again, these results were confirmed with the \emph{Brain Observatory} data (Supplementary Figs.~\ref{fig:analysis_brain_observatory}D--G and~\ref{fig:bayes_hierarchy_score_model_slope_posterior}). 
%conclusion sentence

% would make this its own par.
In sum, the Bayesian model comparison implied that the increase in timescales is better described by a linear increase with anatomical hierarchy score than a group difference between primary visual cortex and higher cortical areas. 
This suggests that temporal processing in cortex is organized gradually along the anatomical hierarchy, and not in a two-stage processing architecture.

\subsection*{Predictability depends on visual stimulation and stimulus selectivity, while timescales are rather invariant} 

Until now, it is open which mechanisms might underlie the observed hierarchy of temporal processing. For rodents, a prominent hypothesis is that feedforward processing leads to more persistent stimulus representations for higher areas~\cite{piasini2021temporal}, thereby causing the longer timescales under visual stimulation. An alternative hypothesis is that stronger recurrence enables longer integration of visual information, similar to what is thought to cause longer timescales in primates~\cite{chaudhuri_2015}.
To test these hypotheses, we note that feedforward processing predicts overall higher timescales and a more pronounced hierarchy under stimulation with a natural movie (with long stimulus timescales) when compared to spontaneous activity in the absence of a time-varying visual stimulus. In contrast, recurrent integration predicts that timescales are rather invariant to the stimulus condition. 

\begin{figure}
    \centering
    \includegraphics[width=16cm]{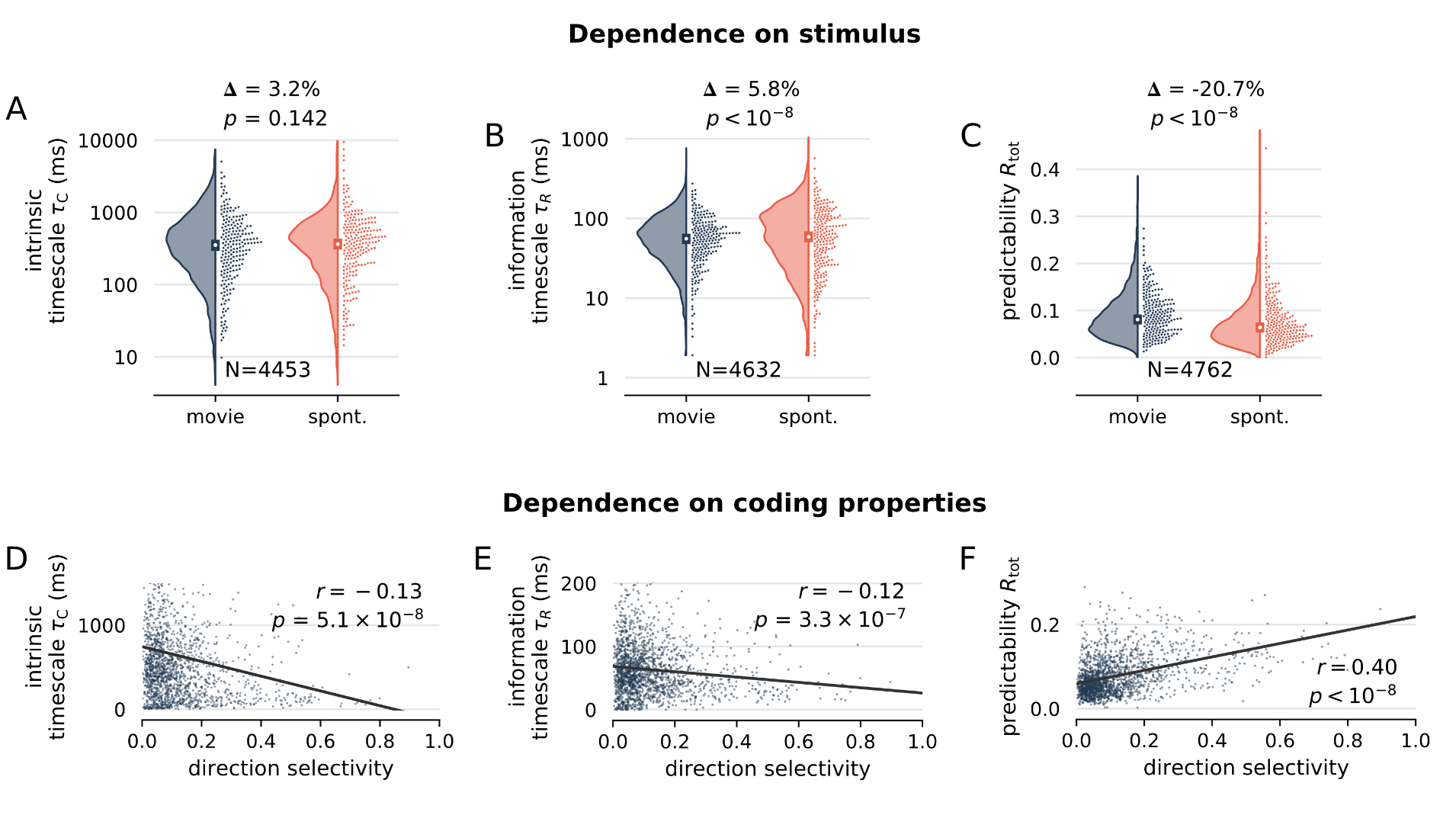}
    %  Functional Connectivity, cortex only \\
    %  \begin{subfigure}[t]{.32\textwidth}
    % \subcaption[]{}
    % \vspace{-9pt}
    % \includegraphics[width=.92\textwidth]{main_figs/allen_violin_tau_C.pdf}
    % \end{subfigure}   \hfill 
    % \begin{subfigure}[t]{.32\textwidth}
    % \subcaption[]{}
    % \vspace{-9pt}
    % \includegraphics[width=.85\textwidth]{main_figs/allen_violin_tau_R.pdf}
    % \end{subfigure} \hfill   
    % \begin{subfigure}[t]{.32\textwidth}
    % \subcaption[]{}
    % \vspace{-9pt}
    % \includegraphics[width=.81\textwidth]{main_figs/allen_violin_R_tot.pdf}
    % \end{subfigure}
    %     \centering
    %  \begin{subfigure}[t]{.32\textwidth}
    % \subcaption[]{}
    % \vspace{-9pt}
    % \includegraphics[width=.94\textwidth]{main_figs/measures_vs_allen_metrics_scatter_g_dsi_dg_tau_C.pdf}
    % \end{subfigure}   \hfill 
    % \begin{subfigure}[t]{.32\textwidth}
    % \subcaption[]{}
    % \vspace{-9pt}
    % \includegraphics[width=.93\textwidth]{main_figs/measures_vs_allen_metrics_scatter_g_dsi_dg_tau_R.pdf}
    % \end{subfigure} \hfill   
    % \begin{subfigure}[t]{.32\textwidth}
    % \subcaption[]{}
    % \vspace{-9pt}
    % \includegraphics[width=.91\textwidth]{main_figs/measures_vs_allen_metrics_scatter_g_dsi_dg_R_tot.pdf}
    % \end{subfigure}
    \caption{\textbf{Predictability depends on visual stimulation and stimulus selectivity, while timescales are rather invariant.}
    % overall: differences may or may not be significant, but effect size is tiny.
    \textbf{(A)}~Comparison of $\tau_C$ under stimulation with a natural movie (blue) and spontaneous activity under grey screen illumination (orange). $N$ units pooled from all cortical areas. Timescales for spontaneous activity do not differ significantly.
    \textbf{(B)}~$\tau_R$, in contrast, are more broadly distributed for spontaneous activity, with slightly larger median ($\Delta = 5.8\%$).
    \textbf{(C)}~$R_{\text{tot}}$ shows the largest effect between stimulus conditions  ($\Delta = - 20.7\%$).
    For \textbf{A--C}, p-values are obtained using Wilcoxon signed-rank tests.
    \textbf{(D--F)}~Timescales and predictability as a function of direction selectivity, where each dot represents a unit. Timescales and predictability were computed under natural stimulation in the \textit{Brain Observatory 1.1} data set,  and direction selectivity was calculated from a separate stimulus condition (drifting gratings).
    \textbf{(D, E)} Timescales $\tau_C$ and $\tau_R$ show a significant yet weak negative correlation with direction selectivity (linear regression in black, Pearson correlation $r = -0.13$ and $-0.12$, respectively).
    \textbf{(F)}~In contrast, predictability is positively correlated and has a larger correlation coefficient ($r=0.4$).}
    \label{fig:comparison_conditions_and_selectivity_measures}
\end{figure}

When comparing timescales between the previous natural movie condition and spontaneous activity that was recorded while showing the animal a grey screen (\nameref{sec:methods}), we found that median $\tau_C$ and $\tau_R$ were similar between stimulus conditions (pooled from all units in all cortical areas, Fig.~\ref{fig:comparison_conditions_and_selectivity_measures}A,B).
Moreover, we found that a cortical hierarchy of timescales also exists for spontaneous activity (Supplementary Fig.~\ref{fig:analysis_spontaneous},~\ref{fig:bayes_hierarchy_score_model_slope_posterior} and \ref{fig:bayes_structure_groups_model_offset_posterior}).
In contrast, median predictability differed between stimuli by $\approx 20 \%$
(Fig.~\ref{fig:comparison_conditions_and_selectivity_measures}C).
These trends were consistent when pooling and comparing units for each cortical area separately (Supplementary Fig.~\ref{fig:nm_vs_spontaneous_areas}).
Moreover, when applying the Bayesian regression to predictability for spontaneous activity, we neither found a credible decrease from V1 to higher cortical areas, nor a credible negative slope with the anatomical hierarchy score (Supplementary Figs.~\ref{fig:bayes_hierarchy_score_model_slope_posterior}I and \ref{fig:bayes_structure_groups_model_offset_posterior}I). 
Thus, while the hierarchy of timescales is found for both stimulus condition, differences in predictability are more pronounced under a time-varying visual stimulus.

To further test the relation between the measures and stimulus encoding, we assessed how timescales and predictability related to stimulus selectivity measures for individual neurons (e.g., direction and image selectivity).
% how $\tau_C$, $\tau_R$ and $R_{\rm tot}$ behave as a function direction selectivity and image selectivity (Fig.~\ref{fig:comparison_conditions_and_selectivity_measures}D--F, and
% Supplementary Figs.~\ref{fig:measures_vs_image_selectivity_all_cortical}).
To this end, we considered the \textit{Brain Observatory 1.1} data set,
which contains recordings under natural movie stimulation and a range of other conditions (drifting gratings, static images) required to quantify stimulus selectivity.
From the natural-movie recordings we again estimated timescales and predictability of each unit, and compared them to each unit's direction-selectivity, which was previously computed from drifting-gating recordings~\cite{siegle_2021}.
When pooling units across cortical areas, we found that timescales were only weakly and negatively correlated with the direction selectivity ($r=-0.13$ and $r=-0.12$, Fig.~\ref{fig:comparison_conditions_and_selectivity_measures}D,E), whereas predictability was more strongly and positively correlated ($r= 0.40$, Fig.~\ref{fig:comparison_conditions_and_selectivity_measures}F).
When repeating the analysis for each cortical area individually, timescales were not significantly correlated with direction-selectivity,
but predictability was (Supplementary Fig.~\ref{fig:measures_vs_direction_selectivity_per_area}).
Moreover, we found the same overall dependence as on direction-selectivity also on image-selectivity (Supplementary Figs.~\ref{fig:measures_vs_image_selectivity_all_cortical},~\ref{fig:measures_vs_image_selectivity_per_area}).

% add reference to the model part
% Together, these results provide further evidence that single-neuron predictability is enhanced by visual coding (Fig.~\ref{fig:comparison_conditions_and_selectivity_measures}C,F), whereas intrinsic and information timescales are mostly uncorrelated, and rather tend to be smaller for neurons that are more selective to visual stimulus features (Fig.~\ref{fig:comparison_conditions_and_selectivity_measures}A,B,D,E).
Together, these results imply that the hierarchy in timescales seems to rely mostly on network-intrinsic mechanisms (independent of stimulus conditions or visual coding properties), whereas single-unit predictability is at least partly induced by visual input. This also suggests that the decrease in predictability along the hierarchy might reflect an active cancellation of predictability in visual stimuli, e.g., through stronger adaptation in these areas.

%%%%%%%%%%%%%%%%%%%%%%%%%%%%%%%%%%%%%%%%%%%%%%%%%%%%%%%%%%%%%%%%%%%%%%%%%%%%%%%%%%%%%%%%%%%%%%%%%%%%%%%%%%%%%%%%%%%%%
% PS [date: 23-02-24]
% \subsection*{Experimental timescales and predictability are consistent with an increase of recurrence in network models with correlated external input}
\subsection*{Gradients of timescales and predictability are consistent with an increase of recurrence in network models with correlated external input}
% % Intro part describing the model
% The increase of timescales along the anatomical hierarchy suggests that they are an intrinsic property of cortical networks,
% and it has been hypothesized that long $\tau_C$ in neural spiking data are a result of recurrent activity propagation in the network~\cite{wilting_2019, wilting_2019a, chaudhuri_2015, zeraati_2021}.
It has been hypothesized that long intrinsic timescales are a result of enhanced recurrent activity propagation in cortical networks~\cite{wilting_2019, wilting_2019a, chaudhuri_2015, zeraati2023intrinsic}, which is in line with the observed stimulus independence of the timescale hierarchy. 
However, it is not known how higher recurrence affects predictability and information timescales. In particular, one might expect that higher recurrence also leads to more predictable spiking, thus contradicting the observed decrease of predictability for higher areas. 

\begin{figure}[t!]
    \centering
    \includegraphics[]{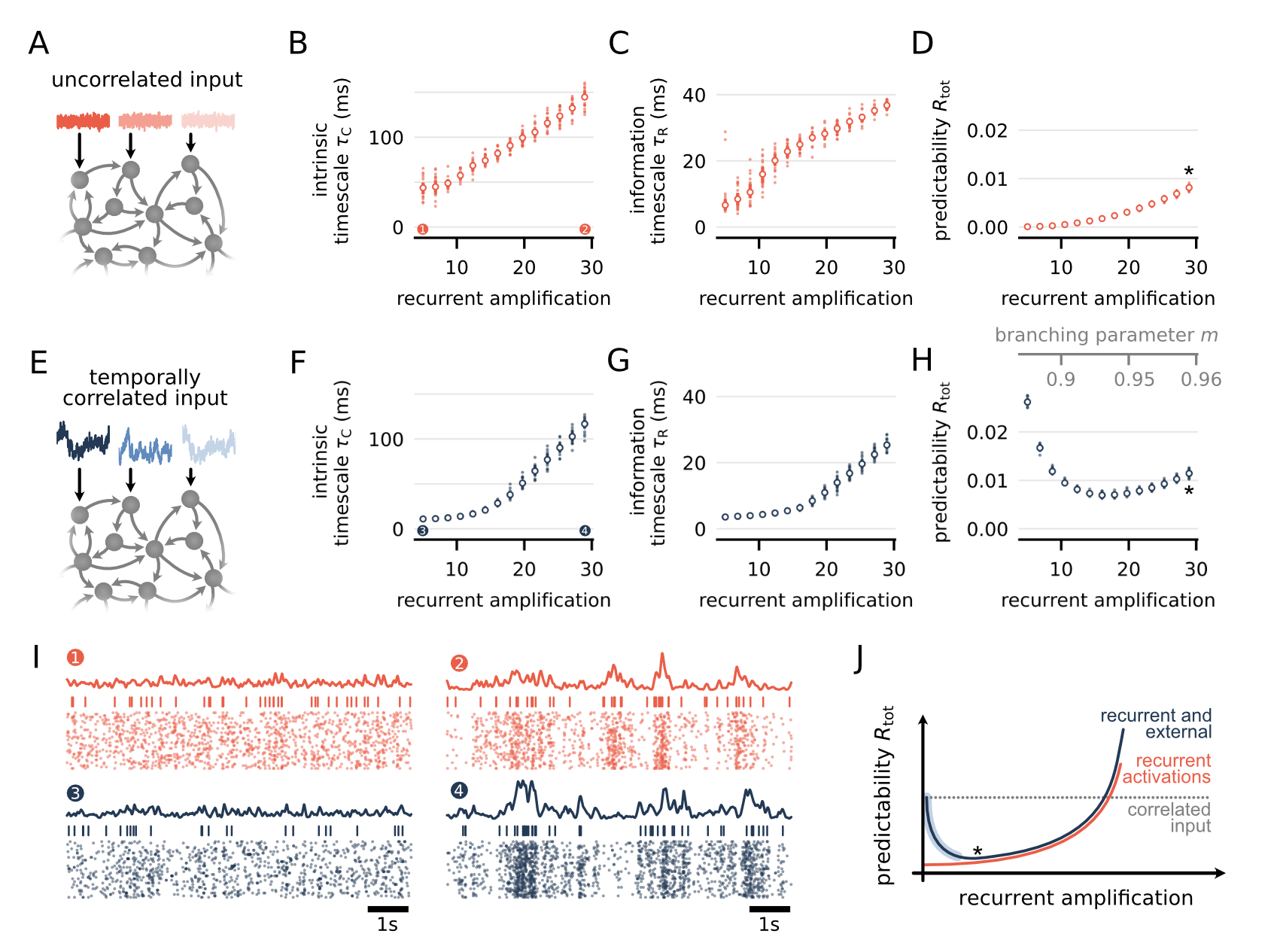}
    \caption{\textbf{Gradients of timescales and predictability are consistent with an increase of recurrence in network models with correlated external input.}
    \textbf{(A)} Schematic of a branching network: Each of the $N=1000$ units is connected randomly to $k=10$ other units (arrows). In a given time step, units are either active or inactive, and, if active, can activate each connected neighbour with fixed probability $m/k$ for the next time step.
    In addition to recurrent activations, each unit gets a temporally uncorrelated external input (red traces). 
    \textbf{(B--D)} Timescales and predictability increase with recurrent amplification $a/h = 1/(1-m)$ in the network, i.e., the mean number of generated spikes per external activation. 
    \textbf{(E--H)} Same setup as before, but each unit receives temporally correlated external input that follows an Ohrnstein-Uhlenbeck process with timescale $\tau_{\text{ext}} = 30\,\text{ms}$ (blue traces). 
    \textbf{(F-H)} For correlated input, the timescales behave similarly as before, but predictability is higher and decreases with recurrent amplification until $a/h\approx 20$.
    All measures are consistently about an order of magnitude smaller than for the experimental data, which we attribute to the simplicity of the model (see main text).
    Small dots indicate the median over $20$ randomly selected units for $10$ simulations. Big dots indicate median values over all simulations. 
    \textbf{(I)} Example raster plots for uncorrelated (top) and correlated input (bottom), at $a/h = 5$ (left) and $a/h = 30$ (right). Each panel shows the population rate, spiking of a single unit, and population raster of 40 units.
    \textbf{(J)} Sketch of single-unit predictability for different sources of temporal correlations. Recurrent activations yield sizable predictability only for very high recurrent amplification (red line, c.f. panel D). Otherwise, predictability through correlated input is higher (grey dashed line). Hence, in the presence of correlated input, increasing recurrent amplification decreases predictability, because more and more spikes are caused by recurrent activations with lower predictability (blue line, c.f. panel H). When recurrent amplification is high enough (star), most activations are recurrent, and both models give the same behavior.
    }
    \label{fig:model_results}
\end{figure}

To test this we considered a minimal model of recurrent activity propagation. The branching network~\cite{larremore_inhibition_2014, zierenberg_tailored_2020} consists of sparsely connected units that are either active or silent in each discrete time bin. 
Activity propagation in the network is governed by the branching parameter $m$, which gives the mean number of units that an active unit activates in the next time step. Moreover, independent of recurrent activations, each unit is activated by temporally uncorrelated external input with mean rate $h$ (Fig.~\ref{fig:model_results}A). Recurrence in the network is then characterized by the recurrent amplification $a/h = 1/(1-m)$, which gives the mean number of generated spikes $a$ per external input activation $h$ (c.f.~\cite{london_sensitivity_2010}). 
To test how recurrent amplification $a/h$ affects timescales and predictability, we simulated branching networks consisting of $N=1000$ units with $k=10$ outgoing connections each, and estimated $\tau_C$, $\tau_R$ and $R_{\rm tot}$ using the same procedure as for the experimental data (\nameref{sec:methods}).

%with $k=10$ outgoing connections per neuron, each connection activating the connection unit with probability $m/k$.

% Results for uncorrelated input
In branching networks, the intrinsic timescale $\tau_C$ is known to scale linearly with the recurrent amplification $a/h$, which is caused by long-lasting reverberations of activity in the network~\cite{wilting_2018, wilting_2019}.
We could reproduce this linear increase by fitting $\tau_C$ for individual units in the network, and found a similar increase of $\tau_R$ with $a/h$ (Fig.~\ref{fig:model_results}B,C).
Notably, also the predictability $R_{\rm tot}$ increased with $a/h$, which constitutes an important difference to the experimental data.
However, $R_{\rm tot}$ was more than an order of magnitude smaller than in the experimental data, indicating that a source of predictability is missing in the model (Fig.~\ref{fig:model_results}D).

% Description of the result for correlated input 
Since neurons in cortex receive temporally correlated inputs rather than uncorrelated Poisson noise~\cite{pozzorini_2013},
we replaced the constant external activation rate $h$ for each unit by a fluctuating rate $h(t)$, which was modeled by an Ornstein-Uhlenbeck process with autocorrelation time $\tau_{\rm ext} = 30\,\text{ms}$ (Fig.~\ref{fig:model_results}E).
As before, the external input is unique for each unit, and thus induces temporal correlations without increasing spatial correlation between units.
With temporally correlated input, and intermediate values of recurrent amplification ($ 0 < a/h \leq 20$), we recovered qualitative agreement between model and experimental data;
timescales increased with $a/h$ (Fig.~\ref{fig:model_results}F,G), whereas predictability $R_{\text{tot}}$ decreased until reaching $a/h \approx 20$ (Fig.~\ref{fig:model_results}H).
Moreover, when comparing the two models, timescales share the same overall trend, but predictability differs systematically, and, at small $a/h$, is much higher with correlated than with uncorrelated input (Fig.~\ref{fig:model_results}H vs.~D). 
This is consistent with the observation that the Bayesian analysis that did not find a credible decrease in predictability for spontaneous activity (Supplementary Figs.~\ref{fig:bayes_hierarchy_score_model_slope_posterior}I and \ref{fig:bayes_structure_groups_model_offset_posterior}I), and that predictability is higher under a temporally correlated (movie) stimulus (Fig.~\ref{fig:comparison_conditions_and_selectivity_measures}C).

In sum, the combined treatment of recurrence and temporally correlated input suggests that both, an increase in timescales and a decrease of predictability, are consistent with increasing recurrence along the anatomical hierarchy in cortex.
Although timescales and predictability in the model are almost an order of magnitude smaller than in the data, the models' simplicity and few mechanistic components enable a systematic explanation of the emergent effect (see \nameref{sec:discussion} and Fig.~\ref{fig:model_results}I,J).
Moreover, we confirmed that the same effect also occurs in a leaky integrate-and-fire (LIF) model with excitatory and inhibitory neurons, which was implemented on a neuromorphic chip~\cite{cramer_2020, pehle2022brainscales} (Supplementary Fig.~\ref{fig:measures_from_neuromrophic}).
% For example, we found the same increase in timescales and decrease in predictability in a plastic leaky integrate-and-fire (LIF) model with excitatory and inhibitory neurons~\cite{cramer_2020} (Supplementary Fig.~\ref{fig:measures_from_neuromrophic}). 
In this case, recurrent amplification was adapted through synaptic plasticity,
while the membrane dynamics and refractoriness of the LIF neurons provided the additional source of predictability (inputs were temporally uncorrelated).
We conclude that the increase in timescales and decrease in predictability is a general effect that occurs when (i) recurrent activity propagation leads to stronger and longer-lasting fluctuations of activity, and (ii) some other source of single-neuron predictability exists, which then has a diminishing effect on single-neuron predictability with increasing recurrence.

\section*{Discussion}\label{sec:discussion}
Here, we used information theory and an autocorrelation analysis to characterize the temporal statistics of spike trains across the mouse visual system. We found that these statistics differ systematically between processing stages from thalamus to higher cortical areas. 
In particular, the information and intrinsic timescales increase along an anatomical hierarchy in cortex, whereas predictability of neural spike trains decreases. 
Moreover, we found the same gradients and median values of timescales under spontaneous activity, whereas a decrease in predictability is only observed under stimulation with a natural movie.
We concluded that the gradients in timescales are therefore likely caused by intrinsic mechanisms, like recurrent coupling that could serve an enhanced temporal integration for higher areas~\cite{wilting_2018, wilting_2019, chaudhuri_2015, zeraati2023intrinsic, runyan2017distinct}, whereas a reduction of predictability might be the result of stronger cancellation of redundant, predictable stimuli, in line with hierarchical predictive and efficient coding~\cite{rao1999predictive, simoncelli2003vision}. 
% Mention model 
In a basic model of recurrent activity propagation, as well as a plastic leaky integrate-and-fire model on a neuromorphic chip, we verified that information and intrinsic timescales indeed increase with recurrent coupling, whereas single-neuron predictability can even decrease with recurrence. 
In sum, our results therefore suggest that enhanced temporal integration in higher areas could be complementary to the observed increase of adaptation along a cortical processing hierarchy~\cite{kaliukhovich2018hierarchical}, where adaptation attenuates responses to redundant, predictable stimuli~\cite{price2022efficient, vinken2017recent}.

% MORE RESULTS FOCUSSED SUMMARY HERE!
% pplicable tool to characterize the temporal spike statistics of
% individual neurons, and we have found that it reveals key differences in the spiking statistics
% we have applied this approach to spike recordings from the mouse visual system, and found that
% predictability and the information timescale differ systematically between processing stages from
% thalamus to different cortical areas. Specifically, we found that the information timescale, as well
% as the intrinsic timescale based on autocorrelation, increase along an anatomical hierarchy in
% cortex, whereas the predictability decreases. Moreover, the same gradients in timescales are
% found under spontaneous activity, whereas a decrease in predictability is only observed under
% stimulation with a natural movie. We concluded that the gradients in timescales are therefore
% likely caused by intrinsic mechanisms, like recurrent coupling that could serve an enhanced
% temporal integration for higher areas [Lr8, 55, 60, 61], whereas a reduction of predictability might
% be the result of stronger cancellation of redundant, predictable stimuli, in line with hierarchical
% predictive and efficient coding [37, 62]. Our results also suggest that enhanced temporal
% integration in higher areas could be orthogonal to the observed increase of adaptation along a
% cortical processing hierarchy [63], where adaptation attenuates responses to redundant,
% predictable stimuli [64, 65].

\textbf{Hierarchy of timescales in rodents}\\\noindent
An increase in timescales along the hierarchy was surprising, because the original study by Siegle et al. did not find a hierarchy in intrinsic timescales for spontaneous activity (Extended Data Fig.~9 in~\cite{siegle_2021}).
% while a recent study on (anesthesized) rats found a hierarchy... then again, spontaneous vs stimulated. piasini?
% - Murray approach, movie, 
However, a later study analysing the same data found a hierarchy of intrinsic timescales~\cite{piasini2021temporal}---but focused on natural movie stimulation instead of spontaneous activity, and used a different fitting approach~\cite{murray_2014}. 
% Yet, even regarding the intrinsic timescale $\tau_C$, the presented results were unexpected, since the original study did not find a hierarchy in timescales for the spontaneous activity condition of the analyzed data set ~\cite{siegle_2021}. 
Therefore, we carefully assessed the robustness of our fitting procedure of the intrinsic timescale $\tau_C$, and found that the discrepancy to the results of Siegle et al.~is most likely caused by differences in the fitting range (\nameref{sec:methods}). 
In particular, we found that when the maximal time lag $T_{\rm max}$ used for fitting is less than 1000 ms, then estimated timescales are generally much smaller (Supplementary Fig.~\ref{fig:mre_Tmax_comparison}A), and no correlation between timescales and anatomical hierarchy score is found (Supplementary Figs.~\ref{fig:mre_Tmax_comparison}B,C and \ref{fig:hierachy_fitting_ranges_tau_C_tmax500}). In contrast, for $T_{\rm max} > 1000\,\text{ms}$ the inferred hierarchy was robust to the exact choice of $T_{\rm max}$ (Supplementary Fig.~\ref{fig:mre_Tmax_comparison}B,C). 
The presented results are further supported by a similar hierarchy in information timescales~(Fig.~\ref{fig:main_analysis}E), which are estimated with an entirely different approach, and do not rely on fitting an exponential decay rate (\nameref{sec:methods}). 

% - Is there really a single hierarchy? Discuss the gradual For example, RL sticks out. 
In primates, a hierarchy of timescales was found along a global hierarchy of cortical areas, spanning different modalities and different levels of cognitive abstraction~\cite{murray_2014, badre_is_2009}. 
Here, for mice, we found a hierarchy of timescales specifically for the visual sensory pathway.
% This was possible due to the huge number of recorded neurons per area, which reveals systematic differences between areas, despite the high heterogeneity of timescales within an area. 
This is an important finding, since cortex in rodents is thought to have fewer hierarchical stages, which allows visual information to be provided more readily (and in parallel) to multimodal interactions or motor outputs~\cite{marshel_2011}.
This is supported by direct V1 input to essentially all extrastriate (higher cortical) visual areas~\cite{coogan1993hierarchical, olavarria1989organization, wang2007area}, as well as evidence that several extrastriate areas process information related to other sensory modalities~\cite{miller1984direct,sanderson1991prosencephalic, wagor1980retinotopic}.
At the same time, anatomical feedforward- and feedback-connectivity motifs as well as several functional properties also point to a hierarchical ordering of extrastriate areas~\cite{siegle_2021}.
Therefore, we tested whether the observed increase in timescales is gradual (following the anatomical hierarchy), or parallel (reflecting an organization into two stages, V1 vs.~extrastriate visual areas).
We found that a model with gradual increase had more predictive power for both, the intrinsic and information timescale (Fig.~\ref{fig:main_analysis}G, and Supplementary Figs.~\ref{fig:analysis_brain_observatory}G and~\ref{fig:analysis_spontaneous}G),
suggesting a hierarchical organization of temporal processing in mouse visual cortex.

However, we also found deviations from a single gradient in timescales. 
In particular, the rostro-lateral area (RL) consistently had the smallest median timescales across all data sets and stimulus conditions, and the anteromedial area (AM) had a smaller median timescale than PM despite its higher anatomical hierarchy score.
This might be due to the functional specialization of the posterior parietal areas (RL, AL, AM), which are thought to process information related to motion and behavioral actions, whereas latero- and posteromedial areas LM and PM are thought to play a stronger role in object perception, analogous to the dorsal and ventral streams described in other species~\cite{marshel_2011, wang2011gateways, wang2012network}.
Thus, our results might also support a grouping of areas into two parallel yet hierarchically organized pathways.
This was similarly found in a functional connectivity study that suggested a sensory-to-motor and a transmodal pathway~\cite{huntenburg2021gradients}.

\textbf{Recurrence could shape hierarchy of timescales and predictability}\\\noindent
After identifying the gradient of timescales and predictability along the anatomical hierarchy, we uncovered potential mechanisms behind it in an easy-to-interpret model of recurrent activity propagation.
In line with previous studies~\cite{chaudhuri_2015, wilting_2019, zeraati2023intrinsic}, the branching network model illustrates how recurrent activity propagation leads to long timescales of single unit spiking activity, even though individual units in the branching network are memory-less (Fig.~\ref{fig:model_results}~B,C).
% Could leave this part out
% Moreover, since long timescales in the model are mainly caused by recurrent activations, they rather decrease when adding correlated input (Fig.~\ref{fig:model_results}F,G), consistent with the slight decrease in timescales for visual stimulation and higher stimulus selectivity (Fig.~\ref{fig:comparison_conditions_and_selectivity_measures}).
Whereas recurrent activations clearly increase the timescale, the \emph{magnitude} of single-unit correlation through recurrent activations is small~\cite{wilting_2018} --- and in fact much smaller than that induced by correlated input (Fig.~\ref{fig:model_results}D vs.~H). 
Hence, in the presence of correlated input, increasing recurrent amplification
decreases predictability, because more and more spikes are caused by recurrent activations with lower
predictability (Fig.~\ref{fig:model_results}J).
These model results are consistent with the experimental observation that predictability is higher under visual stimulation than for spontaneous activity, and positively correlated with stimulus selectivity (Fig.~\ref{fig:comparison_conditions_and_selectivity_measures}C,F).

% PS: 23-04-24: below does not answer the questions that is raised above: why do they think that this (high rec. amp) is not the regime where cortex operates?
Since recurrent network structures are a general property of cortex, we expect the same qualitative behavior also in more complex network models. 
For example, we found the same increase in timescales and decrease in predictability in a model with leaky integrate-and-fire (LIF) excitatory and inhibitory neurons~\cite{cramer_2020} (Supplementary Fig.~\ref{fig:measures_from_neuromrophic}). 
In this model, recurrent amplification was adapted through plasticity,
and the membrane dynamics and refractoriness of the LIF neurons provided the additional contributions to predictability.
Indeed, refractoriness, membrane dynamics and spike adaptation are also likely sources of predictability in the experimental data, which is supported by the characteristic negative autocorrelation for small time lags (Supplementary Fig.~\ref{fig:autocorrelation_functions_data}).

Yet, in the data, recurrent amplification might not be the only mechanism causing the observed long timescales.
For example, clustered connectivity is known to lead to long timescales in balanced neural networks~\cite{litwin2012slow}. 
Overall, we do not expect such additional mechanisms to alter the effect of recurrence, but they might explain why timescales and predictability are higher in the data compared to the simple branching network.

One particular mechanism besides recurrence that might give rise to a hierarchy of timescales are long-range projections between cortical areas.
For example, in a model based on measured connectivity in macaque cortex, long-range feedforward and feedback excitatory connections were shown to influence the exact layout of the hierarchy~\cite{chaudhuri_2015}.
However, a more recent study found that a detailed balance between long-range excitatory inputs and local inhibitory inputs could prevent longer or shorter timescales to spill over from other areas, which yields a better segregation, and facilitates functional specialization~\cite{li2022hierarchical}.  
Moreover, in~\cite{chaudhuri_2015}, it was found that recurrence is a necessary requirement to achieve long timescales in the first place, since leaving out recurrent connections in the model led to much smaller timescales in general. 
This supports our perspective that recurrent amplification plays a key role in shaping temporal processing along the anatomical hierarchy.

% VP: the following is a nice paragraph:
However, evidence for stronger recurrent coupling in higher areas is not as clear in mice as it is in primates. 
In particular, the density and numbers of dendritic spines of layer 3 pyramidal neurons (which are interpreted as markers of recurrence~\cite{elston2002cortical}) are much more similar between cortical areas in mice than in primates~\cite{benavides2006dendritic, ballesteros2006density, gilman2017area}. 
On the other hand, mouse cortex exhibits other gradients in microstructural properties~\cite{fulcher2019multimodal}, in particular a decrease in the density of parvalbumin (PV)-containing (inhibitory) interneurons for higher areas. 
This suggests that a gradient of recurrence in mice is rather controlled by the levels of inhibition than the net number (or density) of excitatory synapses~\cite{ding2024cell}.

% INTERPRETING RESULTS ON CODING LEVEL: POPULATION VS SINGLE NEURON CODE
% @jz: add salt-and-pepper comparison, part of the in-meat examples?.

% FF perspective, because dependence of timescales on stimulus timescales BUT only in response timescales. INtrinsic timescales show stronger differences between areas, and not so much between different stimuli!
% Intrinsic timescale is an intrisic effect (e.g., recurrence, not so much coding of stimuli) AGREES with little qualitative change and also quantitative change of median timescales between movie and spontaneous 

\textbf{Coding perspective}\\\noindent
% Could structure this: timescales vs predictability
% Turning to a coding perspective, an alternative interpretation to temporal processing shaped by recurrence is that a hierarchy in timescales is merely a side-product of a representation of increasingly more complex visual features, similar to the emergence of complex versus simple cell receptive fields.
Turning to a coding perspective, a hierarchy in timescales can be interpreted as a representation of increasingly more complex visual features, similar to the emergence of complex versus simple cell receptive fields:
A successive integration of temporally short-lived, simple features into temporally long-lived, complex ones would lead to long timescales in those neurons that represent the complex features~\cite{hasson_2008,hasson2015hierarchical, piasini2021temporal}.
However, it is not clear how such temporal integration is implemented. 
Here, we found that timescales hardly change with visual stimulation and stimulus selectivity of the units (Fig.~\ref{fig:comparison_conditions_and_selectivity_measures}), similar to findings in monkey somatosensory cortex~\cite{rossi2021invariant}.
This is also consistent with findings from rat visual cortex~\cite{piasini2021temporal}, where the response timescales depend on the stimulus timescales, whereas the intrinsic timescales depended more strongly on the area than on the stimulus timescale.
Together, these results suggest that long intrinsic and information timescales rely primarily on properties of the network, e.g., recurrence, rather than temporal properties of the inputs.

Of particular interest from a coding perspective are our results on predictability $R_{\rm tot}$, because they imply that the temporal redundancy in spike trains decreases along the hierarchy. This could reflect efficient coding, where subsequent processing stages in cortex remove more and more temporal redundancy in individual spike trains~\cite{barlow_2012, simoncelli2003vision, price2022efficient}.
On a mechanistic level, efficient-coding is in line with the observed increase of adaption along the cortical shape-processing hierarchy in rats~\cite{kaliukhovich2018hierarchical}, where adaptation attenuates responses to redundant, predictable stimuli~\cite{price2022efficient, vinken2017recent}.
Here, we provide a complementary mechanism that relies on temporal correlations of the stimuli and recurrence that may also lead to a decrease of predictability along the hierarchy.
Notably, both mechanisms are compatible with our results and may co-exist to serve complementary function: Recurrence enables long integration for temporal processing on the network level, while adaptation ensures efficiency of the neural code on a single-neuron level. 

% This paragraph is not super essential and might go if needed
In addition, in thalamus we found that both predictability and timescales are much lower than in cortex (Fig.~\ref{fig:main_analysis}A--C), indicating a stronger degree of temporal decorrelation to remove redundancy~\cite{dong_1995}.
This might be due to the function of LGN and other thalamic areas as a relay of information, or a controller, requiring fast processing with little redundancy, little integration, and fast forgetting~\cite{cramer_2020}. Anatomically, these coding signatures could be related to the lack of excitatory recurrent connections in thalamus. 
% For cortex, in contrast, we argue that coding has to be understood through recurrent processing on the population level, where our findings are consistent with hierarchical temporal processing.
For cortex, in contrast, our results suggest that recurrence is key to enable long temporal integration and hierarchical temporal processing.

% Short summary
\textbf{Conclusion}\\\noindent
% THIS IS A GREAT CONCLUSION PARAGRAPH
Until now it has remained open whether single sensory pathways in cortex show a hierarchical organization of temporal processing, especially in rodents, where cortex exhibits less cognitive processing and fewer levels of abstractions compared to primates. 
Here, we found that the visual pathway in mouse cortex indeed shows signatures of such a hierarchy in the form of a gradual increase in intrinsic and information timescales, and a decrease in predictability along the anatomical hierarchy. 
By computing timescales and predictability for a recurrent network model, and by combining this with data from different stimulus conditions, we found that the gradients in timescales are more likely caused by network-intrinsic mechanisms, and could result from increasing recurrence along the anatomical hierarchy.
In contrast, the stimulus-dependent decrease in predictability is in agreement with efficient coding as well as an observed increase of adaptation in rodents, and might constitute another hallmark of hierarchical temporal processing in mammals. 

% To summarize, by analyzing timescales and predictability derived from spiking activity, we found signatures of hierarchical temporal processing in mouse visual cortex. 
% Further, our modelling supports the hypothesis that such a hierarchy is shaped by network-intrinsic mechanisms such as recurrence in cortical networks. 
% In sum, our results suggest that a hierarchical cortical organization presents a general principle for sensory processing across mammals.

\section*{Methods}\label{sec:methods}

\subsection*{Experimental data sets}
\begin{figure}[H]
    \centering
    \begin{subfigure}[t]{.51\textwidth}
    \subcaption{}
    % \vspace{-9pt}
    % \includegraphics[width = .9\textwidth]{figs/neurons_after_filtering_stages.png}
    \includegraphics[width = \textwidth]{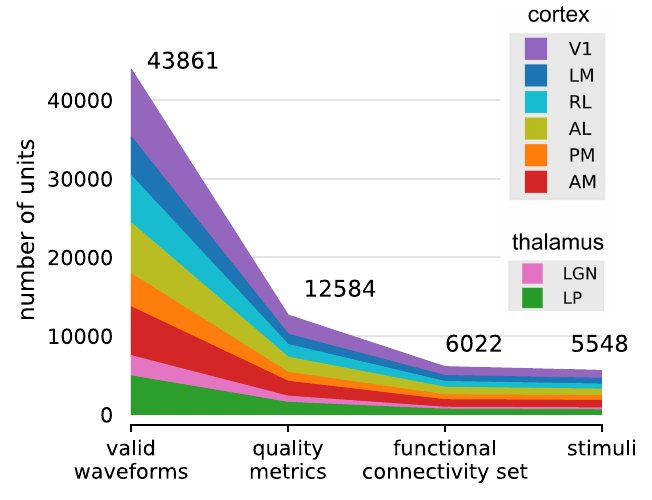}
    \end{subfigure} \hfill
    \begin{subfigure}[t]{.48\textwidth}
    \subcaption{}
    % \vspace{-9.3pt}
% \includegraphics[width = .8\textwidth]{figs/neurons_after_filtering_per_area.png}
    \includegraphics[width = .93\textwidth]{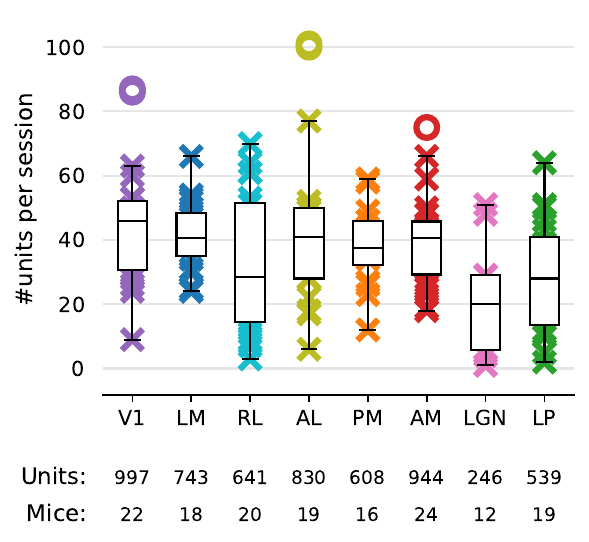}
    \end{subfigure}
    \caption{\textbf{Number of available units for the analysis for the \textit{Functional Connectivity} data set.} 
    \textbf{(A)} Sorted units available for analysis after spike sorting process (“valid wave-forms”), after applying filters of the AllenSDK (“quality metrics”), after selecting only units of the “Functional Connectivity set” and after ensuring that the recordings of the selected “stimuli” for each unit are long enough and do not include invalid spike times (\nameref{sec:methods}). 
    \textbf{(B)} Numbers of units for each session and for each area available for analysis after filtering.}
    \captionlistentry[suppfigure]{\textbf{Fig S1. Number of available units for the analysis for the \textit{Functional Connectivity} data set.}}
    \label{fig:units_filtering}
\end{figure}

\begin{figure}[H] 
    \centering
    \includegraphics[width = .7\textwidth]{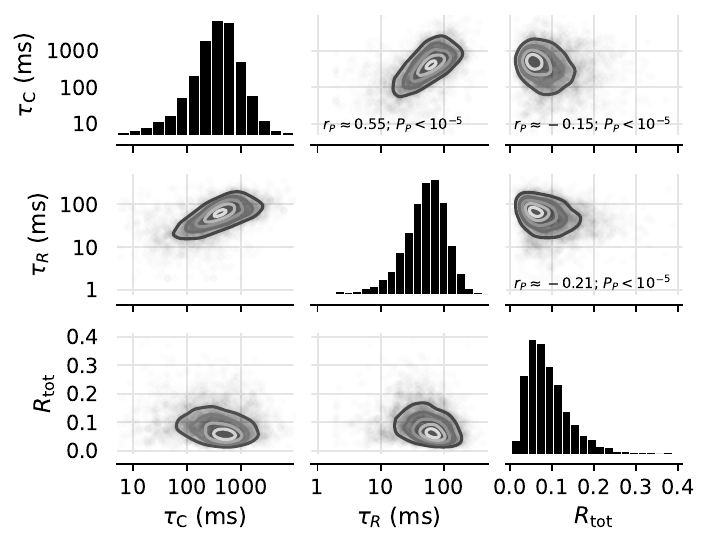}
    \caption{\textbf{Relation between intrinsic and information timescales, as well as predictability across all sorted units.} 
    Histograms of the intrinsic timescale $\tau_C$, the information timescale $\tau_R$ and the predictability $R_{\text{tot}}$ (diagonal), as well as scatter plots of one measure against the other (y-axes refer to scatter plots, no axes shown for histograms) for all analyzed units. Scatter plots are overlaid with kernel density estimations, where lines indicate regions of with equal probability. Intrinsic and information timescales are shown in log scale. Timescales are positively correlated (Pearson correlation), whereas predictability is weakly negatively correlated with the timescales.}
    \label{fig:measures_vs_measures}
    \captionlistentry[suppfigure]{\textbf{Fig S2. Relation between intrinsic and information timescales, as well as predictability across all sorted units.}}
\end{figure}

To investigate temporal processing in the mouse visual system, we analyzed data from the \textit{Visual Coding Neuropixels} data set, which is openly available through the Allen Brain Observatory~\cite{deVries2020large, siegle_2021}.
This data set contains extracellular electro-physiological recordings of mouse brain activity obtained with Neuropixels probes~\cite{jun_fully_2017}. 
This setup enabled to simultaneously record from cortical and sub-cortical structures involved in visual processing (Fig.~\ref{fig:main_analysis}A), with six cortical areas [primary visual cortex (V1), lateromedial area (LM), anterolateral area (AL), rostrolateral area (RL), anteromedial area (AM) and posteromedial area (PM)], and two thalamic areas [(the lateral geniculate nucleus (LGN) and lateral posterior nucleus (LP)], with a minimum of $n=12$ and a maximum of $n = 24$ mice per brain area and experimental setup~(see Suppelementary Fig.~\ref{fig:units_filtering}B for the number of mice and analyzed sorted units per brain area).
The probes record with high temporal precision with 30 kHz sampling rate and sub-millisecond temporal resolution, which is ideal to study temporally precise processing with spikes.
For the experiments, the mice were head-fixed and were shown a range of visual stimuli. 

The data set contains data from two experiments (\textit{Functional Connectivity} and \textit{Brain Observatory 1.1}), which differ in the sequences of stimuli that were shown to the mice. 
To study temporal processing in the mouse visual system in a naturalistic and stationary environment, our main results were obtained from the \textit{Functional Connectivity} experiment, which contains two blocks of around 15 minutes of consecutively recorded spiking activity during the repeated presentation of a 30 second naturalistic movie (termed \verb|natural_movie_one_more_repeats| in the AllenSDK). 
Furthermore, we contrast our findings on the natural movie condition with results obtained from another single block of around 30 minutes of spontaneous activity, where the animal was shown a gray screen (termed \verb|spontaneous|). 
In addition to the \textit{Functional Connectivity} experiment, we also analyzed the \textit{Brain Observatory 1.1} data, with two 10 minute blocks of natural movie stimulation with 120 second clips repeated 5 times per block (termed \verb|natural_movie_three|). This did not only serve as a control, but also enabled the comparison of timescales and predictability to other metrics of visual processing that could be estimated within the \textit{Brain Observatory 1.1} data set (Fig.~\ref{fig:comparison_conditions_and_selectivity_measures}).
To make results comparable across experiments and stimulus conditions, we adjusted the duration of the analyzed snippets of spike recordings to match the shorter recording duration of the natural movie condition of the \textit{Brain Observatory 1.1}. Therefore, we only used the last 10 minutes of each block for the natural movie condition, and the last 20 minutes in the spontaneous activity condition in the \textit{Functional Connectivity} data.
This is important, because different recording lengths can cause systematic biases in the comparison of timescales and predictability~\cite{rudelt_2021}.
Furthermore, at the beginning of each block a transient of 1 minute was removed to improve stationarity of the recording, and natural movie blocks were concatenated to then yield a total recording length of 18 minutes. 

Finally, we only analyzed sorted units that fulfilled certain quality metrics and criteria that are relevant to our analysis. 
Spike sorting was done in~\cite{siegle_2021}, and we used the spike sorted data from there. Moreover, we only selected units based on quality metrics that are provided with the data~\cite{allen_qualitymetrics_url}: a \verb|presence_ratio_minimum| of 0.9 to filter out data corrupted by electrode drift, and an \newline \verb|isi_violations_maximum| of 0.5 to filter out units that contain interspike-intervals that violate a plausible refractory period. While these are the default values from the AllenSDK, we used an \verb|amplitude_cutoff_maximum| of 0.01 (the AllenSDK default is 0.1) to ensure that most spikes are included. 
The number of sorted units after each filtering stage is shown in Supplementary Fig.~\ref{fig:units_filtering}B. 
% XXX: Mention this here? We find a minimal firing rate of approximately 0.02 Hz and a maximal firing rate of approximately 90 Hz, with 95\% of firing rates in the range of 0.19 Hz to 21.11 Hz. The highest firing rates are certainly biologically implausible, but units with these values are few so that they should not distort results in any significant way

\subsection*{Estimation of intrinsic timescale}
The intrinsic timescale $\tau_C$ was estimated as the exponential decay rate of the autocorrelation single neuron spike trains~\cite{siegle_2021, zeraati2022flexible, zeraati2023intrinsic}. 
Spike trains were obtained by binning spiking activity in bins of $\Delta t = 5 \,\text{ms}$, yielding a series of binary activations $a_t$, where $a_t = 0$ if there was no spike, and $a_t = 1 $ if there was one or more spikes in the time interval $[t, t+\Delta t)$. 
The autocorrelation for time lags $T$ was then computed as 
\begin{equation}
    C(T)=\frac{\langle a_t a_{t-T}\rangle_t - \langle a_t \rangle_t^2 }{\langle a_t^2\rangle_t - \langle a_t \rangle_t^2},
\end{equation}
where $\langle \cdot \rangle_t$ is the average over all times $t = T, T+\Delta t, ..., T_{\text{rec}} - \Delta t$, and $T_{\text{rec}}$ is the total recording time.
From this, the intrinsic timescale $\tau_C$ was obtained by fitting an exponential decay $C(T) \propto \exp\left(-\frac{T}{\tau_C}\right)$ to the empirical autocorrelation function. 
Although~\cite{zeraati2022flexible} report that such a naive exponential fit can lead to biased estimates for \emph{short} recordings, the long recording times $T_{\text{rec}}$ of roughly 20 minutes in the analyzed experiment were deemed sufficiently long for an unbiased estimation. 

% For quantifying the intrinsic timescale $\tau_C$, an exponential decay was fit to the autocorrelation function $C(T)$
% \begin{equation}
%     C(T)=\frac{\langle a_t a_{t-T}\rangle_t - \langle a_t \rangle_t^2 }{\langle a_t^2\rangle_t - \langle a_t \rangle_t^2},
% \end{equation} i.e., it was assumed that the autocorrelationdecays exponentially with time lag $T$, i.e. $C(T) \propto \exp\left(-\frac{T}{\tau_C}\right)$. Here, $\langle \cdot \rangle_t$ is the average over all times $t = T, T+\Delta t, ..., T_{\text{rec}} - \Delta t$, where $T_{\text{rec}}$ is the total recording time. 

Moreover, the assumption of an exponentially decaying autocorrelation function is met for a branching process~\cite{wilting_2018a}, and was found to be a good match for most of the neurons in the data (see Supplementary Fig.~\ref{fig:autocorrelation_functions_data}). 
However, we often observed an additional, slow decrease of the autocorrelation function for very long lags $T$, possibly reflecting scale-free fluctuations that might be induced by sensory inputs with scale-free temporal correlations~\cite{dong1995statistics}, or behavioral states~\cite{piasini2021temporal}.
To still estimate an intrinsic timescale $\tau_C$ we thus fitted the measured autocorrelation with a function consisting of two exponential terms
\begin{equation}
    f(T) = A_1 \exp\left(-\frac{T}{\tau_1}\right) + A_2 \exp\left(-\frac{T}{\tau_2}\right).
\end{equation}
Here, we defined the timescale with the larger coefficient $A_i$ as the intrinsic (dominant) timescale $\tau_C=\tau_i$, whereas the other coefficient $A_{j\neq i}$ corresponds to a secondary timescale $\tau_{\mathrm{sec}}=\tau_j$ that accounted for complementary effects. 
Note that for $\approx 80 \%$ of units $\tau_{\mathrm{sec}}$ was larger than $\tau_C$, i.e., it mainly accounted for an additional, slow decay (see Supplementary Fig.~\ref{fig:timescale_fits_distribution} for a distribution of the fitted timescales). 

% Maybe it would be better to mention first the deviations, and hen how we dealt with it. Now it is a bit mixed 

Fitting an additional decay made the analysis of the intrinsic timescale much more robust to the exact choice of the fitting range.
In particular, the two-timescale fit yielded consistent estimates of $\tau_C$ when varying the maximum time lag $T_{\mathrm{max}}$ considered for fitting. In contrast, when fitting a single timescale function with offset $f(T) = A\exp\left(-\frac{T}{\tau_C}\right) + O$, the fitted $\tau_C$ increased monotonously with $T_{\mathrm{max}}$, because considering larger and larger lags $T$ biases estimates towards the slow (possibly scale-free) decay of the autocorrelation~(Supplementary Fig.~\ref{fig:mre_Tmax_comparison}A). 
Yet, we found that considering these large time lags during fitting is crucial because a smaller fitting range (e.g., $T_{\mathrm{min}} = 500\,\mathrm{ms}$) biases estimates towards much shorter timescales~(Supplementary Fig.~\ref{fig:mre_Tmax_comparison}A), and does not reveal a gradient of timescales along the anatomical hierarchy~(Supplementary Fig.~\ref{fig:hierachy_fitting_ranges_tau_C_tmax500}).
For the analysis of experimental data, we therefore used a two-timescale fit with $T_{\mathrm{max}}= 10\,\mathrm{s}$.

In addition, we found that the intrinsic timescale from a two-timescale fit was also less sensitive to deviations from an exponential decaying autocorrelation for small time lags. 
In particular, such deviations consisted of both, negative autocorrelation due to refractoriness and adaptation, or high initial autocorrelation, possibly indicating bursty firing or other short-term facilitating effects (see Supplementary Fig.~\ref{fig:autocorrelation_functions_data}).
To exclude such short-term effects, we excluded autocorrelation coefficients $C(T)$ for short time lags $T < T_{\text{min}}$ during the fit.  
Here, $T_{\text{min}} = 30\,\text{ms}$ was chosen, which gave the best fitting results for most neurons (Supplementary Fig.~\ref{fig:autocorrelation_functions_data}).
% Therefore, only autocorrelation coefficients $C(T)$ for time lags $T > T_{\text{min}}$ where considered for the fit. Here, $T_{\text{min}} = 30\,\text{ms}$ was chosen, which we found was sufficient in most cases to exclude unwanted short-term effects, and gave the best fitting results for most neurons (Supplementary Fig.~\ref{fig:autocorrelation_functions_data}). 
An alternative approach to deal with such short-term effects is to fit an additional shorter timescale as in~\cite{zeraati2022flexible, zeraati2023intrinsic}, but this assumes an exponential decay for these short-term effects, which might not be the case in general.
Moreover, whereas results from the single-timescale analysis were sensitive to the choice of $T_{\text{min}}$, we found that results from the two-timescale fit did not crucially depend on $T_{\text{min}}$ (Supplementary Figs.~\ref{fig:hierachy_fitting_ranges_tau_C_two_timescales_tmax10000} and \ref{fig:mre_Tmax_comparison}). 
Thus, the overall impact of short-term effects on our results is comparably weak, which is also a result of the large fitting range (i.e., $T_{\mathrm{max}}= 10\,\mathrm{s}$) considered here. 

To conclude, the hypothesis underlying this analysis is that between short and long timescales (in a range roughly between 20
to 1000 ms), there exists a decay that is characteristic for a unit’s temporal processing, e.g., due to
recurrent activity propagation in a network~\cite{wilting_2018, wilting_2019, zeraati2023intrinsic}.
Here, we used a two-timescale fit that enables to extract such a characteristic timescale from
the empirical autocorrelation function of individual units in a way that is robust to exact fitting parameters. 

\subsection*{Estimation of predictability and information timescale} 
To not only assess the timescale of time-lagged dependence, but also how predictable spiking is, we estimate a neuron's predictability as the relative spiking information $R(T)$ that can be predicted from past spiking in a past range $T$ (Fig.~\ref{fig:measures}B)~\cite{rudelt_2021}. 
Since more past information can only increase predictability, $R(T)$ increases monotonously with $T$ until it reaches some value $R_\text{tot}$, the \emph{total predictability} of a spike train. 
From this analysis we also obtain an \emph{information timescale} $\tau_R$, which gives something similar to a rise time of the predictability, and indicates a typical timescale on which past information is informative, i.e. adds to the predictability of current spiking. 
Details on the measures and their estimation can be found in~\cite{rudelt_2021}. 

Here, we optimized $d=5$ past bins for the estimation of predictability $R(T)$ for past ranges $T\in [T_{\rm min}, T_{\rm max}]$, with $T_{\mathrm{min}} = 30 \,\mathrm{ms}$ and $T_{\mathrm{max}} = 5 \,\mathrm{s}$.
The minimum past range of $T_{\mathrm{min}} = 30 \,\mathrm{ms}$ was chosen so that the timescale $\tau_R$ only reflects past information \emph{beyond} short-term effects due to a neuron's intrinsic spiking dynamics, similar to the exclusion of short time lags for the intrinsic timescale analysis (see Supplementary Fig.~\ref{fig:hierachy_fitting_ranges_tau_R} for a comparison of different $T_{\mathrm{min}}$).
For the maximum past range $T_{\mathrm{max}} = 5 \,\mathrm{s}$, we found that for all the analyzed units the estimated $R(T)$ reached the maximum value, which is required for the analysis of $R_{\rm tot}$.
% mention in the methods that increasing Tmin leads to a better "resolution" for the hierarchy

\subsection*{Hierarchical Bayesian model for area differences}
\begin{figure}
    \centering
    \includegraphics[width = .7\textwidth]{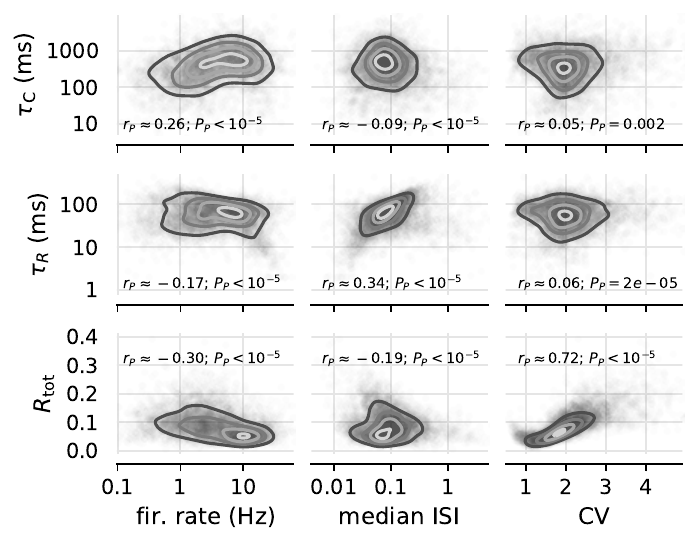}
    \caption{\textbf{Relation of intrinsic and information timescales, as well as predictability to common firing statistics.} 
    Scatter plots of a measure of timescale or predictability versus common firing statistics such as the average firing rate, median inter-spike-interval (ISI) and coefficient of variation (CV) for all analyzed units. Scatter plots are overlaid with kernel density estimations, where lines indicate regions of with equal probability. Intrinsic and information timescales, as well as median ISIs are shown in log scale. The firing rate is mainly positively correlated with the intrinsic timescale, and negatively correlated with the predictability (Pearson correlation). The median ISI is mainly correlated with the information timescale, whereas the CV is strongly correlated with the predictability.
    }
    \label{fig:measures_vs_firing_statistics}
     \captionlistentry[suppfigure]{\textbf{Fig S3. Relation of intrinsic and information timescales, as well as predictability to common firing statistics.}}
\end{figure}

We used a hierarchical Bayesian analysis to investigate whether there are systematic area differences in the estimated timescales or predictability. This analysis enables to consider recordings for each mouse separately, while also incorporating effects that are shared among mice.
The model assumes that the respective measure $y_{i,j} \in \lbrace \tau_C, \tau_R, R_{\mathrm{tot}}\rbrace$ for each unit $i$ that was recorded in mouse $j$ is distributed according to a log-normal distribution, i.e. $\log Y_{i,j}$ is normally distributed
\begin{equation*}
    \log Y_{i,j} \sim \mathcal{N}(\mu_{i,j}, \epsilon).
\end{equation*}
Here, $\mu_{i,j}$ indicates the mean of the log measure (or $\exp(\mu_{i,j})$ the median of $Y_{i,j}$) and contains information about the unit $i$ and mouse $j$, whereas $\epsilon^2$ accounts for the unexplained variance.
For the timescales we found that the distribution of log values was negatively skewed (Supplementary Fig.~\ref{fig:bayes_predictive_checks}), which cannot be accounted for by a normal distribution. Therefore we used a skew normal distribution for the timescales, with 
\begin{equation*}
\log Y_{i,j} \sim \mathrm{SkewNormal}(\mu_{i,j}, \epsilon, \alpha),
\end{equation*}
where the additional shape parameter $\alpha$ controls the skewness of the distribution: $\alpha > 0 $ gives a positively skewed, $\alpha < 0$ a negatively skewed, and $\alpha = 0$ recovers a normal distribution. An important property of this distribution is that, although mean and variance are not the same as for the normal distribution, the scale $\epsilon$ and shape $\alpha$ only add a constant to the mean. Therefore, differences in the log mean are still governed by $\mu_{i,j}$.
For both models, the mean
\begin{equation*}
\mu_{i,j} = \theta_{\log \nu} \log \nu_{i,\text{norm}} + \theta_{\mathrm{rf}} T_{\mathrm{rf},i} + f_j(\mathrm{area}_i) 
\end{equation*}
is the sum of a linear predictor based on the unit's normalized log firing rate $\log \nu_{i,\text{norm}}$, a predictor $\theta_{\mathrm{rf}}$ for visual responsiveness that is added if the unit has a significant receptive field on screen ($T_{\mathrm{rf},i} = 1$, $T_{\mathrm{rf},i} = 0$ otherwise), and a term $f_j(\mathrm{area}_i)$ that models the dependence of the measure on the unit's area, and is specific for each mouse j (see below).  
Here, we included firing rate and the existence of a receptive field in the model, because they were found to be correlated with some of the measures (Supplementary Fig.~\ref{fig:measures_vs_firing_statistics}), thus potentially improving the predictive power of the model. Moreover, not including them might lead to effects that are caused, e.g., by trivial differences in firing rate between different areas. 
Note that we use the logarithm of the firing rate $\nu$ divided by two standard deviations as regression
input, so that its regression coefficient is comparable to the binary ones~\cite{gelman2008scaling}. Hence, a difference of $1$ in $\log \nu_{i,\text{norm}}$ (as between the two states of a binary variable) thus covers 2 standard deviations, e.g. the mean ±1 standard deviation.
As a prior, the standard normal distribution $\mathcal{N}(0,1)$ was chosen for $\alpha$, $\theta_{\nu}$ and $\theta_{\mathrm{rf}}$, whereas for the scale $\epsilon > 0 $ a Half-Chauchy prior distribution $\mathrm{HalfCauchy}(0,\beta)$ was chosen with scale parameter $\beta = 10$. 

We used this approach to test (i) whether there is a systematic difference between structural groups such as thalamus (LGN and LP), primary visual cortex (V1), and higher cortical areas (LM, AL, RL, AM, PM), or (ii) whether cortical areas are correlated with the anatomical hierarchy score, indicating a hierarchy of temporal processing in cortex. 
Therefore, we built the following two models:
\begin{itemize}
    \item[(i)] A \emph{cortical groups} model that incorporates offsets of the log mean for units from areas $\mathrm{area}_i$ of the structural groups thalamus $\theta_{\mathrm{th},j}$ or higher cortical $\theta_{\mathrm{hc},j}$, while the intercept $\theta_{0,j}$ governs the log mean for units from V1: 
    \begin{equation*}
    f_j(\mathrm{area}_i) = \theta_{0,j} + \theta_{\mathrm{th},j} \mathbf{1}_{\mathrm{th}}(\mathrm{area}_i) + \theta_{\mathrm{hc},j}\mathbf{1}_{\mathrm{hc}}(\mathrm{area}_i),
    \end{equation*}
    % $\theta(G_{i}) \in \lbrace\text{thalamus}, \text{V1}, \text{higher cortical}\rbrace$; 
    \item[(ii)] A \emph{cortical hierarchy} model that assumes a linear relation between a unit's log mean and the anatomical hierarchy score of a unit's area $\mathrm{HS}(\mathrm{area}_i)$ with intercept $\theta_{0,j}$ and slope $\theta_{\mathrm{hs},j}$:
    \begin{equation*}
    f_j(\mathrm{area}_i) = \theta_{0,j} + \theta_{\mathrm{HS},j} \mathrm{HS}(\mathrm{area}_i).
    \end{equation*}
    Note that we shifted the hierarchy score to attain zero for V1 so that the intercept $\theta_{0,j}$ again reflects the log mean of units from V1. 
\end{itemize}
All these parameters are modelled hierarchically, where for each mouse there is a parameter set $\boldsymbol{\theta}_j$, and each parameter $\theta_{k,j}$ of that set is drawn from a normal parent distribution with mean $\mu_{\theta_k}$ and variance $\mu_{\theta_k}$. 
This type of modelling respects that data are collected from different mice, but still allows to draw general conclusions based on the means $\mu_{\theta_k}$ of the parent distributions. 
As a prior for the means a standard normal distribution was assumed as a prior $\mu_{\theta_k}\sim \mathcal{N}(0,1)$, whereas for the standard deviation a Half-Cauchy prior distribution was chosen $\sigma_{\theta_k}\sim \mathrm{HalfCauchy}(0,\beta)$ with location 0 and scale parameter $\beta = 1$.
We performed posterior predictive checks for both models  and found that they were both well calibrated for the predictability and information timescale, whereas we found systematic deviations between the observed and modelled distribution of intrinsic timescales (Supplementary Fig.~\ref{fig:bayes_predictive_checks}). 
However, we do not expect this to have a significant effect on the present analysis, since the purpose is not to accurately describe the variability for individual data points, but to make statements about means and average effects of specific predictors.

\subsection*{Branching network model and simulations}
To investigate how intrinsic timescale and predictability of single neurons depends on recurrent amplification, we employ a basic branching network that allows us to individually control external and recurrent activation. 
The network consists of $N=1000$ binary units with state $s_i=\left\{0,1\right\}$ on a random, sparse, directed, and weighted graph with mean degree $k=10$ and connection weights $w_{ij}$ drawn randomly from $[0,1)$.
At discrete time steps ($\Delta t = 5 \,{\rm ms}$), each unit can be activated ($s_i(t)\to 1$) recurrently and externally, and we define the population activity as $A(t)=\sum_i s_i(t)$.
The probability to activate unit $i$ recurrently at time $t$ is 
\begin{equation}
    p_{i,{\rm rec}}(t) = \sum_j w_{ij} \, s_j(t-1).
\end{equation}
To control the mean number of recurrent activations, and to ensure that each unit on average activates the same amount of other units, $w_{ij}$ is normalized such that $\sum_i w_{ij} = m$, where $m\in [0,1)$ is the \emph{neural efficacy}. 
% To control the neural efficacy $m\in [0,1)$, $w_{ij}$ is constructed as sparse matrix with $k$ non-zero entries in each row drawn randomly from $[0,1)$, and normalized such that $\sum_i w_{ij} = m$.
%Note that in the simple case of homogeneous weights, $w_{ij} = m/k$ if connected, $0$ else, however, here we allow heterogeneous weights but apply the same normalisation.
Next to recurrent activations, the probability that unit $i$ is activated \emph{externally} is given by a sigmoidal function
\begin{equation}
    %p_{i,{\rm ext}}(t) = \frac{1}{1+\exp{[-{x_i(t)}/{\sigma} - \gamma}]} \,,
    p_{i,{\rm ext}}(t) = \left[1+e^{-\frac{x_i(t)}{\sigma} - \gamma}\right]^{-1} \,,
\end{equation}
where $x_i(t)\in (-\infty, +\infty)$ is a time-dependent external drive, $\sigma$ sets the sensitivity to the external drive, and $\gamma$ is an offset adjusted to match the desired neural firing rate.
To create temporally correlated input, we model $x_i(t)$ as an Ornstein–Uhlenbeck process with a timescale $\tau_{\rm ext} = 30 \,{\rm ms}$; to create uncorrelated input, we set $x_i(t)=0$.
To match the average neural firing rate to the experimentally observed value $\nu^\ast\approx 3.5\,{\rm Hz}$ (which corresponds to a population activity of $A^\ast = N\nu^\ast\Delta t$), we initialize $\gamma=\ln\left(\frac{1-ma}{1-a}-1\right)$ (the mean-field solution for $x_i(t)=0$) and homeostatically regulate a global $\gamma$ parameter in each time step as
\begin{equation}
    \tau_\gamma\Delta\gamma = \Delta t\left[A^\ast - A(t)\right]/N,
\end{equation}
where $\tau_\gamma=60 \,\text{s}$ is a slow time scale such that $\gamma$ changes very little during the recording.

To estimate the mean and statistical error of intrinsic timescales and predictability, we generate 20 random network realizations for each $m$.
For each realization, the simulation is first equilibrated for $20$ minutes to ensure stationary dynamics, before we record for another $20$ minutes spiking activity from $n=20$ randomly selected neurons.
From the spiking activity, $\tau_C$, $\tau_R$ and $R_{\rm tot}$ are calculated for each neuron using the same tools as for the experimental data.
The only differences are that (i) we removed the lower bound of the fit ranges (starting at $T_{\rm min} = 1$ timestep) and (ii) for fitting $C(T)$ we used a simple exponential with offset (only featuring one timescale) since no separate fast and slow timescales needed to be distinguished in this simple model (see \nameref{sec:discussion}).

\subsection*{Data availability}
All data needed to evaluate the conclusions in the paper are present in the paper and/or the Supplementary Materials. Experimental data were obtained from~\cite{siegle_2021}. Analysis code is available at \url{https://github.com/Priesemann-Group/mouse_visual_timescales}, and pre-processed data as well as simulation data are available at \url{https://gin.g-node.org/pspitzner/mouse_visual_timescales}.

\section*{Acknowledgements}
% Ask if I may mention them in the acknowledgements: Fred Wolf, Davide Zoccolan
We want to thank Davide Zoccolan, Jorge Jaramillo, Brandon Munn and the Priesemann group, especially Jonas Dehning, Fabian Mikulasch and Andreas Schneider, for helpful discussions and comments on the manuscript.

% Add funding information? 
\paragraph{Funding:} All authors received support from the Max Planck Society (Max-Planck-Gesellschaft MPRG-Priesemann). L.R. was supported by the Deutsche Forschungsgemeinschaft (DFG, German Research Foundation) as part of the SPP 2205 - project number 430157073. J.Z.~received financial support from the Joachim Herz Stiftung. F.P.S. and V.P. acknowledge funding from the DFG as part of the SFB 1528 \enquote{Cognition of Interaction}. 
\paragraph{Author contributions:} \mbox{} \\
	Conceptualization: LR, VP\\
	Methodology: LR, DGM, BC, JZ, VP \\
    Software: LR, DGM, BC, PSP, JZ \\
    Formal analysis: LR, DGM, PSP \\
	Investigation: LR, DGM \\
	Visualization: LR, DGM, FPS \\
	Supervision: VP \\
	Writing—original draft: LR, FPS \\
	Writing—review \& editing: LR, FPS, BC, JZ, VP
 \paragraph{Competing interests:} The authors declare that they have no competing interests.

\newpage 
\nolinenumbers
\nocite{gelman_2013}
% \bibliography{main}
\printbibliography

\newpage

\beginsupplement
% \includepdf[pages={1}]{SM_cover.pdf}
% \setcounter{page}{1}
\section*{Supplementary Material} 
% \listofsuppfigures
% \subsection*{List of supplementary figures}
% TODO: How do other papers include SI figures., is there a list of these figures? How did I do it in the PLOS CB?

% \pagebreak

% \clearpage

% \clearpage

% \clearpage

\begin{figure}
    \centering
    \begin{subfigure}[t]{.277\textwidth}
    \subcaption[]{}
    % \vspace{-9pt}
    \includegraphics[width=.93\textwidth]{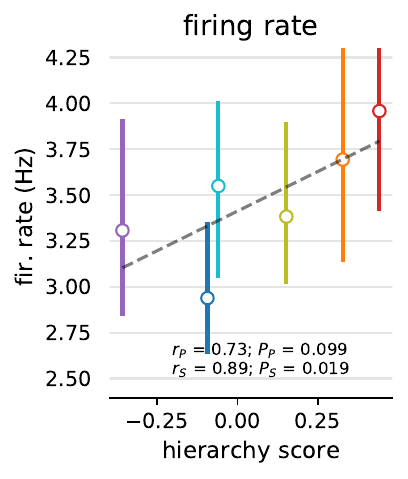}
    \end{subfigure}   \hfill 
    \begin{subfigure}[t]{.272\textwidth}
    \subcaption[]{}
    % \vspace{-9pt}
    \includegraphics[width=.98\textwidth]{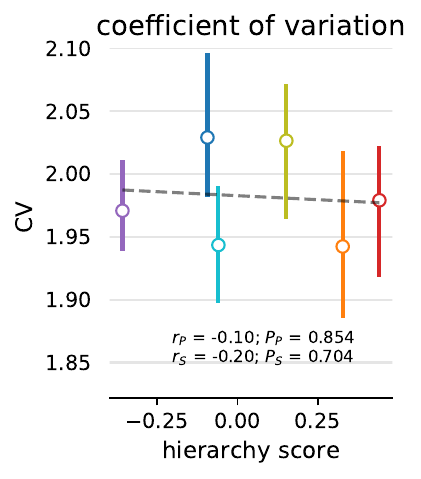}
    \end{subfigure} \hfill   
    \begin{subfigure}[t]{.30\textwidth}
    \subcaption[]{}
    % \vspace{-9pt}
    \includegraphics[width=.84\textwidth]{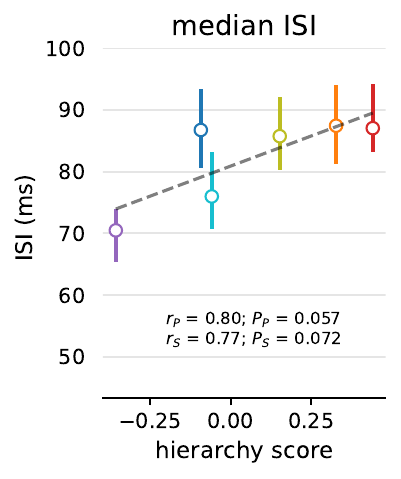}
    \end{subfigure} \hfill
    \begin{subfigure}[t]{.075\textwidth}
    \vspace{13pt}
     \includegraphics[width=\textwidth]{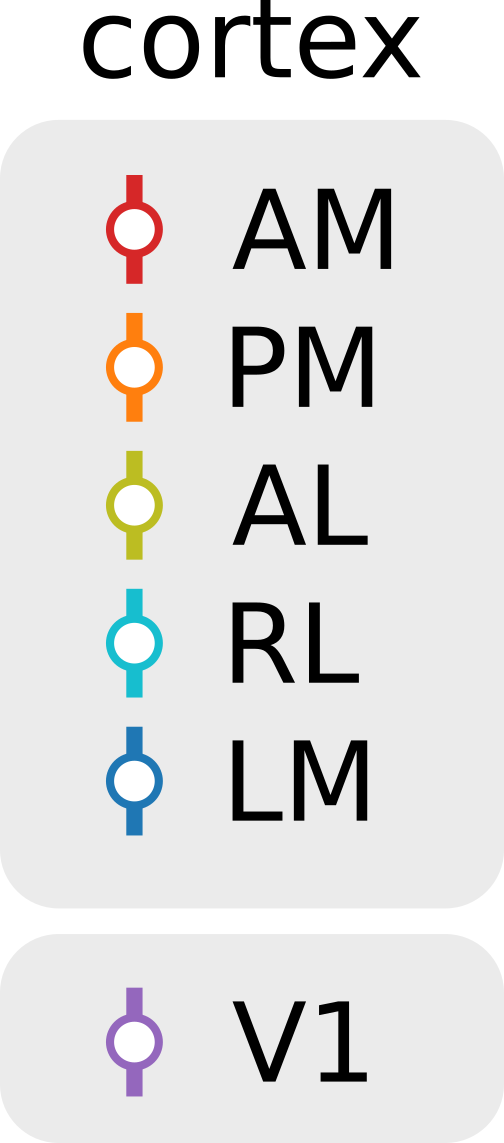}
    \end{subfigure}
    \caption{\textbf{Correlation between single neuron firing statistics and hierarchy score for natural movie stimulation in the \textit{Functional Connectivity} data set.} 
    Firing rate and median interspike-interval (ISI) of sorted units tend to increase with hierarchy score, but have lower Pearson correlation coefficients $r_P$ (with higher p-values $P_P$), as well as Spearman correlation coefficients $r_S$, compared to timescales or predictability on the same data set in Fig.~\ref{fig:main_analysis}. 
    The coefficient of variation (CV), in contrast, is not correlated with the hierarchy score.
    }
 \captionlistentry[suppfigure]{\textbf{Fig S4. Correlation between single neuron firing statistics and hierarchy score for natural movie stimulation in the \textit{Functional Connectivity} data set.}}
    \label{fig:stats_vs_hierarchy_score}
\end{figure}

% \clearpage

% INTRINSIC TIMESCALE ANALYSIS
\begin{figure}
\centering 
  \begin{subfigure}[t]{.3\textwidth}
    \subcaption[]{}
    % \vspace{-9pt}
    \includegraphics[width=.96\textwidth]{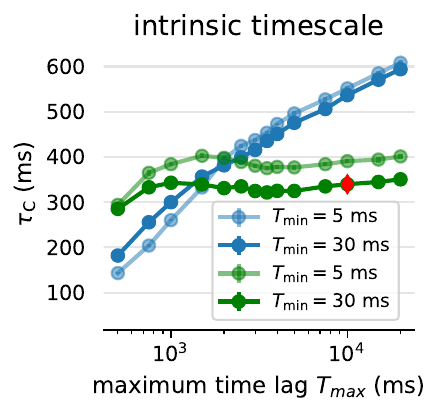}
    \end{subfigure}   \hfill 
    \begin{subfigure}[t]{.33\textwidth}
    \subcaption[]{}
    % \vspace{-9pt}
    \includegraphics[width=.95\textwidth]{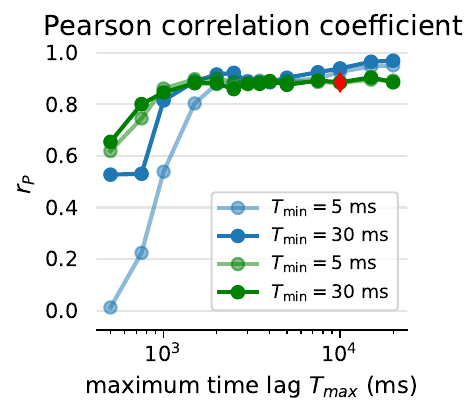}
    \end{subfigure} \hfill   
    \begin{subfigure}[t]{.33\textwidth}
    \subcaption[]{}
    % \vspace{-9pt}
    \includegraphics[width=.95\textwidth]{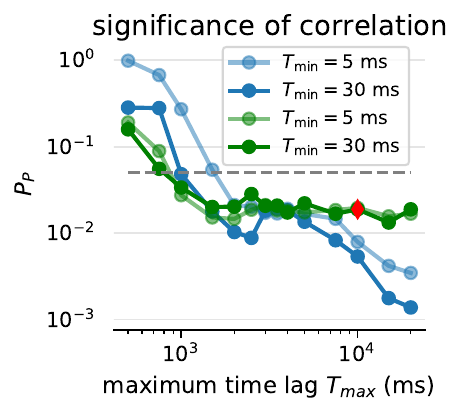}
    \end{subfigure}
    \caption{\textbf{For single timescale fit, median intrinsic timescale and hierarchy score correlation are sensitive to the maximum time lag used for fitting.} 
    \textbf{(A)} For a single timescale fit, the median estimated timescale (over all cortical units) increases monotonously with maximum lag $T_{\text{max}}$, independent of the minimum time lag $T_{\text{min}}$ used for fitting (light and dark blue lines). In contrast, when using a two-timescale fit, the median estimated intrinsic timescale remains consistent for sufficiently large max lags $T_{\text{max}}> 1000 \,\mathrm{ms}$, for both, $T_{\text{min}}= 5 \,\mathrm{ms}$ and $T_{\text{min}}= 30 \,\mathrm{ms}$ (light and dark green lines). However, timescales are larger for $T_{\text{min}}= 5 \,\mathrm{ms}$, because often fits are flattened by negative autocorrelation for short time lags, which are mostly excluded with $T_{\text{min}}= 30 \,\mathrm{ms}$. 
    \textbf{(B)} The Pearson correlation coefficient $r_P$ between an area's median intrinsic timescale and anatomical hierarchy score (c.f.~Fig.~\ref{fig:main_analysis}D) is consistently high if $T_{\text{max}}$ is chosen sufficiently high. 
    \textbf{(C)} Similarly, for sufficiently large $T_{\text{max}}$, the p-value of the fit is consistently below 0.05. For all plots, red dot indicates the $T_{\text{max}}=10000 \,\text{ms}$ used for the main analyses. Timescales were estimated for the natural movie condition of the \emph{Functional Connectivity} data set. 
    }
    \label{fig:mre_Tmax_comparison}
  \captionlistentry[suppfigure]{\textbf{Fig S5. For single timescale fit, median intrinsic timescale and hierarchy score correlation are sensitive to the maximum time lag used for fitting.}}  
\end{figure}

% \clearpage

\begin{figure}
\centering 
    \includegraphics[width=.4\textwidth]{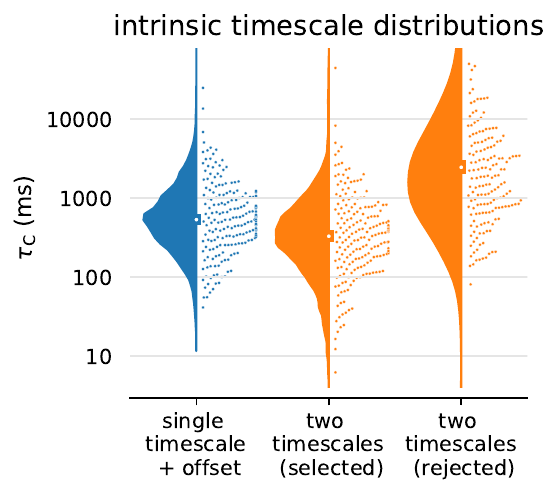}
    \caption{\textbf{Distribution of fitted timescales for the single and two-timescales fit.} 
    To assess how estimated intrinsic timescales differ for the single and two-timescale fits, we here show distribution of the fitted timescales for the single timescale fit (blue), and the two fitted timescales for the two-timescale fit (orange). 
    For the latter, we show both, the selected timescale (see \nameref{sec:methods}), which yields the estimate of the intrinsic timescale, and the rejected timescale. 
    The single timescale is generally larger than the selected timescale of the two-timescale fit, because it also accounts for a potential slow decay of autocorrelation on long timescales. For the two-timescale fit, in contrast, it is mostly the rejected timescale that accounts for the long timescales, since the rejected timescales are generally much larger. 
    }
    \label{fig:timescale_fits_distribution}
     \captionlistentry[suppfigure]{\textbf{Fig S6. Distribution of fitted timescales for the single and two-timescales fit.}}
\end{figure}

% \clearpage

\begin{figure}
\centering
     \includegraphics[width=\textwidth]{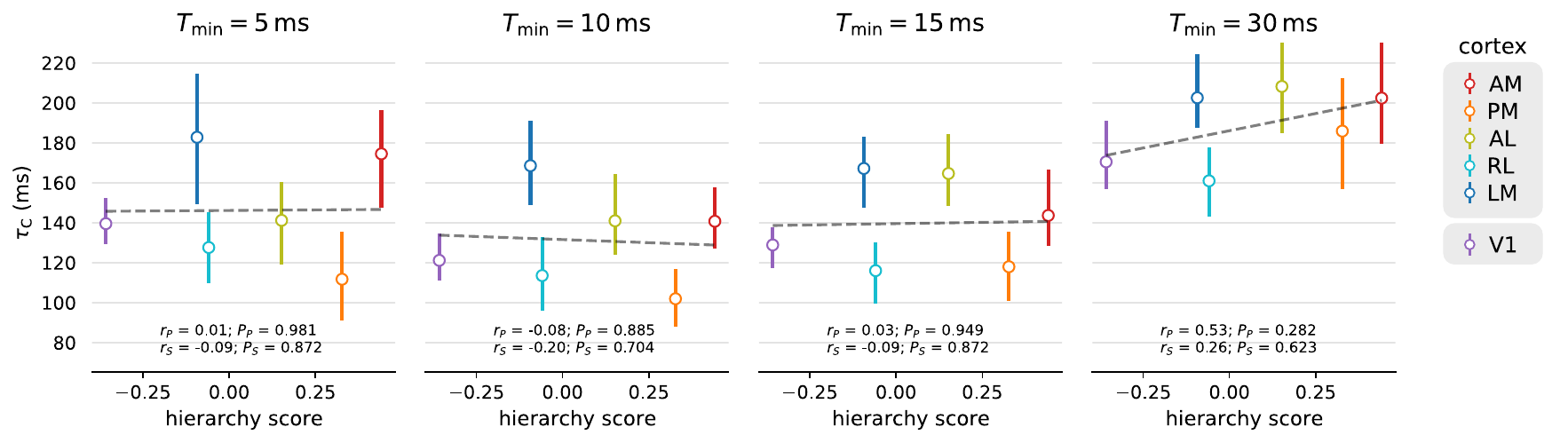}
    \caption{\textbf{No hierarchy in intrinsic timescales is found for a small fitting range.}
    To demonstrate the effect of the fitting range on the inferred hierarchy of intrinsic timescales $\tau_C$, we repeated the same analysis from Fig.~\ref{fig:main_analysis}D for a smaller maximum time lag $T_{\rm max} = 500 \,\text{ms}$, and different minimum time lags $T_{\rm min}$. 
    For this choice of $T_{\rm max}$, no hierarchy is found, and median values of $\tau_C$ are in general much lower (dots, bars indicate bootstrapping confidence intervals on median).
    This is found for all choices of $T_{\rm min}$, indicating that this is primarily caused by omitting larger time lags $T > T_{\rm max}$ during fitting. 
    Here, intrinsic timescales were computed for spiking activity under natural movie stimulation in the \emph{Functional Connectivity} data set.
    Moreover, $\tau_C$ was obtained from a single timescale fit to enable a comparison to previous analyses (c.f. Extended Data Fig.~9 in~\cite{siegle_2021}), but we obtained a similar result also for the two-timescale fit, although this approach is generally much more robust to the choice of fitting range (c.f. Supplementary Fig.~\ref{fig:mre_Tmax_comparison}).
    }
    \label{fig:hierachy_fitting_ranges_tau_C_tmax500}
     \captionlistentry[suppfigure]{\textbf{Fig S7. No hierarchy in intrinsic timescales is found for a small fitting range.}}
\end{figure}

% \clearpage

\begin{figure}
\centering
     \includegraphics[width=\textwidth]{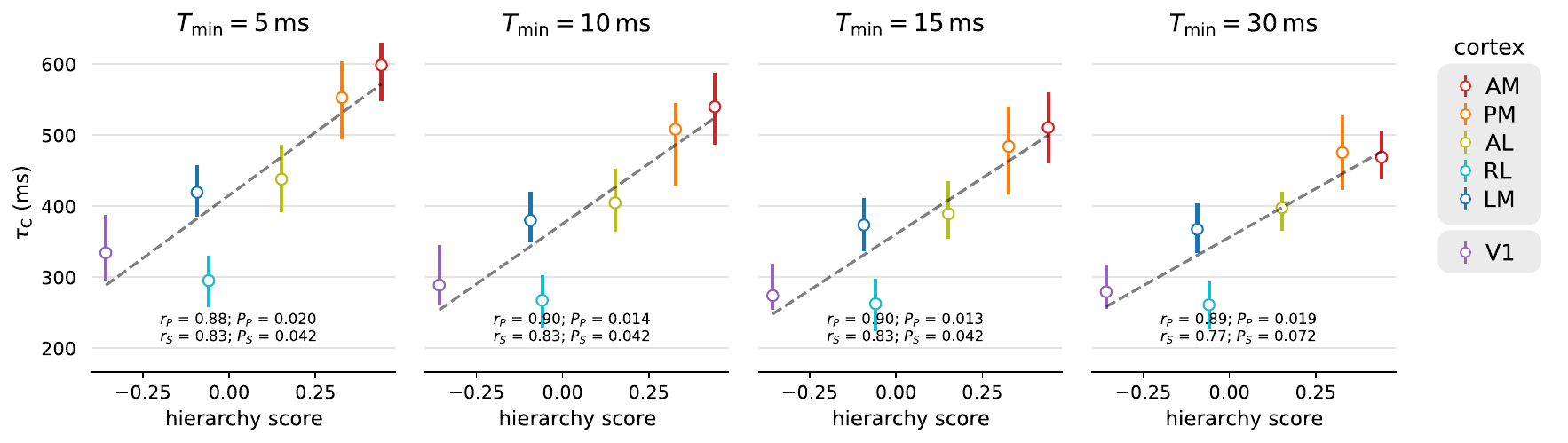}
   \caption{\textbf{For a sufficiently large fitting range, the hierarchy in timescales is found independent of the exclusion of small time lags during fitting.}
   For a large fitting range with $T_{\rm max} = 10000 \,\text{ms}$, we consistently find a hierarchy in intrinsic timescales $\tau_C$, whereas $T_{\rm min}$ only slightly affects the exact layout of the hierarchy, and the median values of $\tau_C$ (c.f. Supplementary Fig.~\ref{fig:mre_Tmax_comparison}).
   Here, intrinsic timescales were computed for spiking activity under natural movie stimulation in the \emph{Functional Connectivity} data set.
   Moreover, timescales where obtained using the two-timescale fitting procedure (\nameref{sec:methods}), because this is the analysis used for all main results obtained in this paper.}
    \label{fig:hierachy_fitting_ranges_tau_C_two_timescales_tmax10000}
     \captionlistentry[suppfigure]{\textbf{Fig S8. For a sufficiently large fitting range, the hierarchy in timescales is found independent of the exclusion of small time lags during fitting.}}
\end{figure}

% \clearpage

% \begin{figure}[t]
% \centering
%     \begin{subfigure}[t]{.9\textwidth}
%     \subcaption[]{natural movie}
%      \includegraphics[width=\textwidth]{SI_figs/allen_T0_hierarchy_Tmax_500_nm_tau_C.pdf}
%     \end{subfigure}
%     \begin{subfigure}[t]{.9\textwidth}
%     \subcaption[]{spontaneous}
%      \includegraphics[width=\textwidth]{SI_figs/allen_T0_hierarchy_Tmax_500_sp_tau_C.pdf}
%     \end{subfigure}
%     \caption{\textbf{No hierarchy in intrinsic timescales when choosing a small fitting range of $T_{\text{max}} = 500\,\text{ms}$}}
%     \label{fig:hierachy_fitting_ranges_tmax500}
% \end{figure}

\begin{figure}
    \centering
    \includegraphics{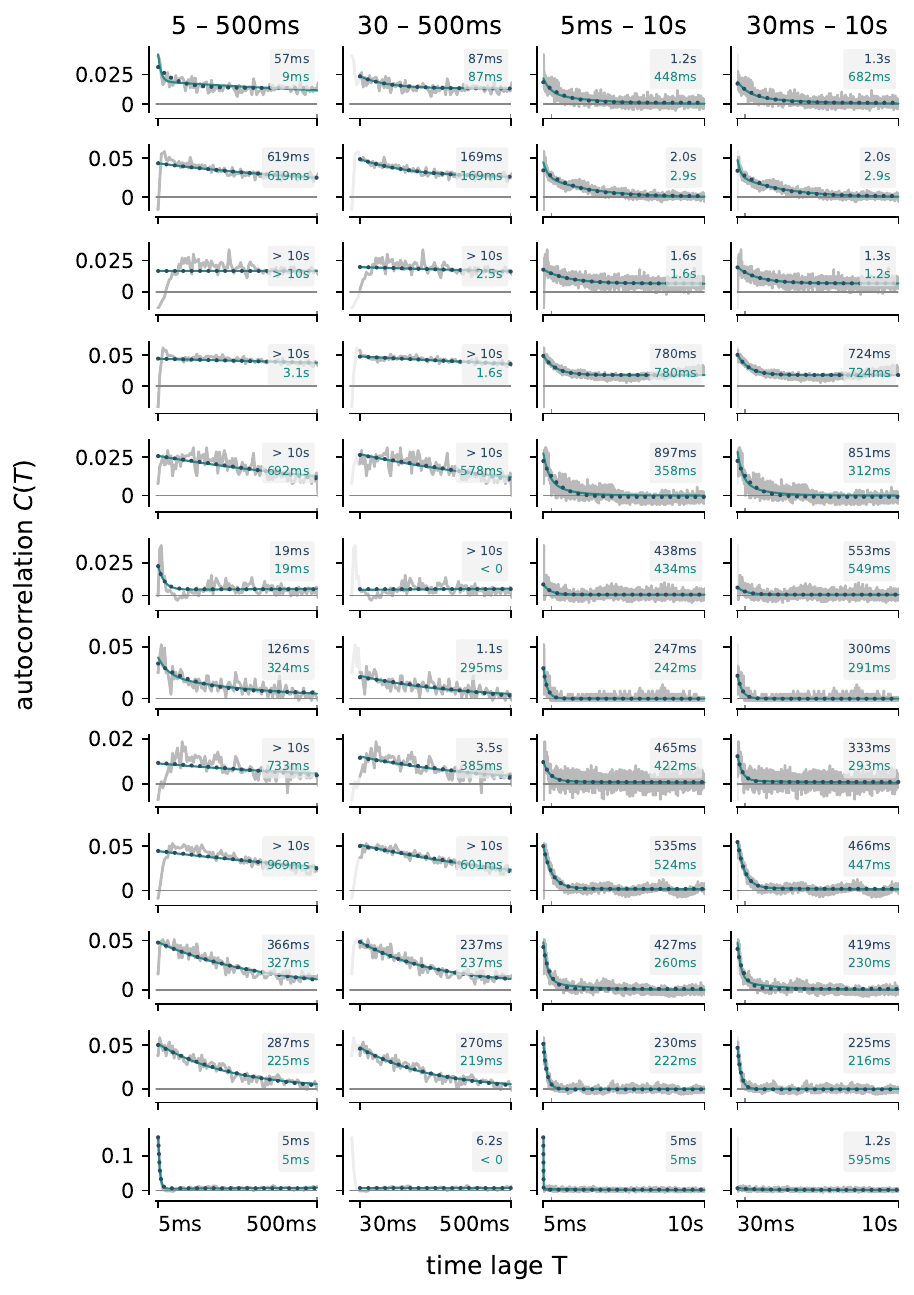}
    \caption{\textbf{Randomly selected examples of autocorrelation functions and single and two-timescale fits for different fitting ranges.} 
    Measured autocorrelation functions (grey line) for units in the \emph{Functional Connectivity} data set under natural movie stimulation.
    Black dots and green lines indicate single and two-timescale fits, respectively, with the inferred timescale stated in the corresponding color.}
    \label{fig:autocorrelation_functions_data}
     \captionlistentry[suppfigure]{\textbf{Fig S9. Example autocorrelation functions and single and two-timescale fits for different fitting ranges.}}
\end{figure}

% \clearpage

\begin{figure}[t]
    \centering
     \includegraphics[width=\textwidth]{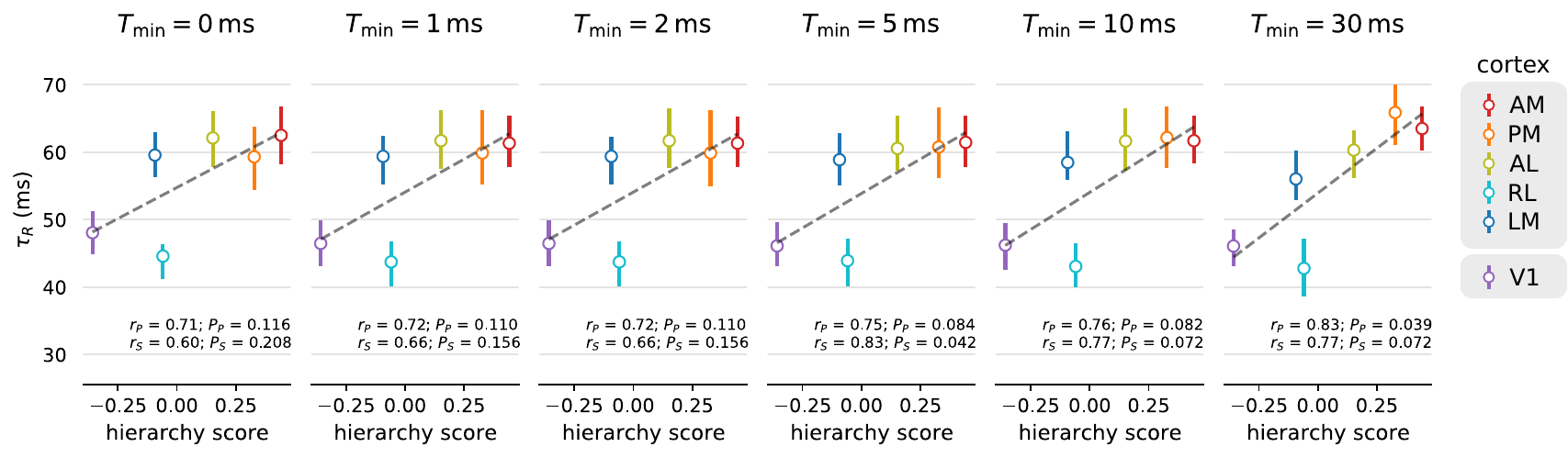}
    \caption{\textbf{Hierarchy in information timescale relies on excluding small past ranges from the analysis.}
    Similar to the intrinsic timescale, we excluded past ranges smaller than some $T_{\rm min}$ when computing the information timescale $\tau_R$ to exclude short-term effects like refractoriness and tonic firing (\nameref{sec:methods}). 
    Here, we show median information timescales for cortical areas, each time for a different choice of minimal past range $T_{\rm min}$.
    Notably, differences between higher cortical areas (LM, AL, PM, AM) are only visible when excluding past ranges smaller than $T_{\rm min}= 30\,\text{ms}$.
    Information timescales were computed for spiking activity under natural movie stimulation in the \emph{Functional Connectivity} data set.
}
    \label{fig:hierachy_fitting_ranges_tau_R}
     \captionlistentry[suppfigure]{\textbf{Fig S10. Hierarchy in information timescale relies on excluding small past ranges from the analysis.}}
\end{figure}

\begin{figure}
    \centering 
    \includegraphics[width=.9\textwidth]{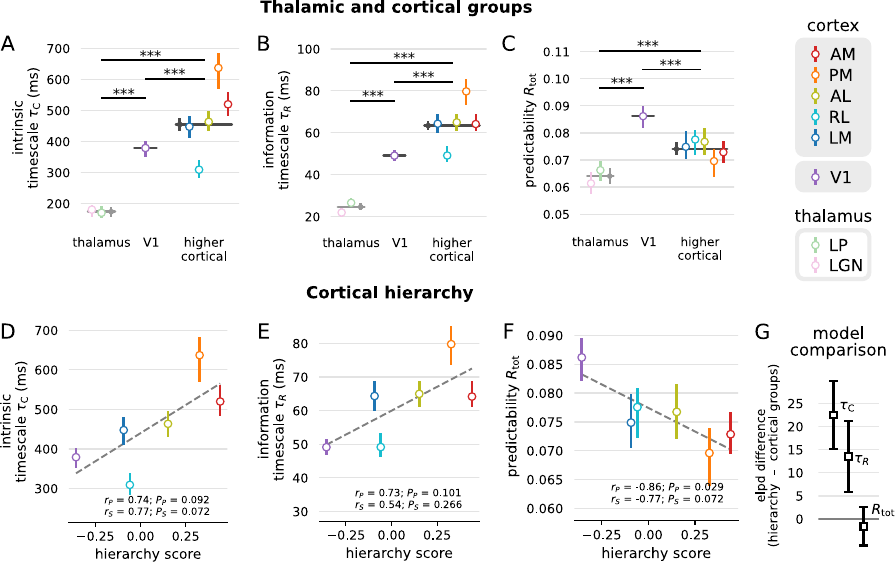}
    \vspace{.5em}
    % Brain Observatory, natural movie \\
    % \begin{subfigure}[t]{.284\textwidth}
    % \subcaption[]{}
    % \vspace{-7.3pt}
    % \includegraphics[width=.95\textwidth]{SI_figs/allen_grouped_bo_nm_tau_C.pdf}
    % \end{subfigure}   \hfill 
    % \begin{subfigure}[t]{.274\textwidth}
    % \subcaption[]{}
    % \vspace{-7.3pt}
    % \includegraphics[width=.95\textwidth]{SI_figs/allen_grouped_bo_nm_tau_R.pdf}
    % \end{subfigure} \hfill   
    % \begin{subfigure}[t]{.29\textwidth}
    % \subcaption[]{}
    % \vspace{-7.3pt}
    % \includegraphics[width=.95\textwidth]{SI_figs/allen_grouped_bo_nm_R_tot.pdf}
    % \end{subfigure} \hfill
    % \begin{subfigure}[t]{.085\textwidth}
    % \vspace{25pt}
    %  \includegraphics[width=.95\textwidth]{main_figs/area_legend_with_thalamus.png}
    % \end{subfigure} 
    % \begin{subfigure}[t]{.277\textwidth}
    % \subcaption[]{}
    % \vspace{-9pt}
    % \includegraphics[width=.95\textwidth]{SI_figs/allen_hierarchy_bo_nm_tau_C.pdf}
    % \end{subfigure}   \hfill 
    % \begin{subfigure}[t]{.272\textwidth}
    % \subcaption[]{}
    % \vspace{-9pt}
    % \includegraphics[width=.95\textwidth]{SI_figs/allen_hierarchy_bo_nm_tau_R.pdf}
    % \end{subfigure} \hfill   
    % \begin{subfigure}[t]{.30\textwidth}
    % \subcaption[]{}
    % \vspace{-9pt}
    % \includegraphics[width=.95\textwidth]{SI_figs/allen_hierarchy_bo_nm_R_tot.pdf}
    % \end{subfigure} \hfill
    % \begin{subfigure}[t]{.085\textwidth}
    % \vspace{25pt}
    %  \includegraphics[width=.96\textwidth]{main_figs/area_legend_cortex.png}
    % \end{subfigure}
    \caption{\textbf{Timescale and predictability of different brain areas under natural movie stimulation in the \textit{Brain Observatory 1.1} data set.} \textbf{(A--C)}~As for recorded activity in the \textit{Functional Connectivity} data set, the medians of all measures differ significantly for different structural groups (thalamus, primary visual cortex and higher cortical), with the same ordering as before. Black boxes indicate the median over sorted units of the different structural groups, whereas coloured dots indicate the median for individual areas. Bars indicate standard deviation on the median obtained from bootstrapping. 
    \textbf{(D--F)}~Measures across the cortical hierarchy show the same general increase as for the \textit{Functional Connectivity} data set. However, information timescales show higher variability, and thus we find smaller correlation coefficients (dashed line, Pearson and Spearman correlation coefficients and p-values shown at the bottom). This might reflect the existence of two pathways, one along posterior parietal areas (RL, AL, AM), which are thought to process information related to motion and behavioral actions, and one along latero- and posteromedial areas LM and PM (see~\nameref{sec:discussion}).}
    \label{fig:analysis_brain_observatory}
     \captionlistentry[suppfigure]{\textbf{Fig S11. Timescale and predictability of different brain areas under natural movie stimulation in the \textit{Brain Observatory 1.1} data set.}}
\end{figure}

% \clearpage

\begin{figure}
    \centering 
    \includegraphics[width=.9\textwidth]{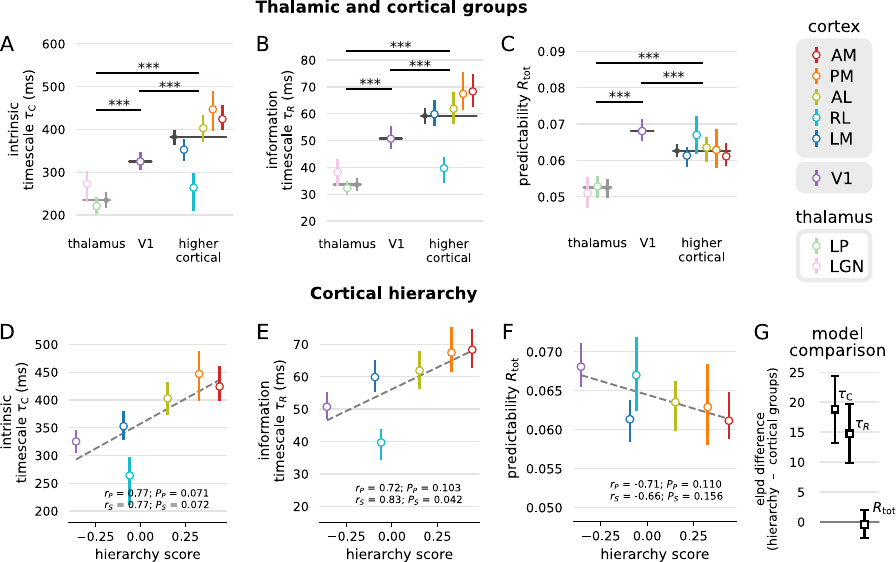}
    \vspace{.5em}
    % Functional Connectivity, spontaneous activity \\
    % \begin{subfigure}[t]{.284\textwidth}
    % \subcaption[]{}
    % \vspace{-7.3pt}
    % \includegraphics[width=.95\textwidth]{SI_figs/allen_grouped_sp_tau_C.pdf}
    % \end{subfigure}   \hfill 
    % \begin{subfigure}[t]{.274\textwidth}
    % \subcaption[]{}
    % \vspace{-7.3pt}
    % \includegraphics[width=.95\textwidth]{SI_figs/allen_grouped_sp_tau_R.pdf}
    % \end{subfigure} \hfill   
    % \begin{subfigure}[t]{.29\textwidth}
    % \subcaption[]{}
    % \vspace{-7.3pt}
    % \includegraphics[width=.95\textwidth]{SI_figs/allen_grouped_sp_R_tot.pdf}
    % \end{subfigure} \hfill
    % \begin{subfigure}[t]{.085\textwidth}
    % \vspace{25pt}
    %  \includegraphics[width=.95\textwidth]{main_figs/area_legend_with_thalamus.png}
    % \end{subfigure} 
    % \begin{subfigure}[t]{.277\textwidth}
    % \subcaption[]{}
    % \vspace{-9pt}
    % \includegraphics[width=.95\textwidth]{SI_figs/allen_hierarchy_sp_tau_C.pdf}
    % \end{subfigure}   \hfill 
    % \begin{subfigure}[t]{.272\textwidth}
    % \subcaption[]{}
    % \vspace{-9pt}
    % \includegraphics[width=.95\textwidth]{SI_figs/allen_hierarchy_sp_tau_R.pdf}
    % \end{subfigure} \hfill   
    % \begin{subfigure}[t]{.30\textwidth}
    % \subcaption[]{}
    % \vspace{-9pt}
    % \includegraphics[width=.95\textwidth]{SI_figs/allen_hierarchy_sp_R_tot.pdf}
    % \end{subfigure} \hfill
    % \begin{subfigure}[t]{.085\textwidth}
    % \vspace{25pt}
    %  \includegraphics[width=.96\textwidth]{main_figs/area_legend_cortex.png}
    % \end{subfigure}
    \caption{\textbf{Timescales and predictability of different brain areas for \emph{spontaneous activity} in the \textit{Functional Connectivity} data set.} \textbf{(A--C)}~As for recorded activity under natural stimulation, the medians of all measures differ significantly for different structural groups (thalamus, primary visual cortex and higher cortical), with the same ordering as before. Black boxes indicate the median over sorted units of the different structural groups, whereas coloured dots indicate the median for individual areas. Bars indicate standard deviation on the median obtained from bootstrapping. 
    \textbf{(D--F)}~Measures across the cortical hierarchy show the same general trend as for natural movie stimulation. However, in general, measures for different areas are more similar, leading to smaller correlation values with the hierarchy score and higher p-values (dashed line, Pearson and Spearman correlation coefficients and p-values shown at the bottom).}
    \label{fig:analysis_spontaneous}
     \captionlistentry[suppfigure]{\textbf{Fig S12. Timescales and predictability of different brain areas for \emph{spontaneous activity} in the \textit{Functional Connectivity} data set.}}
\end{figure}

\begin{figure}
\centering
    { \textbf{Functional Connectivity (natural movie)}} \\
  \begin{subfigure}[t]{.31\textwidth}
    \subcaption[]{}
        \centering
    \vspace{-9pt}
    intrinsic timescale
    \includegraphics[width=\textwidth]{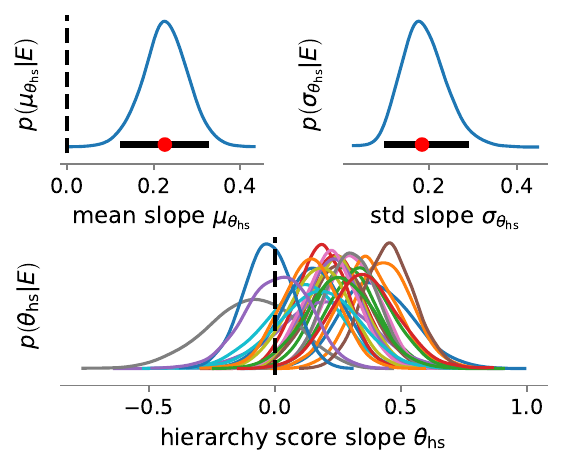}
    \end{subfigure}
    \hfill 
    \begin{subfigure}[t]{.31\textwidth}
    \subcaption[]{}
        \centering
    \vspace{-9pt}
    information timescale
    \includegraphics[width=\textwidth]{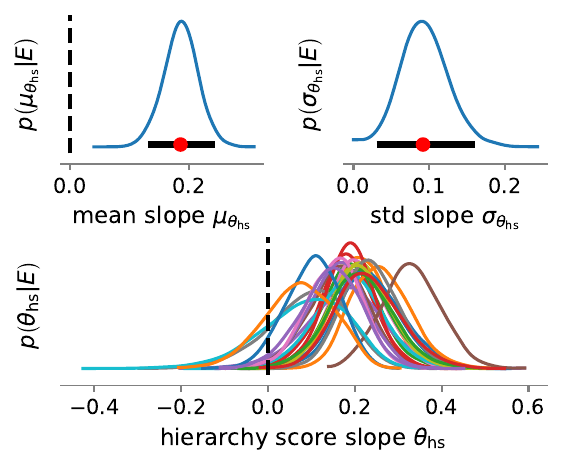}
    \end{subfigure} \hfill   
        \begin{subfigure}[t]{.31\textwidth}
    \subcaption[]{}
    \centering
    \vspace{-9pt}
    predictability
    \includegraphics[width=\textwidth]{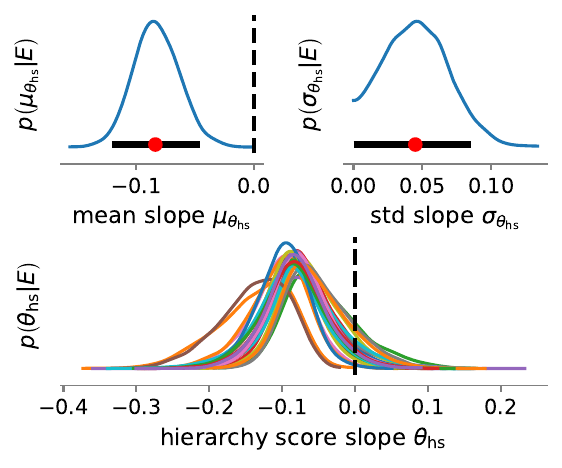}
    \end{subfigure}
     { \textbf{Brain Observatory 1.1 (natural movie)}} \\
  \begin{subfigure}[t]{.31\textwidth}
    \subcaption[]{}
        \centering
    \vspace{-9pt}
    intrinsic timescale
    \includegraphics[width=\textwidth]{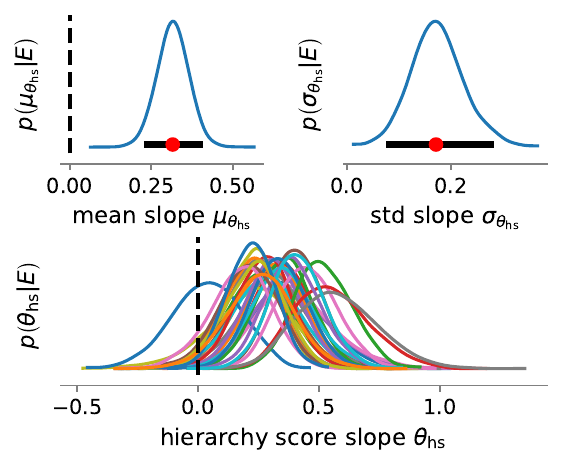}
    \end{subfigure}
    \hfill 
    \begin{subfigure}[t]{.31\textwidth}
    \subcaption[]{}
        \centering
    \vspace{-9pt}
    information timescale
    \includegraphics[width=\textwidth]{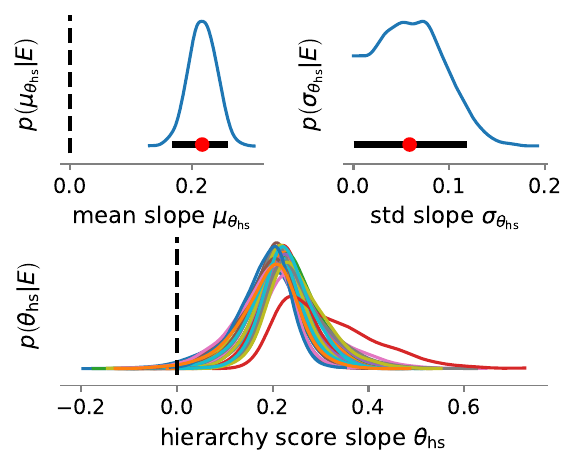}
    \end{subfigure} \hfill   
        \begin{subfigure}[t]{.31\textwidth}
    \subcaption[]{}
    \centering
    \vspace{-9pt}
    predictability
    \includegraphics[width=\textwidth]{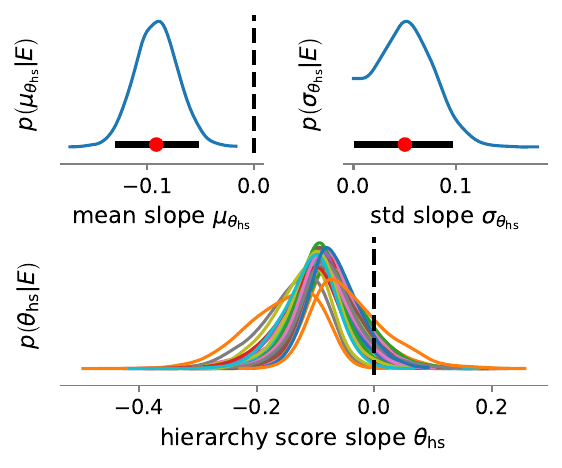}
    \end{subfigure}
       {\textbf{Functional Connectivity (spontaneous)}} \\
  \begin{subfigure}[t]{.31\textwidth}
    \subcaption[]{}
        \centering
    \vspace{-9pt}
    intrinsic timescale
    \includegraphics[width=\textwidth]{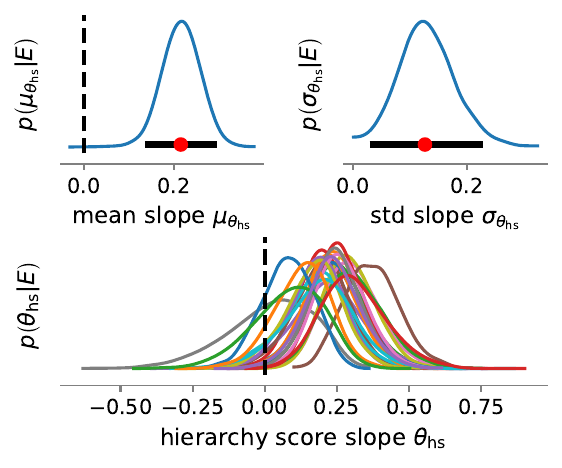}
    \end{subfigure}
    \hfill 
    \begin{subfigure}[t]{.31\textwidth}
    \subcaption[]{}
        \centering
    \vspace{-9pt}
    information timescale
    \includegraphics[width=\textwidth]{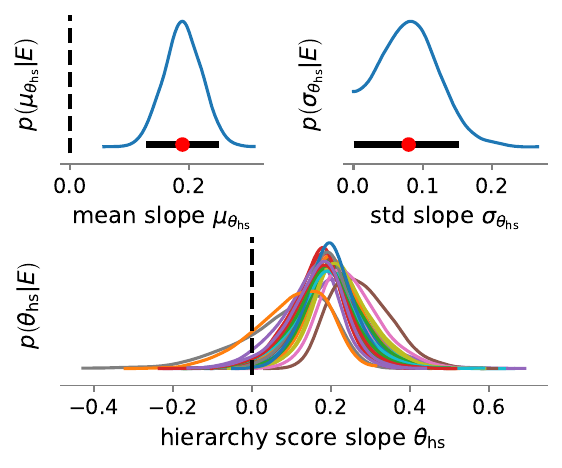}
    \end{subfigure} \hfill   
        \begin{subfigure}[t]{.31\textwidth}
    \subcaption[]{}
    \centering
    \vspace{-9pt}
    predictability
    \includegraphics[width=\textwidth]{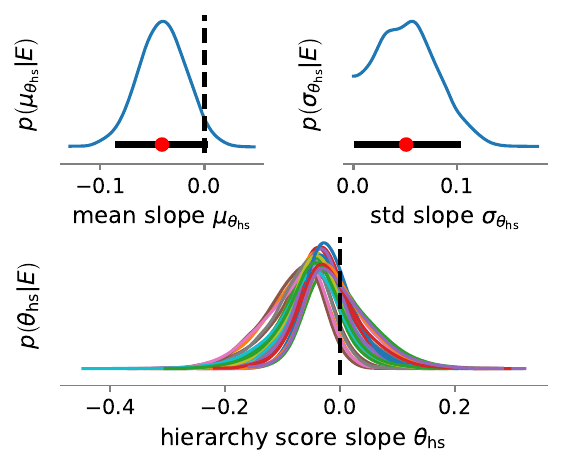}
    \end{subfigure}
    \caption{\textbf{Posterior distributions of the hierarchy score slope reveal a significant increase of timescales and decrease of predictability with hierarchy score.}
    To assess whether timescales and predictability relate to the anatomical cortical hierarchy, a linear relationship between the measures and anatomical hierarchy score was modelled with slope $\theta_{\mathrm{hs}}$. 
    \textbf{(A)} (Top) For the intrinsic timescale, the $95\%$ posterior credible interval of the mean hierarchy score slope $\mu_{\theta_{\mathrm{hs}}}$ across all mice is positive (black bar, red dot indicates median), indicating that there is an increase of intrinsic timescales with the hierarchy score. 
    % The posterior of the standard deviation of slopes $\sigma_{\theta_{\mathrm{hs}}}$ indicates a standard deviation over parameters that are comparable to the mean
    (Bottom) On the level of individual mice, posteriors indicate the same effect, but are more diverse (colors indicate different mice). In particular, for some mice the posteriors also attribute probability to zero or negative slopes, which could be either due to increased uncertainty due to the smaller sampling size, or an incomplete sampling of the areas for individual mice. 
    %According to the posterior the standard deviation of slopes $\sigma_{\theta_{\mathrm{hs}}}$ takes values that are comparable to the mean, which is reflected in a diversity of posteriors of slopes $\theta_{\mathrm{hs}}$ for individual mice (colors indicate different mice). In particular, for some mice the posteriors also attribute significant probability to zero or negative slopes, which could be either due to increased uncertainty due to the smaller sampling size, or an incomplete sampling of the areas for individual mice. 
    \textbf{(B)} Same as A, but for the information timescale $\tau_R$, where one similarly finds a positive posterior credible interval for the mean hierarchy score slope $\mu_{\theta_{\mathrm{hs}}}$.
    \textbf{(C)} For the predictability, the posterior credible interval of the mean slope $\mu_{\theta_{\mathrm{hs}}}$ is negative, hence indicating a credible decrease of predictability with hierarchy score. \textbf{(D--F)} Very similar results are obtained for the \emph{Brain Observatory} data set. \textbf{(D--F)} For spontaneous activity, only the posteriors for timescale slopes indicate a credible positive slope. For predictability, the posterior credible interval also contains a zero slope, indicating that, for spontaneous activity, predictability does not necessarily decrease along the anatomical hierarchy. 
    }
    \label{fig:bayes_hierarchy_score_model_slope_posterior}
     \captionlistentry[suppfigure]{\textbf{Fig S2. Relation between intrinsic and information timescales, as well as predictability across all sorted units.}}
\end{figure}

% \clearpage

% INTERCEPT POSTERIORS, HIERARCHY SCORE MODEL
\begin{figure}
\centering
    { \textbf{Functional Connectivity (natural movie)}} \\
  \begin{subfigure}[t]{.31\textwidth}
    \subcaption[]{}
        \centering
    \vspace{-9pt}
    intrinsic timescale
    \includegraphics[width=\textwidth]{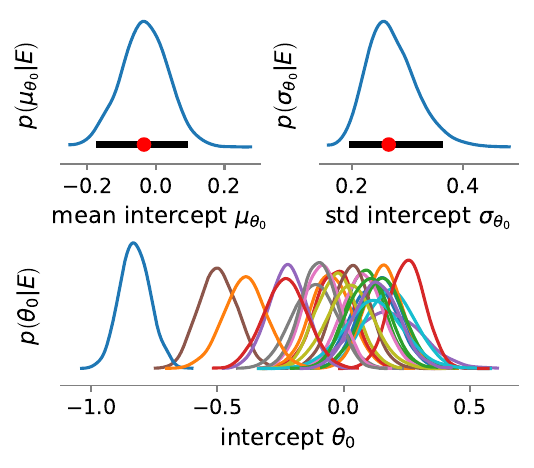}
    \end{subfigure}
    \hfill 
    \begin{subfigure}[t]{.31\textwidth}
    \subcaption[]{}
        \centering
    \vspace{-9pt}
    information timescale
    \includegraphics[width=\textwidth]{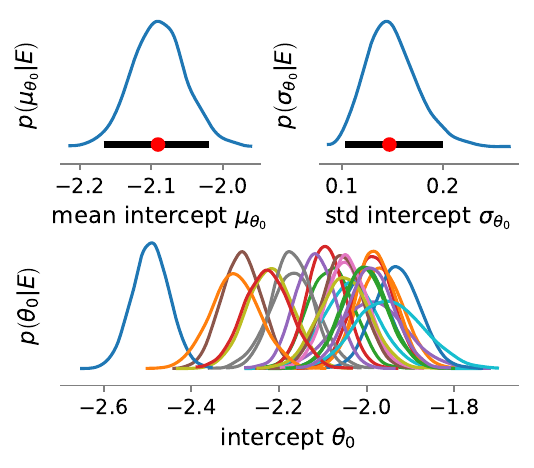}
    \end{subfigure} \hfill   
        \begin{subfigure}[t]{.31\textwidth}
    \subcaption[]{}
    \centering
    \vspace{-9pt}
    predictability
    \includegraphics[width=\textwidth]{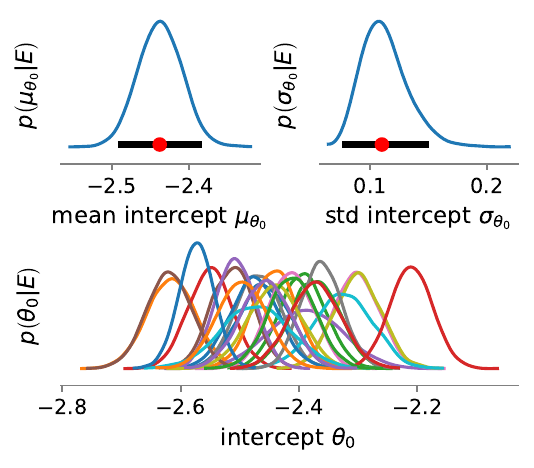}
    \end{subfigure}
     { \textbf{Brain Observatory 1.1 (natural movie)}} \\
  \begin{subfigure}[t]{.31\textwidth}
    \subcaption[]{}
        \centering
    \vspace{-9pt}
    intrinsic timescale
    \includegraphics[width=\textwidth]{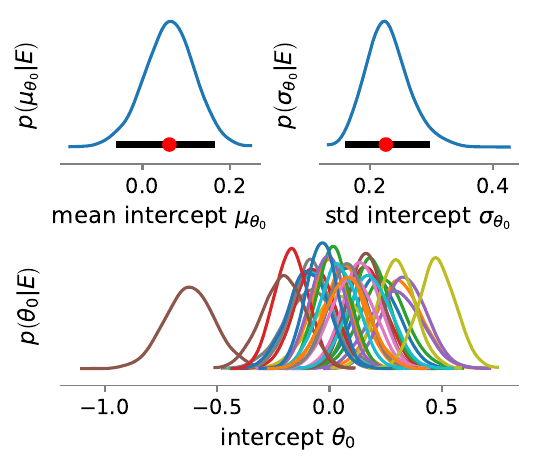}
    \end{subfigure}
    \hfill 
    \begin{subfigure}[t]{.31\textwidth}
    \subcaption[]{}
        \centering
    \vspace{-9pt}
    information timescale
    \includegraphics[width=\textwidth]{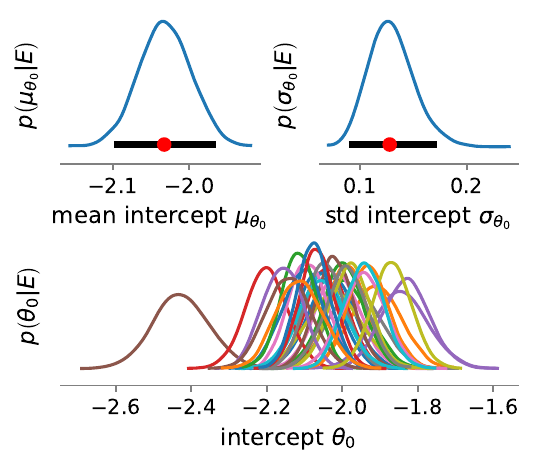}
    \end{subfigure} \hfill   
        \begin{subfigure}[t]{.31\textwidth}
    \subcaption[]{}
    \centering
    \vspace{-9pt}
    predictability
    \includegraphics[width=\textwidth]{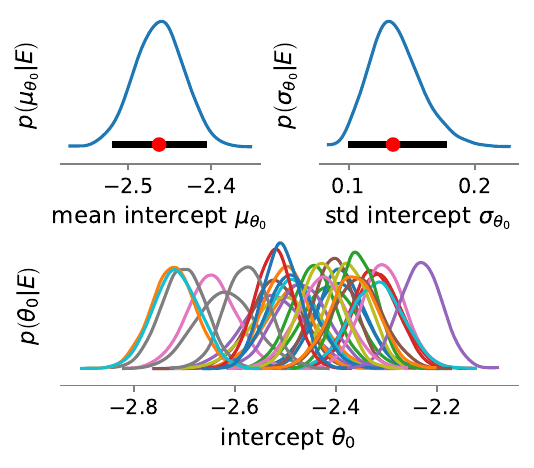}
    \end{subfigure}
       { \textbf{Functional Connectivity (spontaneous)}} \\
  \begin{subfigure}[t]{.31\textwidth}
    \subcaption[]{}
        \centering
    \vspace{-9pt}
    intrinsic timescale
    \includegraphics[width=\textwidth]{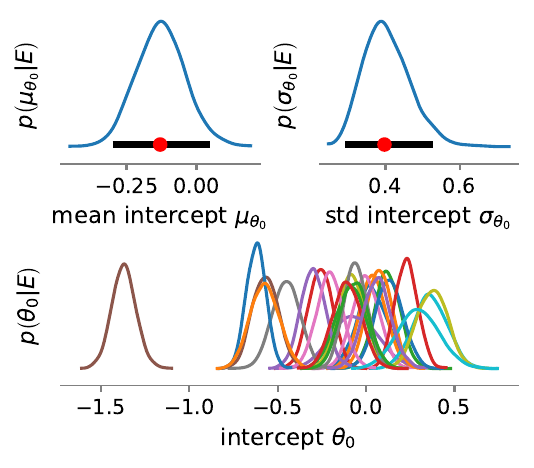}
    \end{subfigure}
    \hfill 
    \begin{subfigure}[t]{.31\textwidth}
    \subcaption[]{}
        \centering
    \vspace{-9pt}
    information timescale
    \includegraphics[width=\textwidth]{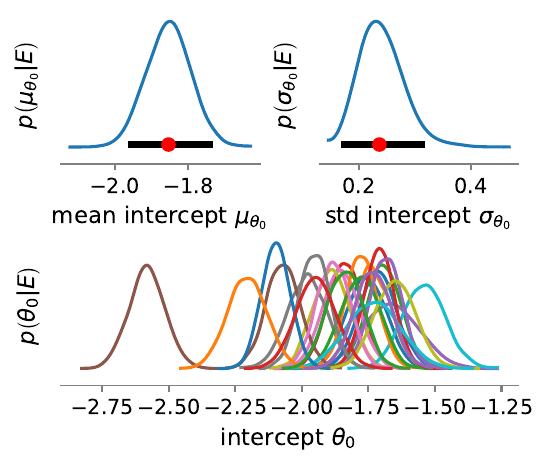}
    \end{subfigure} \hfill   
        \begin{subfigure}[t]{.31\textwidth}
    \subcaption[]{}
    \centering
    \vspace{-9pt}
    predictability
    \includegraphics[width=\textwidth]{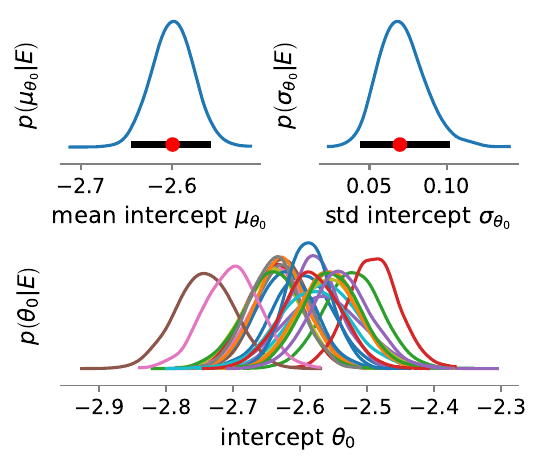}
    \end{subfigure}
    \caption{\textbf{Posterior distributions of intercepts in the cortical hierarchy model.}
    Overall, posteriors over intercepts of the cortical hierarchy model reveal a high diversity among mice, whereas slopes are more similar between different mice (Supplementary Fig.~\ref{fig:bayes_hierarchy_score_model_slope_posterior}).
    \textbf{(A)} (Top) Posterior distributions of the mean $\mu_{\theta_0}$ and standard deviation $\sigma_{\theta_0}$ of the model intercept $\theta_0$ for the intrinsic timescale. (Bottom) Posterior distributions of $\theta_0$ for individual mice (colors indicate different mice). 
    \textbf{(B,C)} Same as A, but for information timescale and predictability.
    \textbf{(D--F)} Same as A--C, but for the \emph{Brain Observatory} data set. 
    \textbf{(G--I)} Same as A--C, but for spontaneous activity in the Functional Connectivity data set. \textbf{(I)} For predictability, intercepts are much more similar between mice, because the slope is much closer to 0 (Supplementary Fig.~\ref{fig:bayes_hierarchy_score_model_slope_posterior}), and thus may partially account for differences in median predictability between mice. 
    For all panels red dots and black bars indicate the median and 95 $\%$ highest-density-interval of the posterior distribution. 
    }
    \label{fig:bayes_hierarchy_score_model_intercept_posterior} 
     \captionlistentry[suppfigure]{\textbf{Fig S2. Relation between intrinsic and information timescales, as well as predictability across all sorted units.}}
\end{figure}

% \clearpage

% OFFSET POSTERIORS, STRUCTURE GROUPS MODEL
\begin{figure}
\centering
    { \textbf{Functional Connectivity (natural movie)}} \\
  \begin{subfigure}[t]{.31\textwidth}
    \subcaption[]{}
        \centering
    \vspace{-9pt}
    intrinsic timescale
    \includegraphics[width=\textwidth]{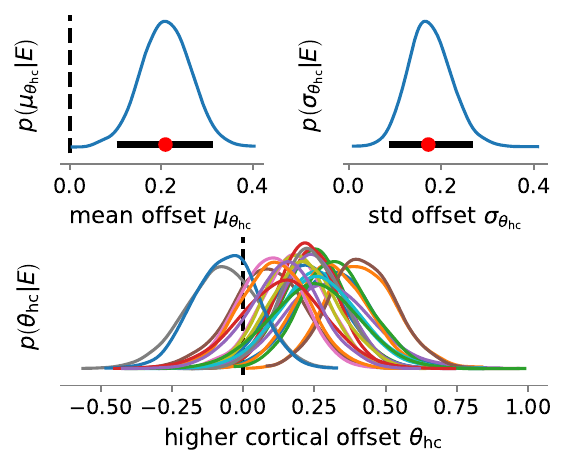}
    \end{subfigure}
    \hfill 
    \begin{subfigure}[t]{.31\textwidth}
    \subcaption[]{}
        \centering
    \vspace{-9pt}
    information timescale
    \includegraphics[width=\textwidth]{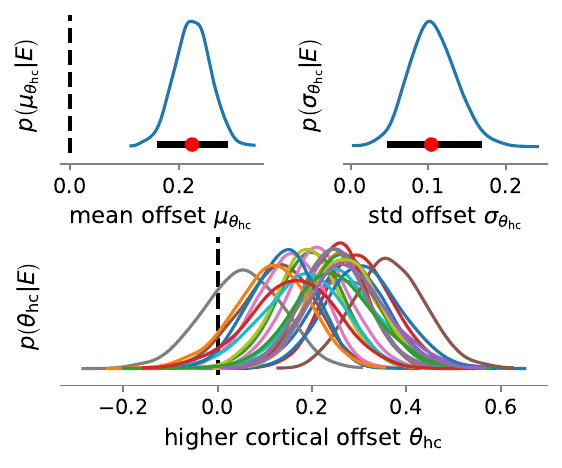}
    \end{subfigure} \hfill   
        \begin{subfigure}[t]{.31\textwidth}
    \subcaption[]{}
    \centering
    \vspace{-9pt}
    predictability
    \includegraphics[width=\textwidth]{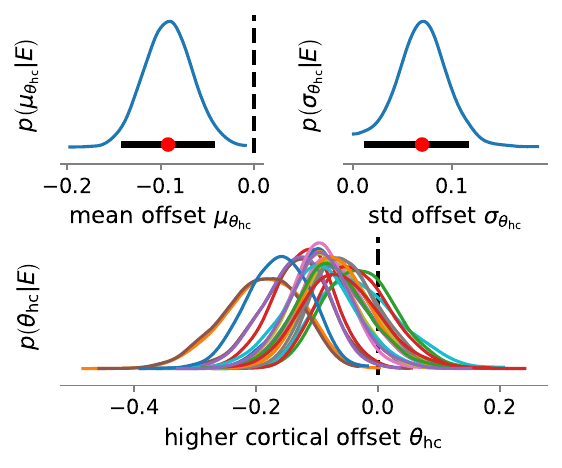}
    \end{subfigure}
     { \textbf{Brain Observatory 1.1 (natural movie)}} \\
  \begin{subfigure}[t]{.31\textwidth}
    \subcaption[]{}
        \centering
    \vspace{-9pt}
    intrinsic timescale
    \includegraphics[width=\textwidth]{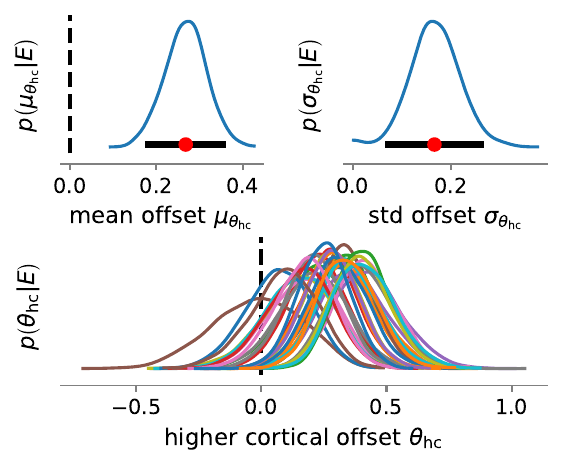}
    \end{subfigure}
    \hfill 
    \begin{subfigure}[t]{.31\textwidth}
    \subcaption[]{}
        \centering
    \vspace{-9pt}
    information timescale
    \includegraphics[width=\textwidth]{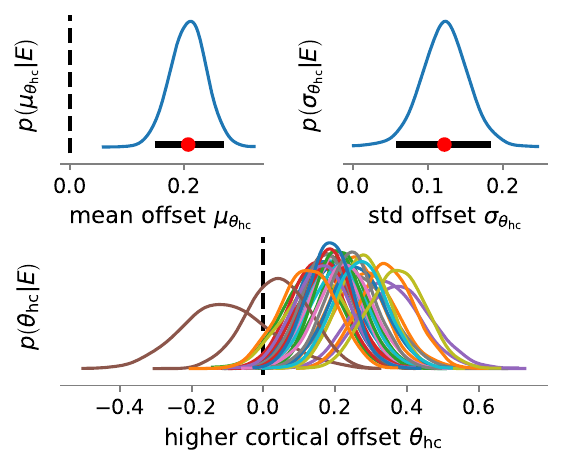}
    \end{subfigure} \hfill   
        \begin{subfigure}[t]{.31\textwidth}
    \subcaption[]{}
    \centering
    \vspace{-9pt}
    predictability
    \includegraphics[width=\textwidth]{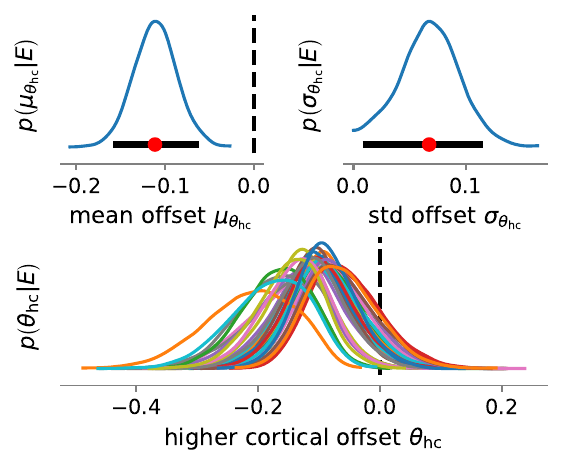}
    \end{subfigure}
    {\textbf{Functional Connectivity (spontaneous)}} \\
    \begin{subfigure}[t]{.31\textwidth}
    \subcaption[]{}
        \centering
    \vspace{-9pt}
    intrinsic timescale
    \includegraphics[width=\textwidth]{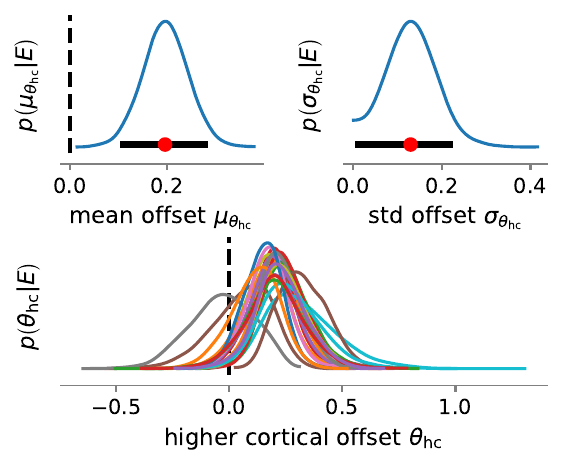}
    \end{subfigure}
    \hfill 
    \begin{subfigure}[t]{.31\textwidth}
    \subcaption[]{}
        \centering
    \vspace{-9pt}
    information timescale
    \includegraphics[width=\textwidth]{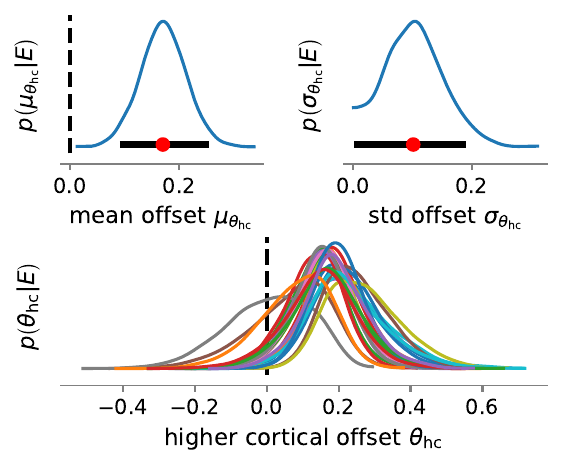}
    \end{subfigure} \hfill   
        \begin{subfigure}[t]{.31\textwidth}
    \subcaption[]{}
    \centering
    \vspace{-9pt}
    predictability
    \includegraphics[width=\textwidth]{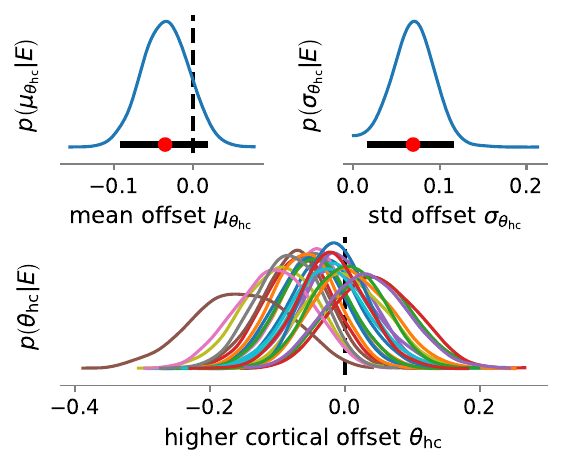}
    \end{subfigure}
    \caption{\textbf{Posterior distributions of the higher cortical offset reveal a significant increase of timescales and decrease of predictability for higher cortical areas.}
    To assess the difference in temporal processing between higher cortical areas and V1, the mean values of timescales and predictability in higher cortical areas were modelled with an offset $\theta_{\mathrm{hc}}$. 
    \textbf{(A)} (Top) For the intrinsic timescale, the  95 $\%$ posterior credible interval of the mean offset $\mu_{\theta_{\mathrm{hc}}}$ across all mice is positive (black bar, red dot indicates median), indicating a credible increase of timescales for higher cortical areas. 
    (Bottom) On the level of individual mice, posteriors indicate the same effect, but are more diverse (colors indicate different mice). In particular, for some mice the posteriors also attribute probability to zero or negative offsets, which could be either due to increased uncertainty due to the smaller sampling size, or an incomplete sampling of the areas for individual mice. 
    %According to the posterior the standard deviation of slopes $\sigma_{\theta_{\mathrm{hs}}}$ takes values that are comparable to the mean, which is reflected in a diversity of posteriors of slopes $\theta_{\mathrm{hs}}$ for individual mice (colors indicate different mice). In particular, for some mice the posteriors also attribute significant probability to zero or negative slopes, which could be either due to increased uncertainty due to the smaller sampling size, or an incomplete sampling of the areas for individual mice. 
    \textbf{(B)} Same as A, but for the information timescale $\tau_R$, where one similarly finds a positive posterior credible interval for the mean offset $\mu_{\theta_{\mathrm{hc}}}$.
    \textbf{(C)} For the predictability, the posterior credible interval of the mean offset $\mu_{\theta_{\mathrm{hc}}}$ is negative, hence indicating a credible decrease of predictability for higher cortical areas. \textbf{(D--F)} Very similar results are obtained for the \emph{Brain Observatory} data set. \textbf{(D--F)} For spontaneous activity, only the posterior credible intervals for timescale offsets are strictly positive. For predictability, the posterior credible interval also contains a zero offset, indicating that, for spontaneous activity, predictability does not necessarily decrease for higher cortical areas.   
    }\label{fig:bayes_structure_groups_model_offset_posterior} 
     \captionlistentry[suppfigure]{\textbf{Fig S2. Relation between intrinsic and information timescales, as well as predictability across all sorted units.}}
\end{figure}

\begin{figure}
\centering
    { \textbf{Functional Connectivity (natural movie)}} \\
  \begin{subfigure}[t]{.31\textwidth}
    \subcaption[]{}
        \centering
    \vspace{-9pt}
    intrinsic timescale
    \includegraphics[width=\textwidth]{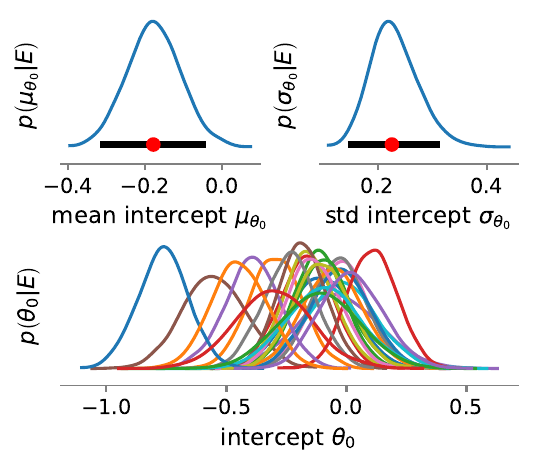}
    \end{subfigure}
    \hfill 
    \begin{subfigure}[t]{.31\textwidth}
    \subcaption[]{}
        \centering
    \vspace{-9pt}
    information timescale
    \includegraphics[width=\textwidth]{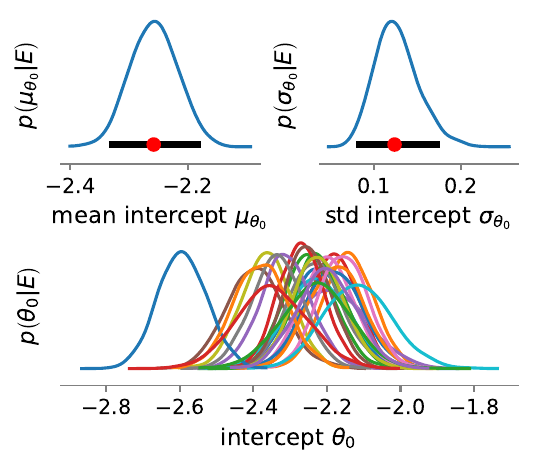}
    \end{subfigure} \hfill   
        \begin{subfigure}[t]{.31\textwidth}
    \subcaption[]{}
    \centering
    \vspace{-9pt}
    predictability
    \includegraphics[width=\textwidth]{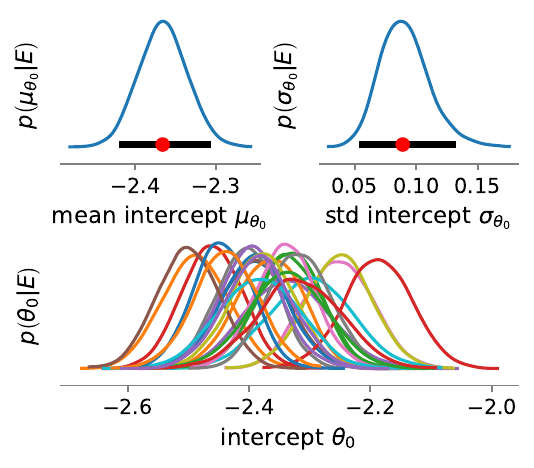}
    \end{subfigure}
     { \textbf{Brain Observatory 1.1 (natural movie)}} \\
  \begin{subfigure}[t]{.31\textwidth}
    \subcaption[]{}
        \centering
    \vspace{-9pt}
    intrinsic timescale
    \includegraphics[width=\textwidth]{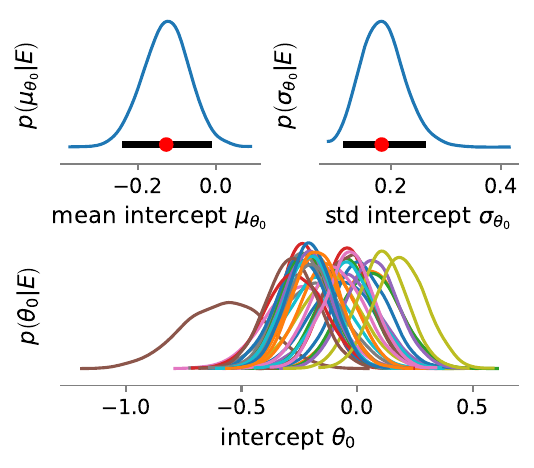}
    \end{subfigure}
    \hfill 
    \begin{subfigure}[t]{.31\textwidth}
    \subcaption[]{}
        \centering
    \vspace{-9pt}
    information timescale
    \includegraphics[width=\textwidth]{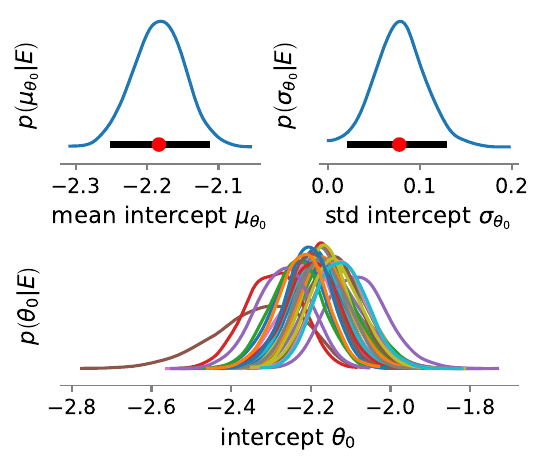}
    \end{subfigure} \hfill   
        \begin{subfigure}[t]{.31\textwidth}
    \subcaption[]{}
    \centering
    \vspace{-9pt}
    predictability
    \includegraphics[width=\textwidth]{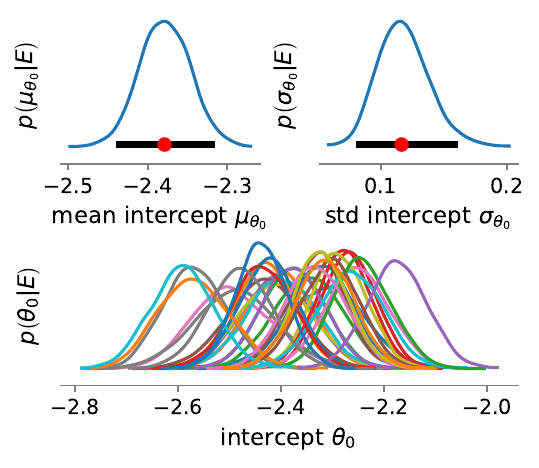}
    \end{subfigure}
    {\textbf{Functional Connectivity (spontaneous)}} \\
    \begin{subfigure}[t]{.31\textwidth}
    \subcaption[]{}
        \centering
    \vspace{-9pt}
    intrinsic timescale
    \includegraphics[width=\textwidth]{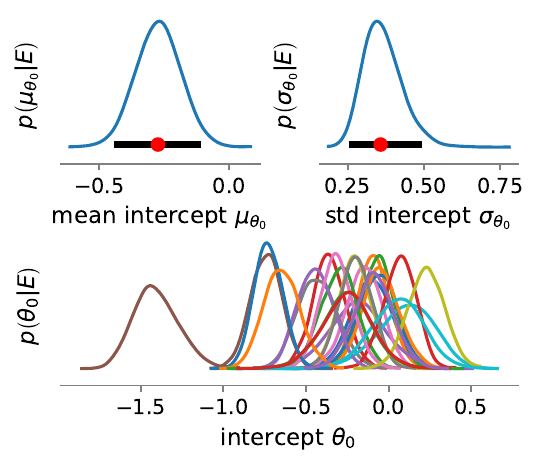}
    \end{subfigure}
    \hfill 
    \begin{subfigure}[t]{.31\textwidth}
    \subcaption[]{}
        \centering
    \vspace{-9pt}
    information timescale
    \includegraphics[width=\textwidth]{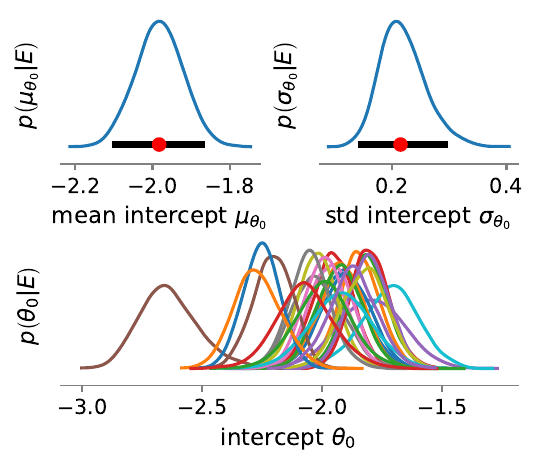}
    \end{subfigure} \hfill   
        \begin{subfigure}[t]{.31\textwidth}
    \subcaption[]{}
    \centering
    \vspace{-9pt}
    predictability
    \includegraphics[width=\textwidth]{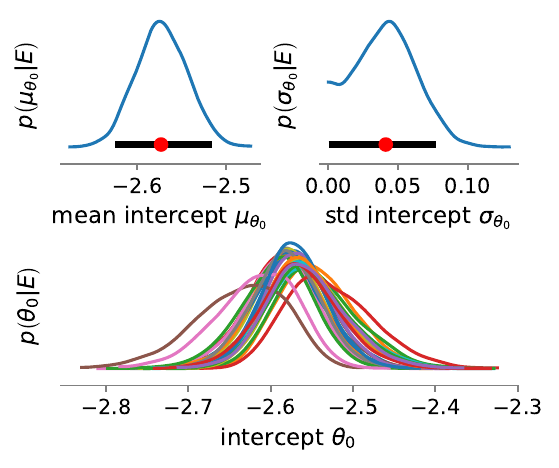}
    \end{subfigure}
     \caption{\textbf{Posterior distributions of intercepts in the cortical groups model.}
    Overall, posteriors over intercepts of the cortical groups model reveal a high diversity among mice, whereas offsets are more similar between different mice (Supplementary Fig.~\ref{fig:bayes_structure_groups_model_offset_posterior}).
    \textbf{(A)} (Top) Posterior distributions of the mean $\mu_{\theta_0}$ and standard deviation $\sigma_{\theta_0}$ of the model intercept $\theta_0$ for the intrinsic timescale. (Bottom) Posterior distributions of $\theta_0$ for individual mice (colors indicate different mice). 
    \textbf{(B,C)} Same as A, but for information timescale and predictability.
    \textbf{(D--F)} Same as A--C, but for the \emph{Brain Observatory} data set. 
    \textbf{(G--I)} Same as A--C, but for spontaneous activity in the \emph{Functional Connectivity} data set. \textbf{(I)} For predictability, intercepts are much more similar between mice, because the higher cortical offset is much closer to 0 (Supplementary Fig.~\ref{fig:bayes_structure_groups_model_offset_posterior}), and thus may partially account for differences in median predictability between mice. 
    For all panels red dots and black bars indicate the median and 95 $\%$ highest-density-interval of the posterior distribution. 
    }
    \label{fig:bayes_structure_groups_model_intercept_posterior} 
     \captionlistentry[suppfigure]{\textbf{Fig S2. Relation between intrinsic and information timescales, as well as predictability across all sorted units.}}
\end{figure}

\clearpage

\begin{figure}
    \centering
    \begin{subfigure}[t]{.32\textwidth}
    \subcaption[]{}
    \centering
    {\tiny \textbf{Functional Connectivity (natural movie)}} \\
    \vspace{-5pt}
    \begin{minipage}[t]{.31\textwidth}
    \vspace{0pt}
    \centering
    \includegraphics[width=\textwidth]{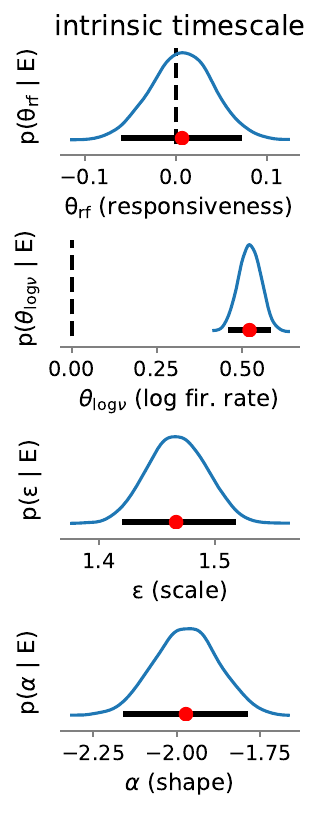}
    \end{minipage} \hfill     
    \begin{minipage}[t]{.31\textwidth}
    \vspace{0pt}
    \centering
     \includegraphics[width=\textwidth]{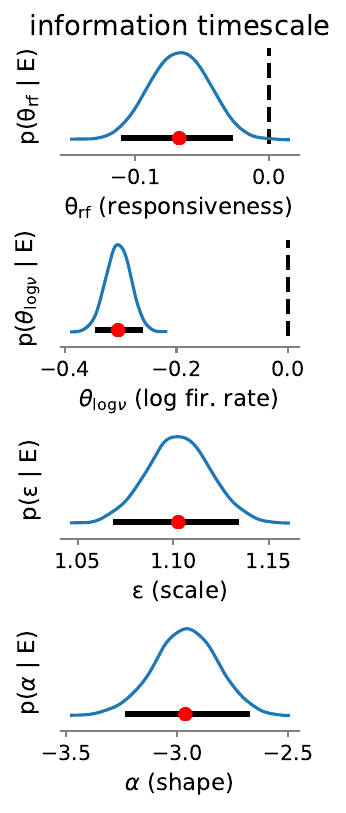}
    \end{minipage} \hfill     
    \begin{minipage}[t]{.31\textwidth}
    \vspace{0pt}
    \centering
    \includegraphics[width=\textwidth]{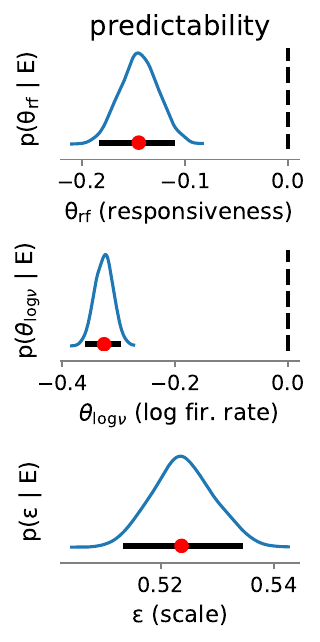}
    \end{minipage} 
    \end{subfigure} \hfill   
      \begin{subfigure}[t]{.32\textwidth}
    \subcaption[]{}
    \centering
    {\tiny \textbf{Brain Observatory 1.1 (natural movie)}} \\
    \vspace{-5pt}
    \begin{minipage}[t]{.31\textwidth}
    \vspace{0pt}
    \centering
    \includegraphics[width=\textwidth]{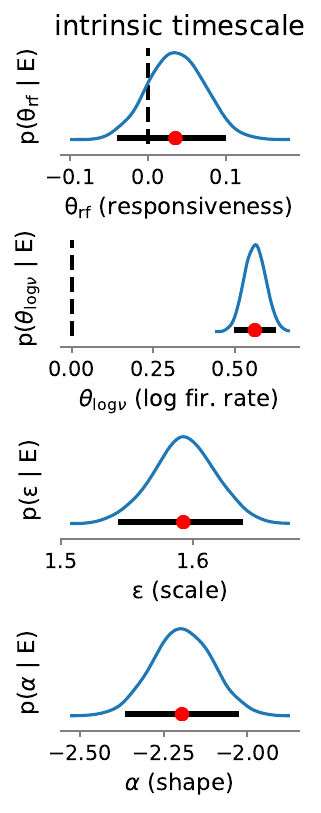}
    \end{minipage} \hfill     
    \begin{minipage}[t]{.31\textwidth}
    \vspace{0pt}
    \centering
     \includegraphics[width=\textwidth]{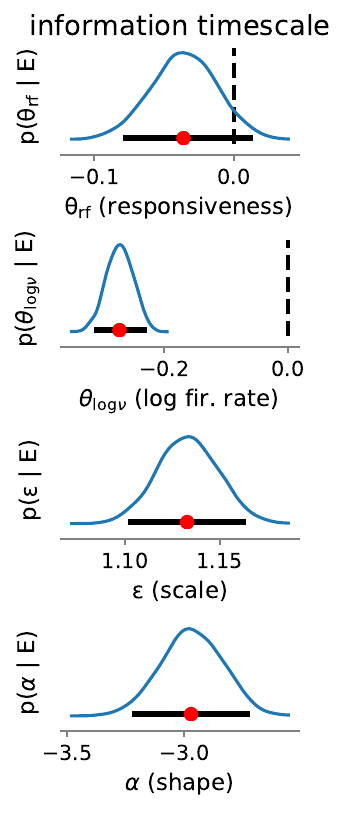}
    \end{minipage} \hfill     
    \begin{minipage}[t]{.31\textwidth}
    \vspace{0pt}
    \centering
    \includegraphics[width=\textwidth]{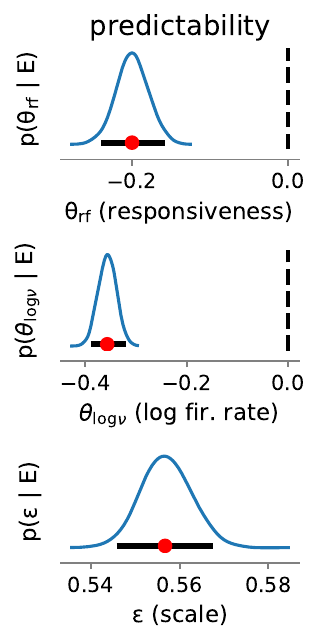}
    \end{minipage} 
    \end{subfigure} \hfill   
      \begin{subfigure}[t]{.32\textwidth}
    \subcaption[]{}
    \centering
    {\tiny \textbf{Functional Connectivity (spontaneous)}} \\
    \vspace{-5pt}
    \begin{minipage}[t]{.31\textwidth}
    \vspace{0pt}
    \centering
    \includegraphics[width=\textwidth]{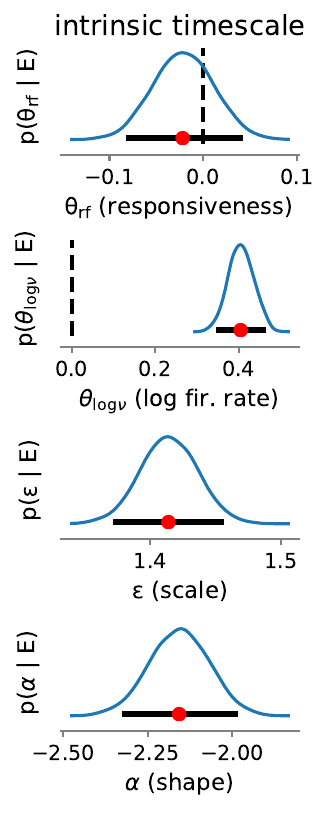}
    \end{minipage} \hfill     
    \begin{minipage}[t]{.31\textwidth}
    \vspace{0pt}
    \centering
     \includegraphics[width=\textwidth]{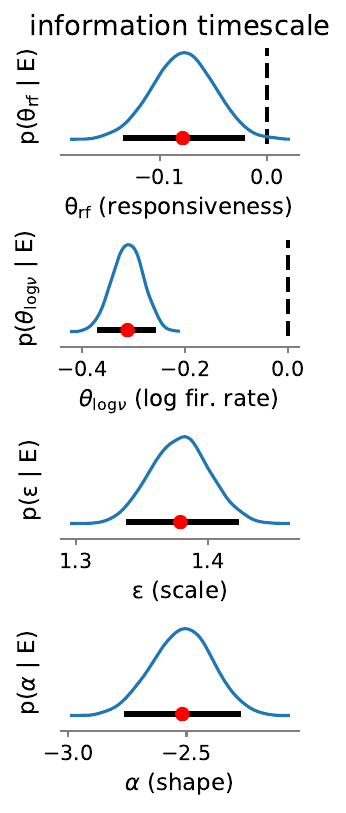}
    \end{minipage} \hfill     
    \begin{minipage}[t]{.31\textwidth}
    \vspace{0pt}
    \centering
    \includegraphics[width=\textwidth]{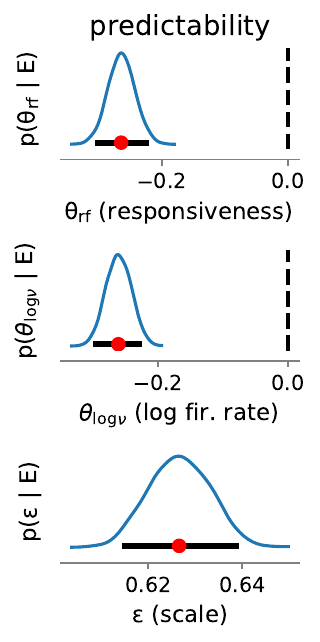}
    \end{minipage} 
    \end{subfigure} \hfill   
    % \begin{subfigure}[t]{.32\textwidth}
    % \subcaption[]{}
    % \vspace{-9pt}
    %     \includegraphics[width=\textwidth]{SI_figs/allen_hierarchical_bayes_hierarchy_score_posterior_nm_tau_R.pdf}
    % \end{subfigure} \hfill   
    % \begin{subfigure}[t]{.32\textwidth}
    % \subcaption[]{}
    % \vspace{-9pt}
    % \includegraphics[width=.9\textwidth]{SI_figs/allen_hierarchical_bayes_hierarchy_score_posterior_nm_R_tot.pdf}
    % \end{subfigure}
    \caption{\textbf{Posteriors for non-hierarchical parameters of the cortical hierarchy model.} \\
%     Moreover, cortical units with a receptive field on screen showed larger intrinsic timescales, but less predictability than units without a receptive field (Supplementary Fig.~\ref{fig:rf_vs_norf_areas}). 
% Although this seems to contradict the idea that visual coding induces single neuron predictability, we think it is more likely that units without a receptive field could represent a different neuron type, or a functional group of neurons with different firing statistics and more predictable spiking.
    \textbf{(A)} Posterior density function for non-hierarchical parameters for the \emph{Functional Connectivity} data set (red dot indicates median and black bar the $95\%$ highest density interval). 
    % For better interpretability, the parameters incorporating unit visual responsiveness $\theta_{\mathrm{rf}}$ and log firing rate $\theta_{\log \nu}$ are transformed from log to linear scale (e.g. $\exp(\theta_{\mathrm{rf}})$), where they correspond to the factor that the mean of $\tau_C$ is multiplied for units with a receptive field on screen, or units with a log firing rate one standard deviation above the mean log firing rate of all units. Hence, a value of 1 indicates no effect of the predictor. 
    For the intrinsic timescale, the credible interval of the receptive field parameter $\theta_{\mathrm{rf}}$ is centered around zero, hence having a receptive field has no credible effect on a unit's intrinsic timescale. In contrast, having a receptive field reduces the mean information timescale and predictability in the model. Although this decrease in predictability seems to contradict the idea that visual coding induces single neuron predictability, we think it is more likely that units without a receptive field could represent a different neuron type, or a functional group of neurons with different firing statistics and overall more predictable spiking.
    The log firing rate has a similar effect on the information timescale and predictability, since the credible interval for log firing rate parameter $\theta_{\log \nu}$ is negative, whereas it is positive for the intrinsic timescale, indicating that the intrinsic timescale is higher for units with higher firing rate. 
    % The scale $\epsilon$ of the skew normal distribution has a narrow posterior around two third relative to the standard deviation of observed log timescales $\log \tau_C$, hence there is considerable residual variance that is not accounted for by the model. 
    The posterior of the shape parameter $\alpha$ is concentrated on negative values, indicating a negatively skewed distribution of log intrinsic timescales. 
    \textbf{(B, C)} Posteriors for the natural movie condition in the \emph{Brain Observatory} data set, and spontaneous activity in the \emph{Functional Connectivity} data set are overall very similar to the posteriors in A. 
    % \textbf{(B)} For the information timescale $\tau_R$, the posterior distribution for $\exp(\theta_{\mathrm{rf}})$ and $\theta_{\log \nu}$  indicate that $\tau_R$ is smaller for units with receptive field and higher firing rate, where the effect is larger for the firing rate. 
    % The posterior for the scale $\epsilon$ has a narrow posterior around one half relative to the standard deviation observed log timescales $\log \tau_R$, hence the model accounts for slightly more residual variance as for the intrinsic timescale. Again, the posterior for $\alpha$ indicates a negatively skewed distribution for log information timescales.
    % \textbf{(C)} Also for the predictability $R_{\mathrm{tot}}$, both the responsiveness as well as the firing rate have a negative effect, where again the effect is strongest for the firing rate. 
    % In contrast to the timescales, the posterior for the scale $\epsilon$ is concentrated on much smaller values of about one quarter relative to the standard deviation of log values of $R_{\mathrm{tot}}$, indicating a much lower residual variance of the model for single unit predictability.
    }
    \label{fig:bayes_hierarchy_score_model_posterior} 
     \captionlistentry[suppfigure]{\textbf{Fig S2. Relation between intrinsic and information timescales, as well as predictability across all sorted units.}}
\end{figure}
% \clearpage

\begin{figure}
    \centering
    \begin{subfigure}[t]{.32\textwidth}
    \subcaption[]{}
    \centering
    {\tiny \textbf{Functional Connectivity (natural movie)}} \\
    \vspace{-5pt}
    \begin{minipage}[t]{.31\textwidth}
    \vspace{0pt}
    \centering
    \includegraphics[width=\textwidth]{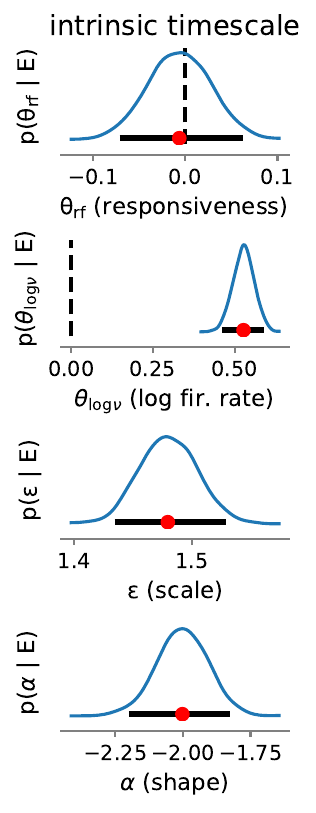}
    \end{minipage} \hfill     
    \begin{minipage}[t]{.31\textwidth}
    \vspace{0pt}
    \centering
     \includegraphics[width=\textwidth]{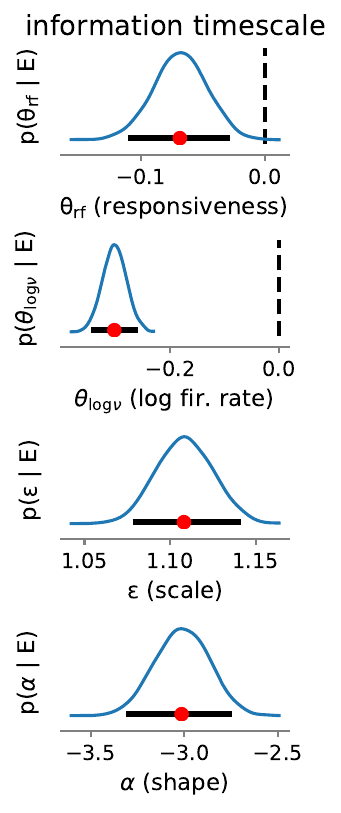}
    \end{minipage} \hfill     
    \begin{minipage}[t]{.31\textwidth}
    \vspace{0pt}
    \centering
    \includegraphics[width=\textwidth]{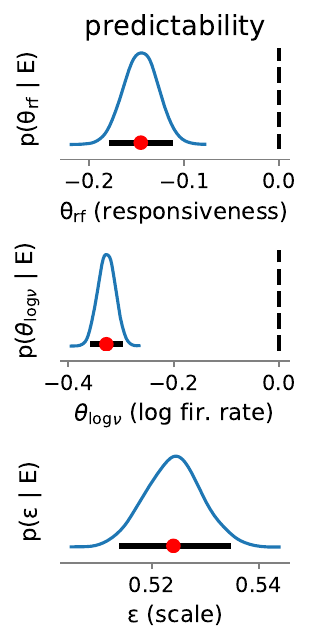}
    \end{minipage} 
    \end{subfigure} \hfill   
      \begin{subfigure}[t]{.32\textwidth}
    \subcaption[]{}
    \centering
    {\tiny \textbf{Brain Observatory 1.1 (natural movie)}} \\
    \vspace{-5pt}
    \begin{minipage}[t]{.31\textwidth}
    \vspace{0pt}
    \centering
    \includegraphics[width=\textwidth]{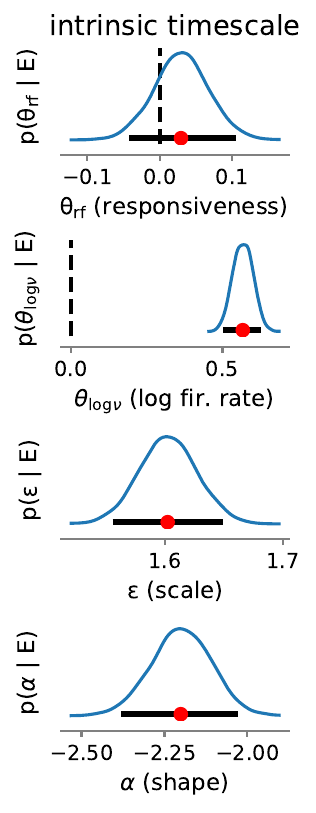}
    \end{minipage} \hfill     
    \begin{minipage}[t]{.31\textwidth}
    \vspace{0pt}
    \centering
     \includegraphics[width=\textwidth]{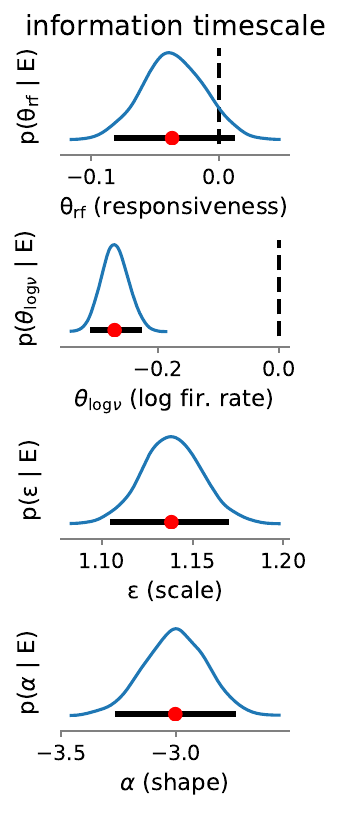}
    \end{minipage} \hfill     
    \begin{minipage}[t]{.31\textwidth}
    \vspace{0pt}
    \centering
    \includegraphics[width=\textwidth]{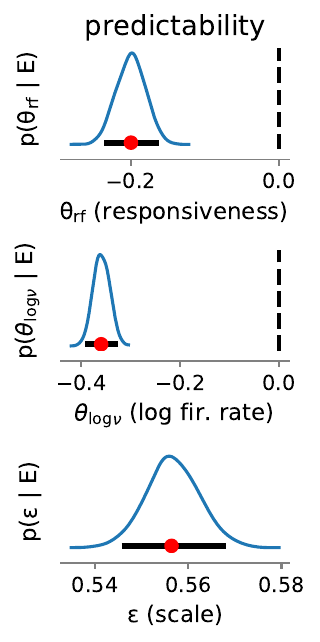}
    \end{minipage} 
    \end{subfigure} \hfill   
      \begin{subfigure}[t]{.32\textwidth}
    \subcaption[]{}
    \centering
    {\tiny \textbf{Functional Connectivity (spontaneous)}} \\
    \vspace{-5pt}
    \begin{minipage}[t]{.31\textwidth}
    \vspace{0pt}
    \centering
    \includegraphics[width=\textwidth]{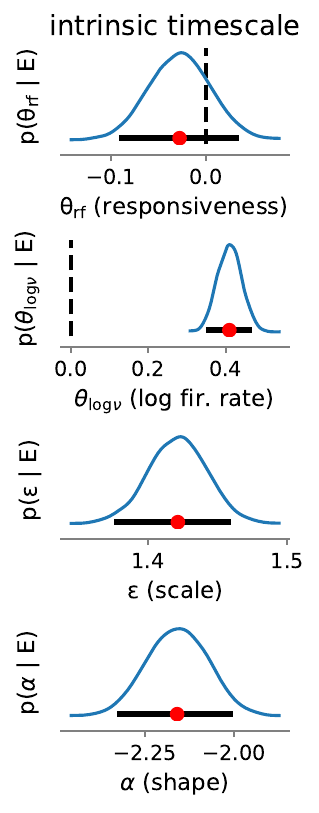}
    \end{minipage} \hfill     
    \begin{minipage}[t]{.31\textwidth}
    \vspace{0pt}
    \centering
     \includegraphics[width=\textwidth]{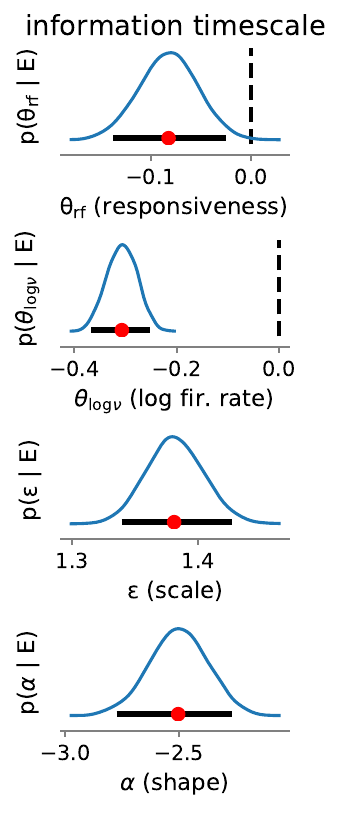}
    \end{minipage} \hfill     
    \begin{minipage}[t]{.31\textwidth}
    \vspace{0pt}
    \centering
    \includegraphics[width=\textwidth]{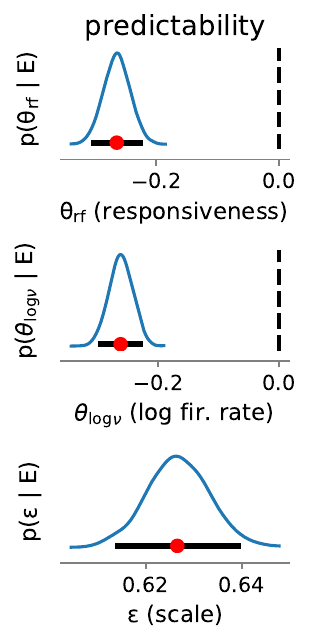}
    \end{minipage} 
    \end{subfigure} \hfill   
    % \begin{subfigure}[t]{.32\textwidth}
    % \subcaption[]{}
    % \vspace{-9pt}
    %     \includegraphics[width=\textwidth]{SI_figs/allen_hierarchical_bayes_hierarchy_score_posterior_nm_tau_R.pdf}
    % \end{subfigure} \hfill   
    % \begin{subfigure}[t]{.32\textwidth}
    % \subcaption[]{}
    % \vspace{-9pt}
    % \includegraphics[width=.9\textwidth]{SI_figs/allen_hierarchical_bayes_hierarchy_score_posterior_nm_R_tot.pdf}
    % \end{subfigure}
    \caption{\textbf{Posteriors for non-hierarchical parameters of the cortical groups model.} \\
    Posterior densities for non-hierarchical parameters of the cortical groups model are very similar to the hierachy score model (c.f.~Supplementary Fig.~\ref{fig:bayes_hierarchy_score_model_posterior}).
    }
    \label{fig:bayes_structure_groups_model_posterior} 
     \captionlistentry[suppfigure]{\textbf{Fig S2. Relation between intrinsic and information timescales, as well as predictability across all sorted units.}}
\end{figure}

\begin{figure}
\centering
    \textbf{Functional Connectivity (natural movie)} \\
    \begin{subfigure}[t]{.49\textwidth}
    \subcaption[]{\quad {\small \textbf{cortical hierarchy model}}}
    \centering
    \includegraphics[width=\textwidth]{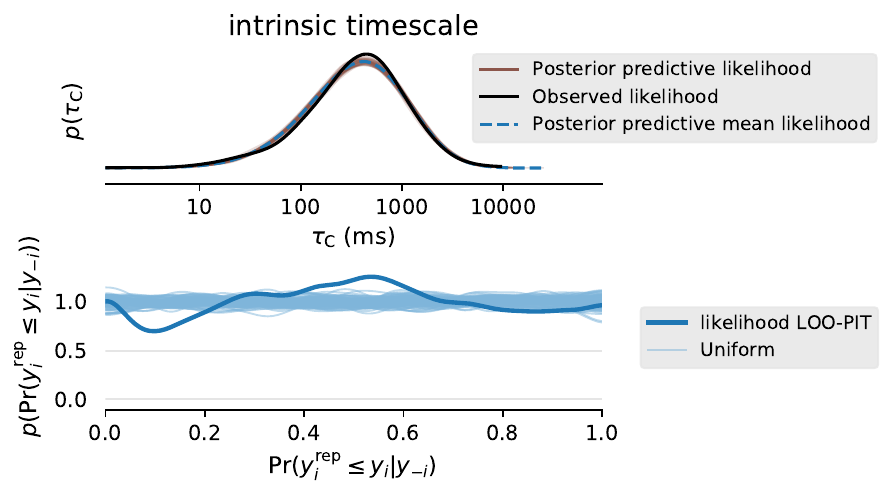}
    \includegraphics[width=\textwidth]{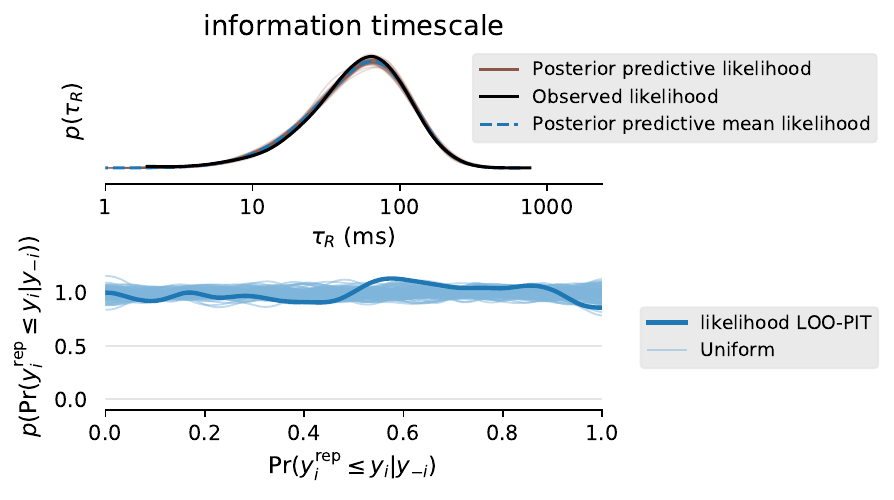}
    \includegraphics[width=\textwidth]{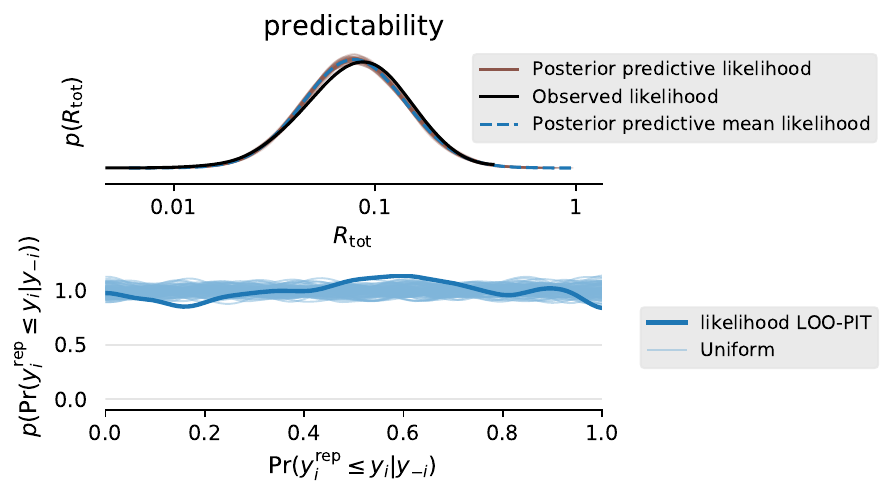}
    \end{subfigure}   \hfill 
    \begin{subfigure}[t]{.49\textwidth}
    \subcaption[]{\quad {\small \textbf{cortical groups model}}}
    \centering
    \includegraphics[width=\textwidth]{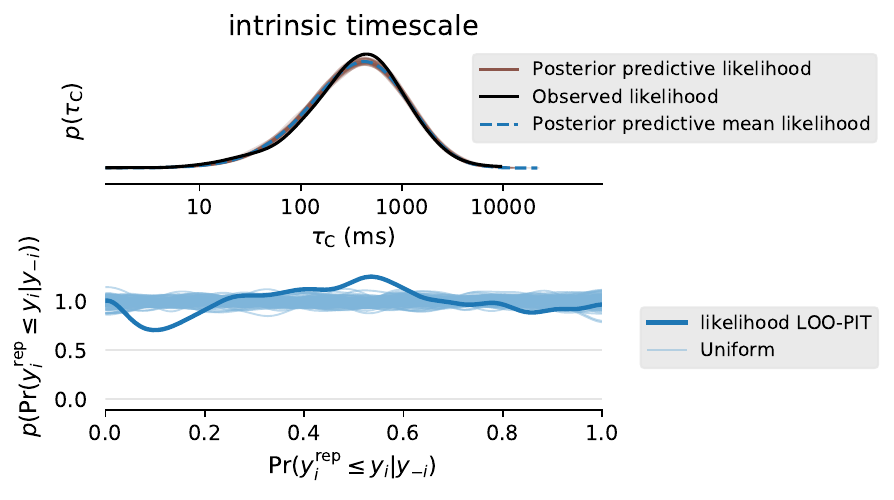}
    \includegraphics[width=\textwidth]{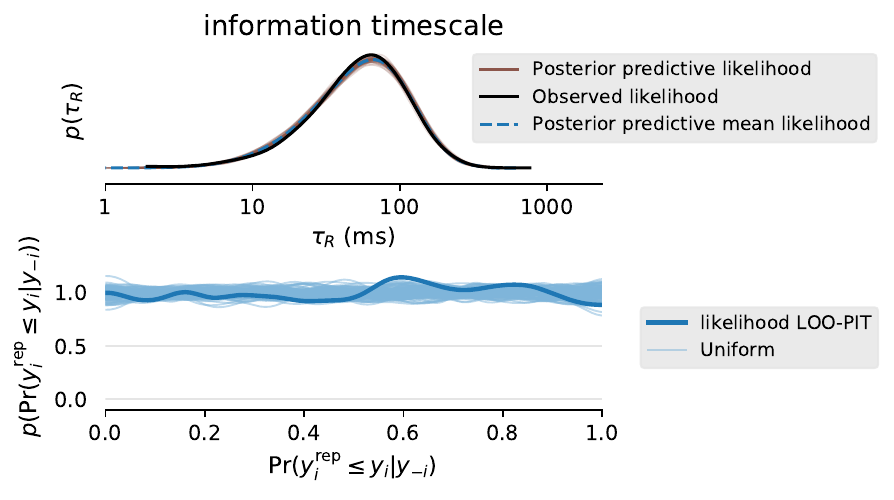}
    \includegraphics[width=\textwidth]{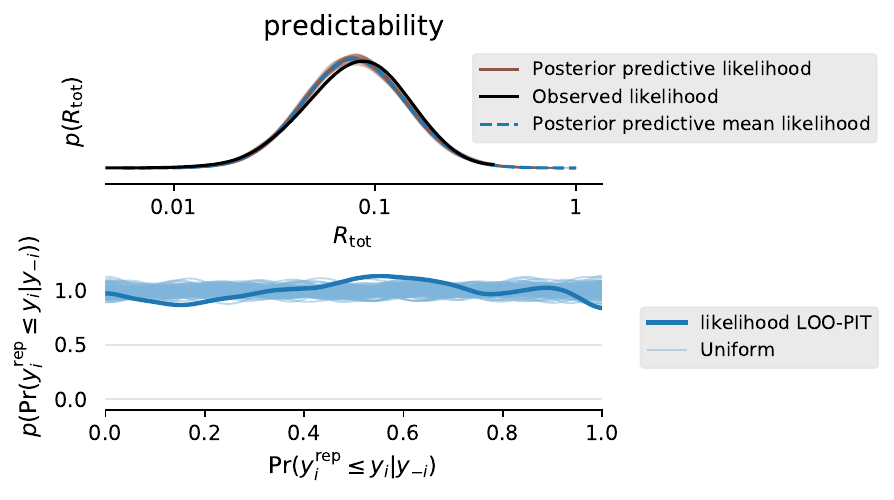}
    \end{subfigure} 
    \caption{\textbf{Posterior predictive checks of the different Bayesian models applied to the \emph{natural movie} condition in the \emph{Functional Connectivity} data set.}
    \textbf{(A)} To test whether the cortical hierarchy model is well calibrated, we compared the posterior predictive likelihood of the model (brown line) with the observed likelihood of timescales and predictability (black line), which overall show a good agreement. Note the skewness in the distribution of log timescales, which led us to use a skew normal distribution to model the residual variability in the timescale data. 
    To obtain a better check for the conditional probabilities of individual data points, and not only the pooled data, we performed LOO cross-validated probability integral transform (PIT) posterior predictive checks~\cite{gelman_2013}. In LOO-PIT,  the model is fitted for each datum $y_i$ to all data except $y_i$, here denoted as $\mathbf{y}^{-i}$. $\mathrm{Pr}(y_i^{\mathrm{model}} \leq y_i |\mathbf{y}^{-i})$ then represents the probability that a value $y_i^{\mathrm{model}}$ simulated from the fitted model is less or equal to $y_i$. If the model and data distributions are the same, then the distribution of these probabilities over all data points $y_i$ (thick blue line) should be uniform~\cite{gelman_2013}, hence we compare it to 100 simulated data sets from a uniform distribution (thin blue lines). The model appears well calibrated for the information timescale and predictability, whereas for the intrinsic timescale the model tends to slightly over-represent intermediate values, and to under-represent smaller values. 
    \textbf{(B)} Same as A, but for the cortical groups model. Notably, the models appear to be equally well calibrated, hence we do not expect the differences in their predictive power to be caused by a sub-optimal calibration of the models.}
% TODO: cite Gelman, A., Carlin, J. B., Stern, H. S., Dunson, D. B., Vehtari, A. and Rubin, D. B. (2013) Marginal predictive checks. In Bayesian Data Analysis, 3rd edn, ch. 6. Boca Raton: Chapman and Hall–CRC.
    \label{fig:bayes_predictive_checks} 
     \captionlistentry[suppfigure]{\textbf{Fig S2. Relation between intrinsic and information timescales, as well as predictability across all sorted units.}}
\end{figure}
% \clearpage

\begin{figure}
\centering
    \textbf{Brain Observatory 1.1 (natural movie)} \\
    \begin{subfigure}[t]{.49\textwidth}
     \subcaption[]{\quad {\small \textbf{cortical hierarchy model}}}
    \centering
    \includegraphics[width=\textwidth]{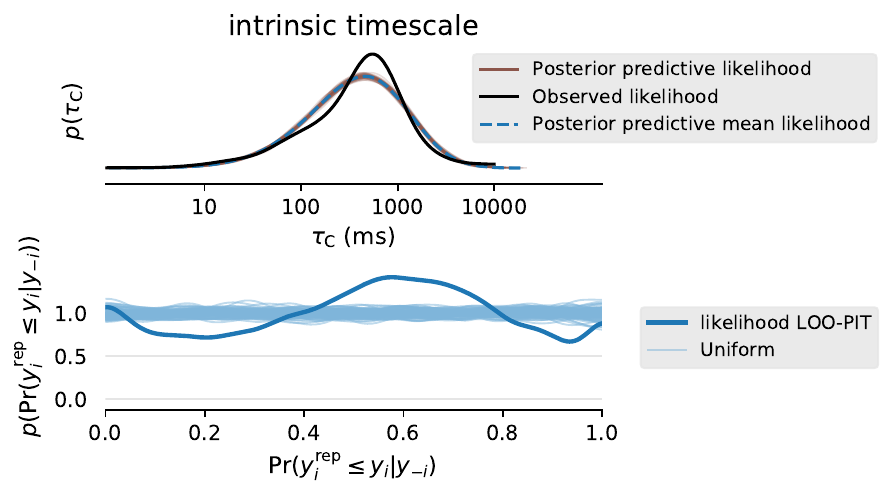}
    \includegraphics[width=\textwidth]{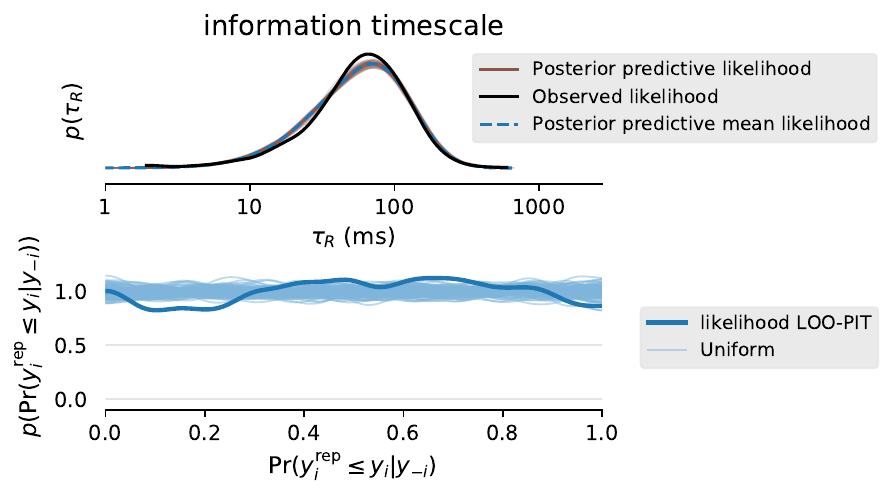}
    \includegraphics[width=\textwidth]{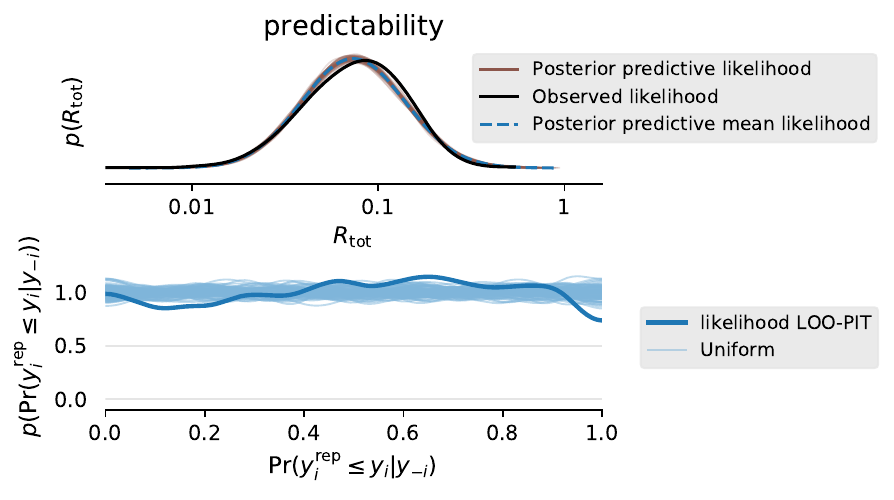}
    \end{subfigure}   \hfill 
    \begin{subfigure}[t]{.49\textwidth}
    \subcaption[]{\quad {\small \textbf{cortical groups model}}}
    \centering
    \includegraphics[width=\textwidth]{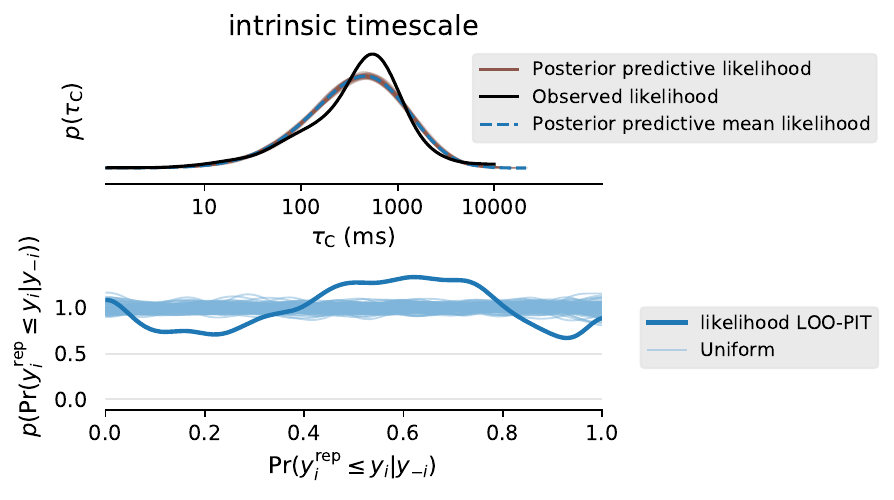}
    \includegraphics[width=\textwidth]{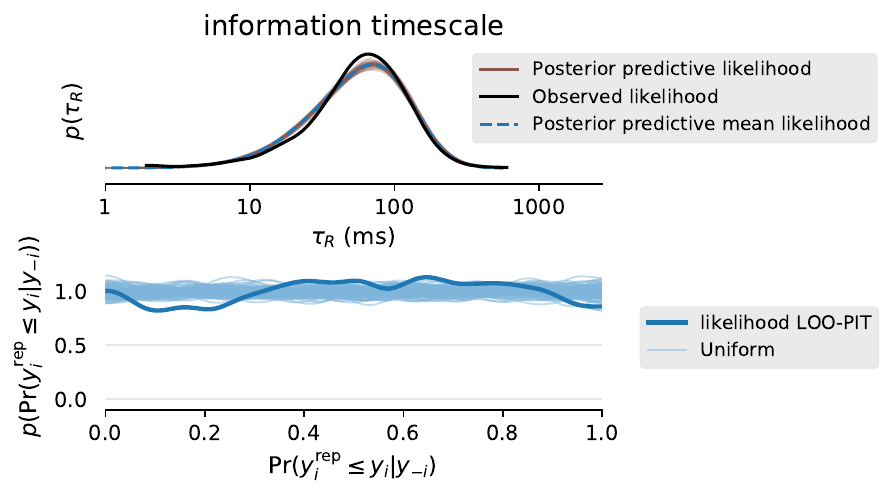}
    \includegraphics[width=\textwidth]{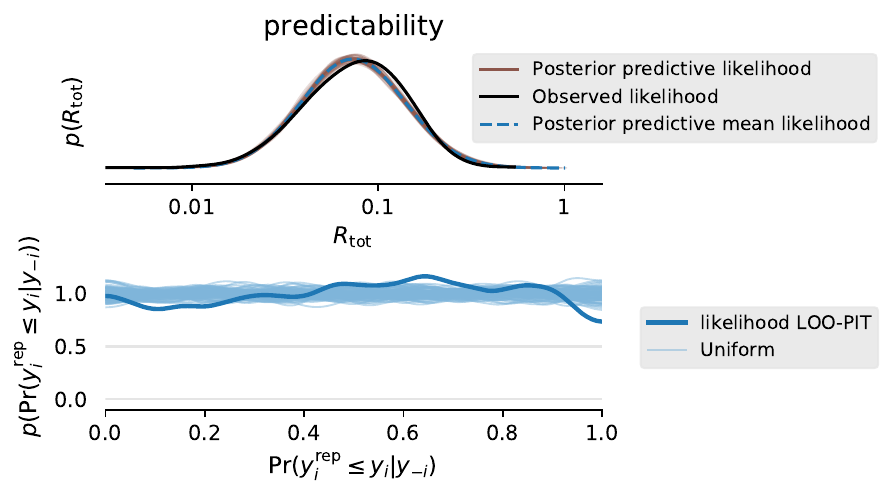}
    \end{subfigure} 
    \caption{\textbf{Posterior predictive checks of the different hierarchical models applied to the natural movie condition in the \emph{Brain Observatory} data set.}
    Same as Fig.~\ref{fig:bayes_predictive_checks}, but for the natural movie condition in the \emph{Brain Observatory 1.1} data set. For these data, the model is slightly worse calibrated for the intrinsic timescale.}
    \label{fig:bayes_predictive_checks_bo_nm} 
     \captionlistentry[suppfigure]{\textbf{Fig S2. Relation between intrinsic and information timescales, as well as predictability across all sorted units.}}
\end{figure}
% \clearpage

\begin{figure}
\centering
    \textbf{Functional Connectivity (spontaneous activity)} \\
    \begin{subfigure}[t]{.49\textwidth}
    \subcaption[]{\quad {\small \textbf{cortical hierarchy model}}}
    \centering
    \includegraphics[width=\textwidth]{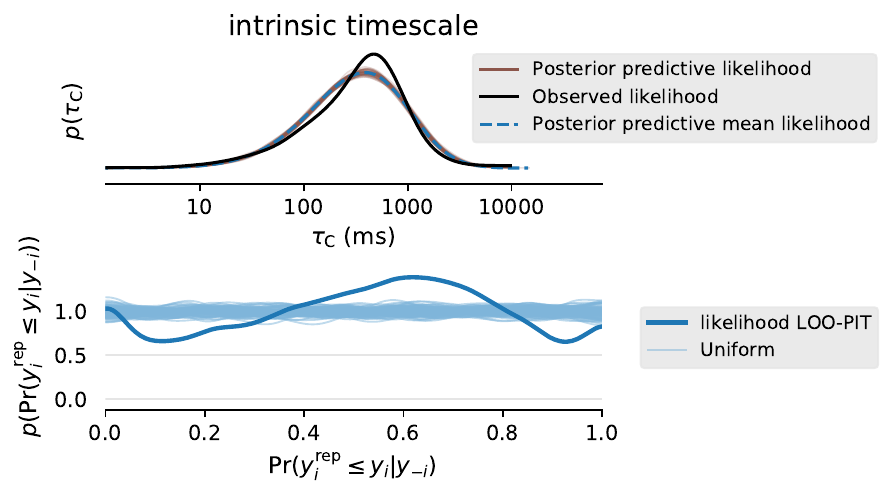}
    \includegraphics[width=\textwidth]{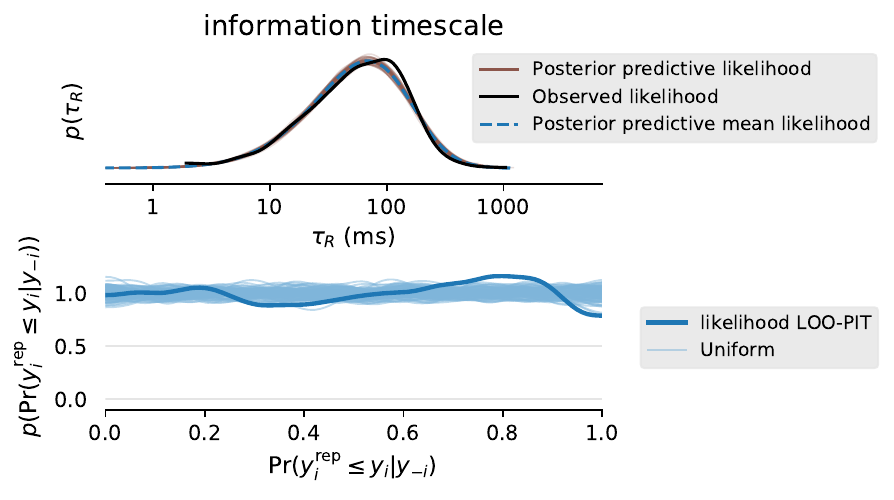}
    \includegraphics[width=\textwidth]{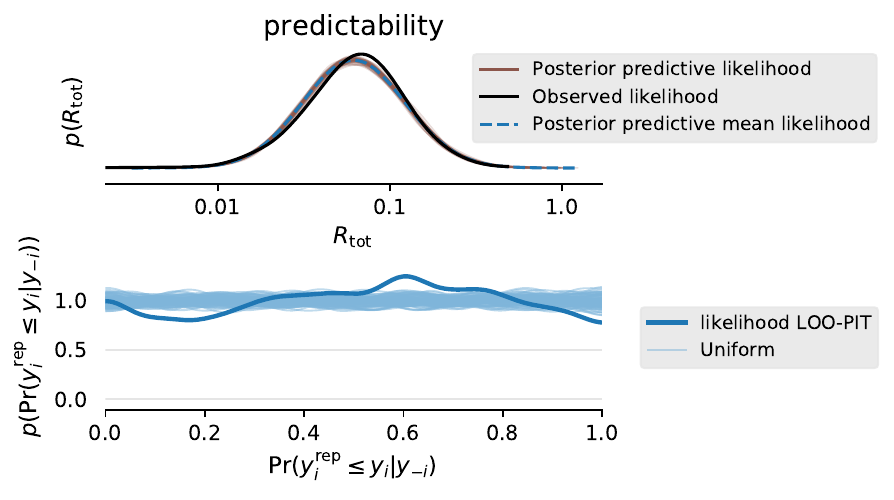}
    \end{subfigure}   \hfill 
    \begin{subfigure}[t]{.49\textwidth}
   \subcaption[]{\quad {\small \textbf{cortical groups model}}}
    \centering
    \includegraphics[width=\textwidth]{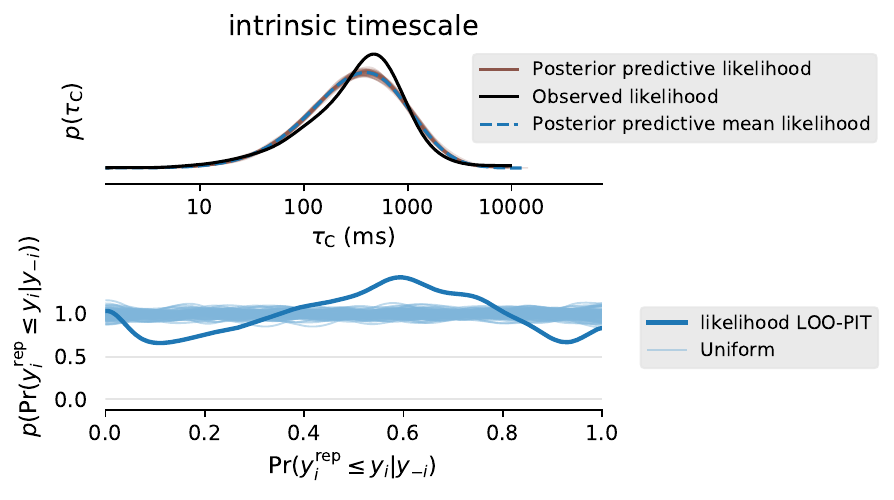}
    \includegraphics[width=\textwidth]{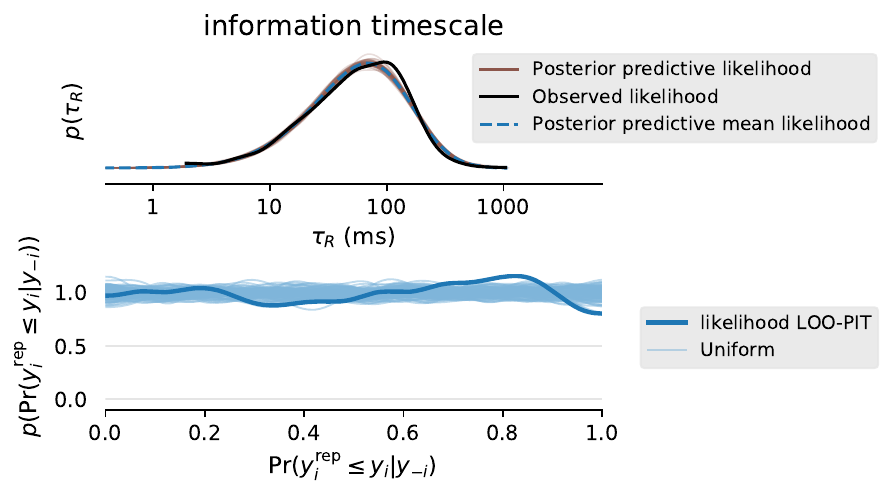}
    \includegraphics[width=\textwidth]{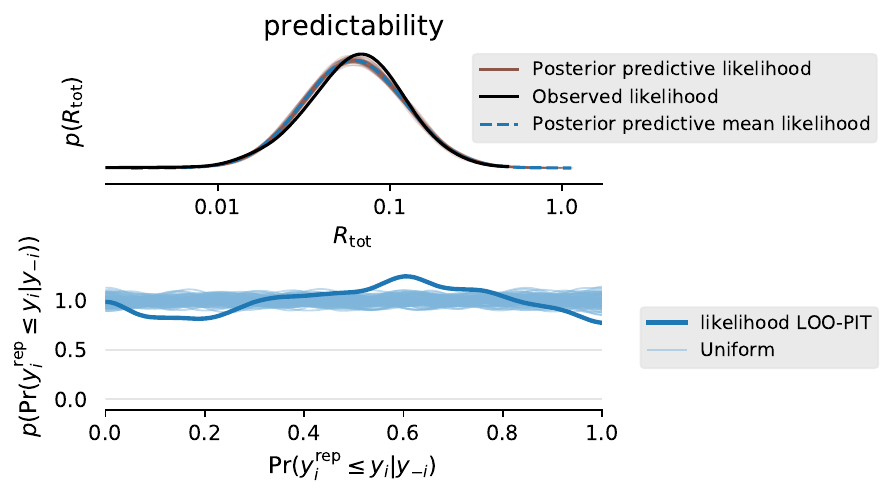}
    \end{subfigure} 
    \caption{\textbf{Posterior predictive checks of the different hierarchical models applied to \emph{spontaneous activity} in the \emph{Functional Connectivity} data set.}
    Same as Fig.~\ref{fig:bayes_predictive_checks}, but for spontaneous activity in the \emph{Functional Connectivity} data set. For these data, the model is slightly worse calibrated, in particular for the intrinsic timescale.}
    \label{fig:bayes_predictive_checks_sp} 
     \captionlistentry[suppfigure]{\textbf{Fig S2. Relation between intrinsic and information timescales, as well as predictability across all sorted units.}}
\end{figure}

\begin{figure}
    \centering
    \includegraphics[width = .9\textwidth]{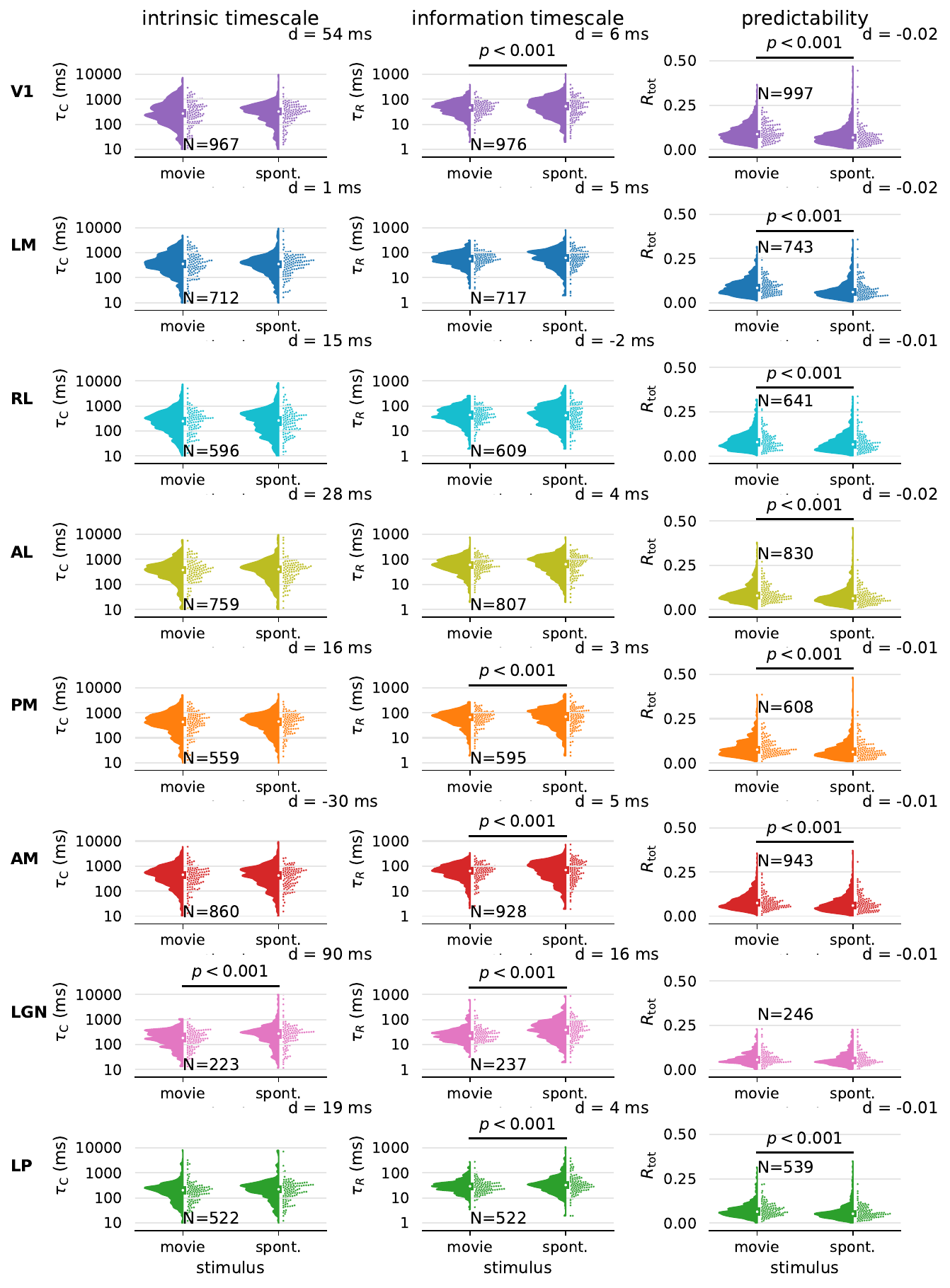}
    \caption{\textbf{Area-wise comparison of timescales and predictability between natural movie and spontaneous activity in the \emph{Functional Connectivity} data set.} 
    We compared for individual areas whether there was a systematic difference in timescales and predictability between stimulation with a natural movie and spontaneous activity with grey screen illumination.
    Both the intrinsic timescales and information timescales tend to be higher for spontaneous activity ($d$ gives difference in medians).
    The predictability is lower for spontaneous activity, with significant differences for all cortical areas (p-values from two-sided Mann-Whitney U tests, only significant differences after Bonferroni correction are shown.)
    }
    \label{fig:nm_vs_spontaneous_areas}
     \captionlistentry[suppfigure]{\textbf{Fig S2. Relation between intrinsic and information timescales, as well as predictability across all sorted units.}}
\end{figure}
% \clearpage

% \begin{figure}
%     \centering
%     \includegraphics[width = .96\textwidth]{SI_figs/allen_violin_per_structure_rf.pdf}
%     \caption{\textbf{Intrinsic timescale is larger and predictability smaller for cortical units that have a significant receptive field.}
% We compared for different areas whether there was a systematic difference in timescales and predictability between units with a receptive field in the area subject to visual stimulation (rf), and units for which no significant receptive field could be found (no rf).  
% While the intrinsic timescale is generally higher for units with a receptive field ($d$ gives difference in medians, p-values from two-sided Mann-Whitney U tests), the opposite holds for the predictability, which, for cortical areas, is higher for units without a receptive field.
% No conclusive results are found for the information timescale, nor for thalamic areas LGN and LP for any of the measures. 
% }
%     \label{fig:rf_vs_no_rf}
% \end{figure}

\begin{figure}
    \centering
    \includegraphics[width = .9\textwidth]{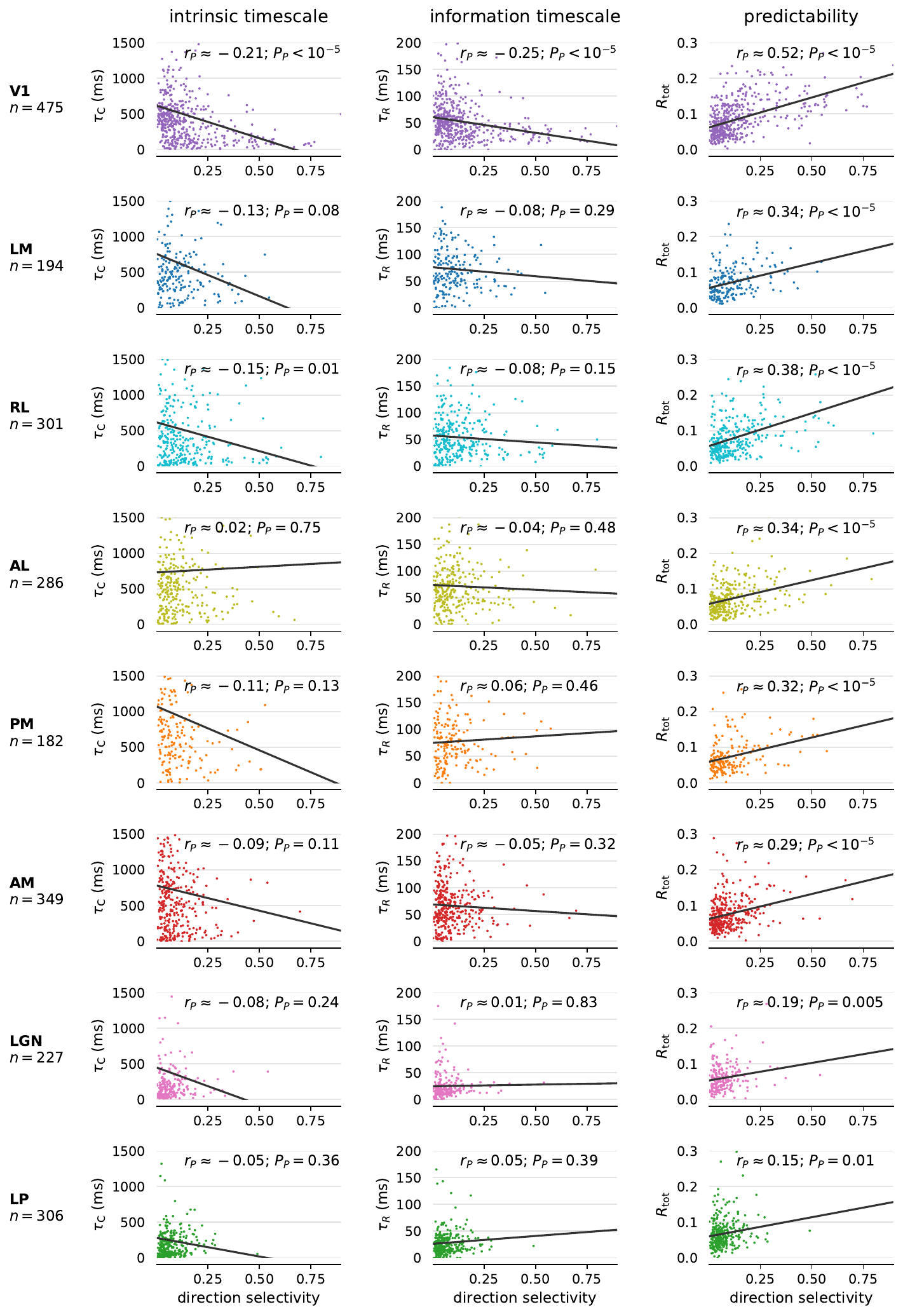}
    \caption{\textbf{Relation of timescales and predictability to direction selectivity of individual units for different visual areas.} Units with higher direction selectivity (measured on drifting gratings shown in 8 different directions~\cite{siegle_2021}) have lower intrinsic and information timescale, and higher predictability (linear regression line shown in black, $r_P$ gives Pearson correlation coefficient with corresponding two sided p-value $P_P$). The effect is most prominent in cortical areas (especially area V1), and for the intrinsic timescale as well as predictability, whereas results are more mixed for the information timescale.}
    \label{fig:measures_vs_direction_selectivity_per_area}
     \captionlistentry[suppfigure]{\textbf{Fig S2. Relation between intrinsic and information timescales, as well as predictability across all sorted units.}}
\end{figure}
% \clearpage

\begin{figure}
    \centering
     \begin{subfigure}[t]{.32\textwidth}
    \subcaption[]{}
    \includegraphics[width=.94\textwidth]{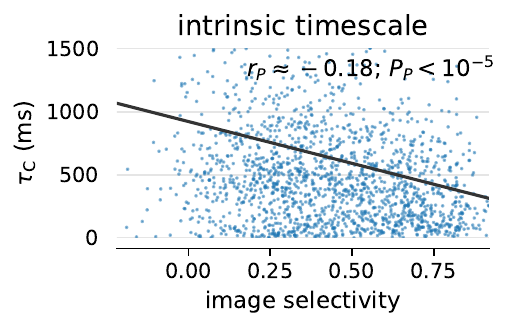}
    \end{subfigure}   \hfill 
    \begin{subfigure}[t]{.32\textwidth}
    \subcaption[]{}
    \includegraphics[width=.93\textwidth]{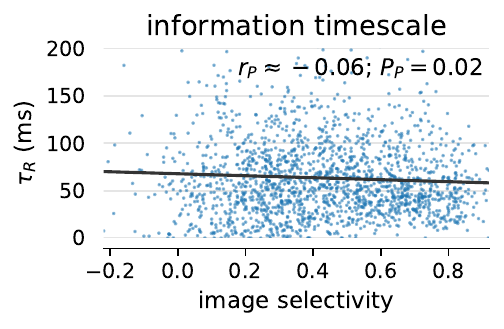}
    \end{subfigure} \hfill   
    \begin{subfigure}[t]{.32\textwidth}
    \subcaption[]{}
    \includegraphics[width=.91\textwidth]{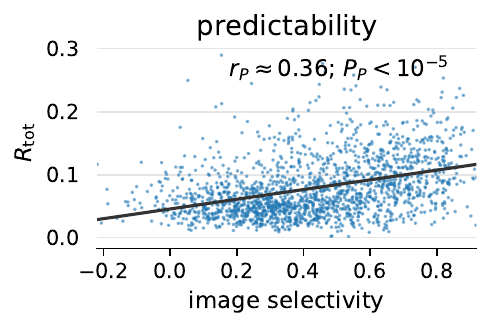}
    \end{subfigure}
    \caption{\textbf{Relation of timescales and predictability to image selectivity of individual units for all cortical areas.} 
    \textbf{(A)}~Intrinsic timescales $\tau_C$ under natural stimulation in the \textit{Brain Observatory 1.1} data set versus the image selectivity index (measured for different static images shown to the mice~\cite{siegle_2021}) of individual cortical units (blue dots). Timescales $\tau_C$ show a weak negative correlation with image selectivity index (linear regression line shown in black, $r_P$ gives Pearson correlation coefficient with corresponding two sided p-value $P_P$).
    \textbf{(B)}~Information timescales $\tau_R$ are very weakly negatively correlated with image selectivity index.
    \textbf{(C)}~In contrast, the predictability $R_{\text{tot}}$ is positively correlated with the image selectivity index index. }
    \label{fig:measures_vs_image_selectivity_all_cortical}
     \captionlistentry[suppfigure]{\textbf{Fig S2. Relation between intrinsic and information timescales, as well as predictability across all sorted units.}}
\end{figure}
% \clearpage

\begin{figure}
    \centering
    \includegraphics[width = .9\textwidth]{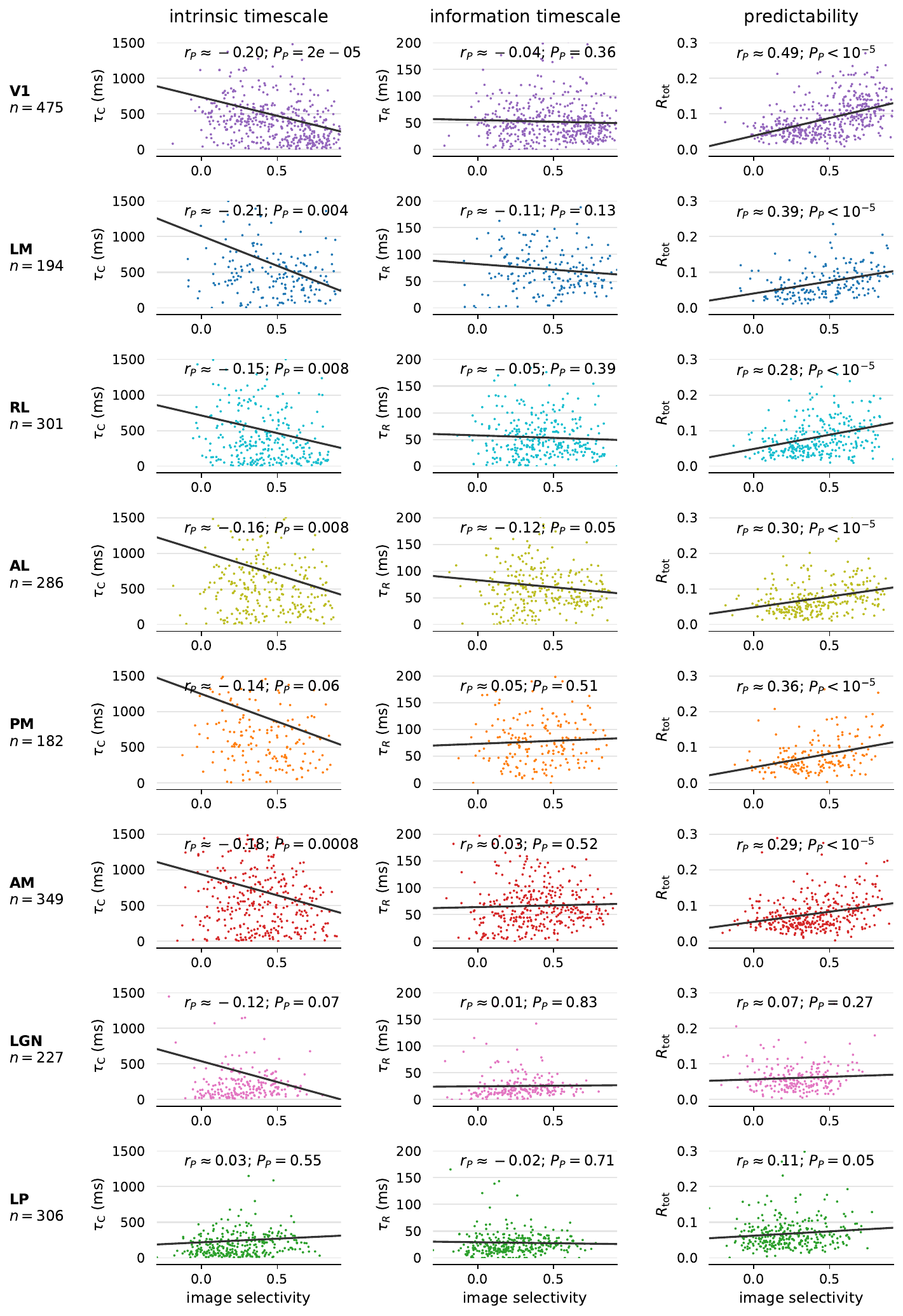}
    \caption{\textbf{Relation of timescales and predictability to image selectivity of individual units for different visual areas.} Units from cortical areas with higher image selectivity (measured for different static images shown to the mice~\cite{siegle_2021}) have lower intrinsic timescale, and higher predictability (linear regression line shown in black, $r_P$ gives Pearson correlation coefficient with corresponding two sided p-value $P_P$). The effect is most prominent in area V1 and strongest for the predictability, whereas no significant correlation is found for the information timescale or thalamic areas.}
    \label{fig:measures_vs_image_selectivity_per_area}
     \captionlistentry[suppfigure]{\textbf{Fig S2. Relation between intrinsic and information timescales, as well as predictability across all sorted units.}}
\end{figure}

\begin{figure}
    \centering
    \includegraphics[width  = .9\textwidth]{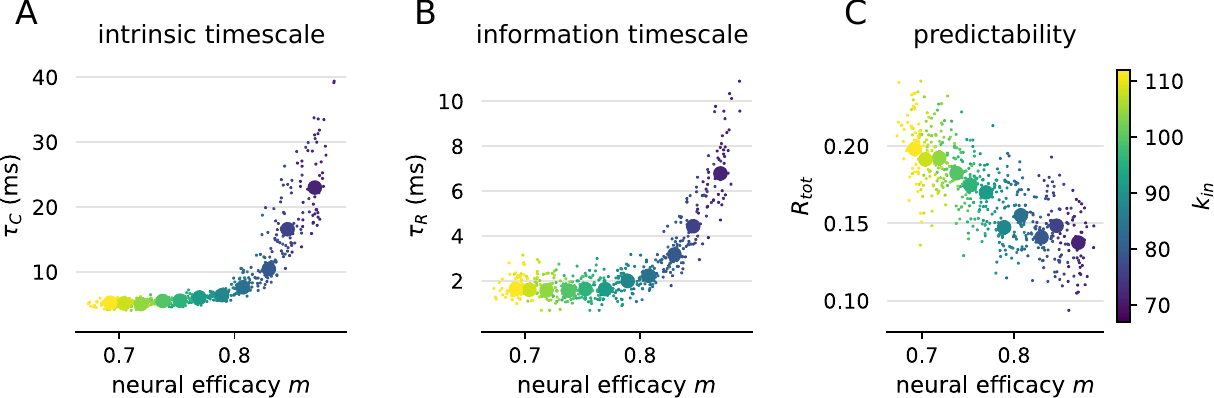}
    \vspace{1em}
    \caption{
    \textbf{Timescales and predictability in a plastic recurrent LIF network implemented on neuromorphic hardware.}
    To compare results from the branching network with a more realistic recurrent network, we consider an implementation of a network of $N=512$ leaky integrate-and-fire neurons with synaptic plasticity on neuromorphic hardware (BrainScaleS-2)~\cite{cramer_2020, pehle2022brainscales}.
    In this implementation, the recurrent amplification (expressed via neural efficacy $m=1-1/a$) cannot be set directly, but is tuned via plasticity, which adapts to the number of input synapses $k_{in}$ from which each unit receives Poisson input. 
    In particular, it has been shown that for less external input, the spike-timing-dependent plasticity tunes the network towards configurations with stronger recurrent coupling, and better integration for complex tasks~\cite{cramer_2020}. To quantify the effective strength of recurrent coupling, we estimated the neural efficacy $m$ via autoregression of the activity time series.
    As in the branching network, an increase of recurrence (here expressed through $m$, and shown for a smaller range) increases $\tau_C$ and $\tau_R$, but decreases $R_{\rm tot}$.
    Notably, in this model the source of single-neuron predictability besides recurrence is not provided through temporal correlations in the input, but by the membrane dynamics, effectively causing single-unit memory and predictability, which then gets diminished by increasing recurrence. Small dots show median values for individual network realizations, and big dots indicate median values over all network realizations for a given $k_{in}$.
    }
    \label{fig:measures_from_neuromrophic}
     \captionlistentry[suppfigure]{\textbf{Fig S2. Relation between intrinsic and information timescales, as well as predictability across all sorted units.}}
\end{figure}

\end{document}

% \caption{\textbf{Signatures of temporal processing for the emulation of a plastic, leaky integrate-and-fire network on a neuromorphic chip.} The intrinsic timescale was computed with fitting interval $[\,\text{ms} ,750 \,\text{ms}]$, and the information timescale with $\tau_R = 10\,\text{ms}$.}

posterior hierarchy groups 
hierarchy score model → Functional Connectivity, natural movie
hierarchy score model → Functional Connectivity, spontaneous activity
hierarchy score model → Brain Observatory, natural movie

posterior cortical groups
cortical groups model → Functional Connectivity, natural movie
cortical groups model → Brain Observatory, natural movie

Posterior
hierarchy score model → Functional Connectivity, natural movie
hierarchy score model → Functional Connectivity, spontaneous activity
hierarchy score model → Functional Connectivity, spontaneous activity
cortical groups model → Functional Connectivity, natural movie